\documentclass[twocolumn,showpacs,preprintnumbers,amsmath,amssymb,superscriptaddress,longbibliography,floatfix,prc,aps,10pt]{revtex4-2}
\usepackage{graphicx}
\usepackage{epsfig}
\usepackage{epstopdf}
\usepackage{dcolumn}
\usepackage{bm}
\usepackage{amsmath,amssymb,amsfonts}
\usepackage{color}
\usepackage[utf8]{inputenc}
\usepackage[T1]{fontenc}
\usepackage{indentfirst}
\usepackage{enumitem}
\usepackage{xspace}
\usepackage[resetlabels]{multibib}
\usepackage[pdfpagelabels]{hyperref}
\hypersetup{breaklinks=true,linktocpage=true,%
bookmarksnumbered=false,%
colorlinks=true,%
linkcolor=blue,%
citecolor=blue,%
urlcolor= blue}

\usepackage{orcidlink}
\usepackage{textcomp}

\newcommand{\disregard}[1]{}

\newcites{sup}{References}

\newcommand{\mytitle}{Electromagnetic moments of ground and excited states calculated in heavy {odd-\boldmath$N$\unboldmath} open-shell nuclei}
\usepackage{colortbl} 
\makeatletter
\def\maketitle{
\@author@finish
\title@column\titleblock@produce
\suppressfloats[t]}
\makeatother

\newcommand{\myabstract}{Within nuclear DFT, we calculated spectroscopic magnetic dipole and electric quadrupole moments for various quasiparticle configurations of odd-$N$, even-$Z$, $83\leq{}N\leq125$ nuclei ranging from gadolinium to osmium. By tagging the blocked quasiparticles with single-particle states of the semi-magic dysprosium isotope, we efficiently computed 22 prolate and 22 oblate states for each of the 154 nuclei and tracked them across the entire major neutron shell. We compared this extensive set of theoretical results with experimental data for 82 states in the region. Breaking rotational, time-reversal, and signature symmetries, we aligned the intrinsic angular momenta along the axis of axial symmetry, thereby enabling full shape- and spin-self-consistent polarizations. The spectroscopic moments were then obtained by restoring rotational symmetry. We conducted a detailed analysis of the pattern of agreement and disagreement between theory and experiment in individual nuclei. For the magnetic dipole moments, agreement with the data varies and is characterized by an overall average and RMS deviation of 0.11\,$\mu_N$ and 0.35\,$\mu_N$, respectively. For the electric quadrupole moments, a good corresponding agreement of 0.16\,b and 0.29\,b was observed.
}

\begin{document}

\preprint{}

\title{\protect\mytitle}

\author{J. Dobaczewski\orcidlink{0000-0002-4158-3770}}
\affiliation{School of Physics, Engineering and Technology, University of York, Heslington, York YO10 5DD, United Kingdom}
\affiliation{Institute of Theoretical Physics, Faculty of Physics, University of Warsaw, ul. Pasteura 5, PL-02-093 Warsaw, Poland}

\author{A.E. Stuchbery\orcidlink{0000-0002-0198-9901}}
\affiliation{Department of Nuclear Physics and Accelerator Applications, Research School of Physics, Australian National University, Canberra ACT 2601, Australia}

\author{G. Danneaux}
\affiliation{Department of Physics, University of Jyv{\"a}skyl{\"a}, PB 35(YFL) FIN-40014 Jyv{\"a}skyl{\"a}, Finland}

\author{A. Nagpal\orcidlink{0000-0001-9206-712X}}
\affiliation{School of Physics, Engineering and Technology, University of York, Heslington, York YO10 5DD, United Kingdom}

\author{P.L. Sassarini\orcidlink{0000-0003-3991-1613}}
\affiliation{School of Physics, Engineering and Technology, University of York, Heslington, York YO10 5DD, United Kingdom}

\author{H. Wibowo\orcidlink{0000-0003-4093-0600}}
\affiliation{School of Physics, Engineering and Technology, University of York, Heslington, York YO10 5DD, United Kingdom}

\date{\today}

\begin{abstract}
\myabstract
\end{abstract}

\maketitle

\section{Introduction}\label{Introduction}

Nuclear moments are fundamental characteristics of an individual nuclear state. The electric quadrupole moment gives a measure of its deformation (or shape). In contrast, the magnetic dipole moment is intimately related to the manner in which the nucleus carries its angular momentum, whether as a collective whole or as individual nucleons with contributions from both their orbital motion and their intrinsic spin~\cite{(Cas90d),(Ney03),(Yan23)}.

There are numerous theoretical approaches to evaluating nuclear moments. In many - if not most - model calculations, it is necessary to invoke effective charges in the calculation of electric quadrupole moments and effective orbital and spin $g$~factors in the evaluation of magnetic dipole moments. Moreover, most models are limited to certain classes of nuclei, for example, those near closed shells or those in deformed regions of the nuclear landscape well away from closed shells. Recent advances in nuclear Density Functional Theory (DFT) have enabled the calculation of magnetic dipole moments and electric quadrupole moments across broad regions of the nuclear chart without the need to invoke effective charges or effective $g$~factors. The focus here is on odd-$A$ nuclei, particularly in the ground state for which extensive data are available. (Both the dipole and quadrupole moments are trivially zero for the $I^{\pi} = 0^+$ ground states of even-even nuclei.)

The present work stems from a novel method we proposed~\cite{(Sas22c)} for calculating nuclear magnetic dipole moments in DFT by breaking the signature and time-reversal symmetries, thereby enabling the alignment of intrinsic angular momenta along the axis of axial symmetry before restoring good angular momentum. The first application of the method~\cite{(Sas22c)} analyzed the electromagnetic moments determined in 32 near-doubly magic nuclei. In subsequent publications~\cite{(Bon23c),(Wib25d)}, the methodology was expanded to paired open-shell systems and determined the electromagnetic moments of selected configurations within long isotopic chains of elements ranging from tin to lead.

In this study, we apply the method to excited quasiparticle configurations in deformed open-shell odd-$N$ isotopes of even-$Z$ elements ranging from gadolinium to osmium. This establishes the methodology for systematic calculations of nuclear electromagnetic moments across the nuclear chart. We calculate both the electric quadrupole moment and the magnetic dipole moment, which provide insight into the nucleus's collectivity (deformation) and the balance between the angular momentum carried by the odd neutron and the core. The focus of our work is on following specific single-particle (s.p.) structures from near spherical to strongly deformed nuclei. This allows for organizing the obtained results into distinct sequences of states that carry similar microscopic s.p.\ contents.

Our approach provides a new perspective on the structure of odd nuclei by examining how odd particles couple with even-even cores. Traditionally, modeling particle-core coupling relies on two contrasting schemes. In the so-called weak-coupling scheme, an odd spherical particle weakly influences a spherical core within a perturbative framework. In the simplest case, coupling the 0$^+$ spherical ground state with the odd particle in a s.p.\ energy level with angular momentum $j$ results in the unique energy level of the odd nucleus having total angular momentum $I=j$. Naturally, different magnetic substates $m$ of the s.p.\ energy level correspond to different magnetic substates $M=m$ of the odd nucleus, and all of these form a single energy level of the odd nucleus with its $2I+1$ degenerate magnetic substates.

In the so-called strong-coupling scheme, which applies to large deformations, each spherical s.p.\ energy level $j$ splits into $2j+1$ axially deformed s.p.\ states with projections of angular momentum $\Omega$ on the axial symmetry axis of $\Omega=-j,\ldots,j$. When coupled with the deformed core representing a rotational band, each deformed s.p.\ state in an odd nucleus leads to a rotational band of energy levels $I=K,K+1,K+2,\dots$ for $K=|\Omega|$, and each of these levels has its $2I+1$ degenerate magnetic substates. Since the rotational degrees of freedom are coupled with s.p.\ degrees of freedom here, there are many more deformed energy levels than spherical ones. Indeed, a unique spherical $I=j$ energy level evolves into a set of $j+1/2$ deformed band-head energy levels, each of which gives rise to a many-level rotational band. The transitional region with moderate deformations is poorly described by either the weak- or strong-coupling scheme.

Since the seminal work of Nilsson~\cite{(Nil55b)}, the strong-coupling scheme has been routinely employed in the phenomenological models based on using specific predefined mean-field potentials like those of Nilsson or Woods-Saxon~\cite{(Rin80)}. In these models, the electromagnetic properties were determined uniquely from the deformed s.p.\ energies and wave-functions, with the quadrupole deformation and electric quadrupole moments fixed by the so-called Strutinsky energy minimization~\cite{(Rin80)}. For the magnetic dipole moments, the collective effects were included in terms of one constant parameter $g_R$, and the angular-momentum core polarization was neglected. Both moments were determined in the intrinsic reference frame, and the rotational symmetry was not restored. Nevertheless, such phenomenological modeling was quite successful, allowing for a good systematic description of data, cf.\ e.g.\ a comprehensive analysis of the odd-$N$, even-$Z$ rare-earth nuclei in Ref.~\cite{(Ben89)}.

In the self-consistent DFT approach used in this work, we have a unique opportunity to cover both weak and strong-coupling schemes and to connect them smoothly. Starting from near semi-magic isotopes with tiny deformations, the nearly spherical self-consistent quasiparticles evolve into deformed ones as neutron numbers increase. The restoration of angular momentum enables us to observe how the unique $I=j$ states split and develop into distinct sets of $j+1/2$ deformed band heads. An inverse process occurs when these band-head sets recombine into nearly spherical, $I=j$ states at the following semi-magic isotopes.

The primary objective of this work is to examine how these configurations split, mix, and recombine across a major shell. This involves not only the ground or near-ground states but also a description of a broad range of excited states, which the methodology used here allows.

In the future, with the upcoming release of the latest version of the computer code {\sc hfodd}~\cite{(Dob21f),(Dob26a)} and detailed guidelines on its usage~\cite{(Res26)}, interested parties will be able to routinely perform such calculations for nuclei of any mass, deformation, or excitation energy.

The paper is organized as follows. Section~\ref{sect:previous-work} reviews previous DFT applications to the calculation of electromagnetic moments and shows the novel features of the present approach.
Section~\ref{sect:deformed} describes the methodology, presents detailed results for the Dy isotopes, and discusses the deformation and angular-momentum dependence on the example of $^{161}$Dy. Section~\ref{sect:deformed-cf-experiment} then follows with a thorough discussion and comparison with the experiment. Concluding remarks are made in Section~\ref{sect:conclusions}.



\section{Nuclear moments from DFT: previous work} \label{sect:previous-work}


\renewcommand{\arraystretch}{1.4}

\begin{table*}
\begin{center}
\scalebox{0.75}{
\begin{tabular}{|c|c|c|c|c|c|c|c|c|c|}
\cline{2-10}
\multicolumn{1}{c|}{}             & \textbf{Bally {\it et al.}}            & \textbf{Bonneau {\it et al.}} & \textbf{Co' {\it et al.}} & \textbf{Borrajo}             & \textbf{Li \& Meng} & \textbf{P\'{e}ru {\it et al.}} & \textbf{Sassarini {\it et al.}}          & \textbf{Ryssens {\it et al.}} & \textbf{Nakada}   \\
\multicolumn{1}{c|}{}             &                                       &                              &                          & \textbf{\& Egido}            &                     &                               &  \textbf{Bonnard {\it et al.}}           &                              & \textbf{\& Iwata} \\
\multicolumn{1}{c|}{}             &                                       &                              &                          &                              &                     &                               &  \textbf{Wibowo {\it et al.}}            &                              &  \\
\multicolumn{1}{c|}{}             & Refs.~\cite{(Bal14b),(Bal22a),(Bal23)} & Ref.~\cite{(Bon15)}          & Ref.~\cite{(Co15)}       & Refs.~\cite{(Bor16),(Bor17)} & Ref.~\cite{(Li18)}  & Ref.~\cite{(Per21a)}          & Refs.~\cite{(Sas22c),(Bon23c),(Wib25d)} & Ref.~\cite{(Rys22b)}         & Ref.~\cite{(Nak26)} \\
\hline
\textbf{Nuclei}                   & $^{25}$Mg,Xe                          & A$\approx$50,100             & Near doubly           & Mg isotopes                  & A$\approx$16,40     & Hg isotopes                   & Near doubly magic                    & In                           & Near doubly    \\
                                  & $^{197}$Au                            & A$\approx$178,236            & magic                    &                              & A$\approx$208       &                               & Open-shell Sn-Pb                        &                              & magic             \\
\hline
\textbf{HF}                       &                                       &                              & \checkmark               &                              & \checkmark          &                               & \checkmark                              &                              & \checkmark         \\
\textbf{HF+BCS}                   &                                       & \checkmark                   &                          &                              & \checkmark          &                               &                                         &                              &  \\
\textbf{HFB}                      & \checkmark                            &                              &                          & \checkmark                   &                     & \checkmark                    & \checkmark                              & \checkmark                   &  \\
\hline
\textbf{Single-particle Operator} & \checkmark                            & \checkmark                   & MEC                      & \checkmark                   & MEC                 & \checkmark                    & \checkmark                              & \checkmark                   & \checkmark        \\
\textbf{Effective $g$~factors}    &                                       & \checkmark                   &                          &                              &                     & \checkmark                    &                                         &                              &  \\
\textbf{Core contribution}        & Microscopic                           & Microscopic                  & Perturbative             & Microscopic                  & Perturbative        & Model                         & Microscopic                             & Microscopic                  & Microscopic       \\
\textbf{Collective Mixing (BMF)}  &                                       &                              &                          & \checkmark                   &                     &                               &                                         & \checkmark                   &  \\
\textbf{Blocking}                 & \checkmark                            & \checkmark                   &                          & \checkmark                   &                     & \checkmark                    & \checkmark                              & \checkmark                   &  \\
\hline
\textbf{Skyrme}                   & SLyMR0                                & SIII, SLyIII.0.8             &                          &                              &                     &                               & UNEDF1, SLy4, SkO$^\prime$              & BSkG                         &  \\
\textbf{Gogny}                    &                                       &                              & D1S, D1M                 & D1S                          &                     & D1M                           & D1S                                     &                              &  \\
\textbf{Yukawa}                   &                                       &                              &                          &                              &                     &                               &                                         &                              &  M3Y-P6           \\
\textbf{Regularized}              &                                       &                              &                          &                              &                     &                               & N$^{3}$LO                               &                              &  \\
\textbf{Relativistic Lagrangian}  &                                       &                              &                          &                              & PK1,PC-F1           &                               &                                         &                              &  \\
\textbf{Spin-spin}                & Native                               & Native                        & None                     & None                         & None                & None                          & $g'_0$ adjusted                         & Native                     & Native             \\
\hline
\textbf{HO Basis}                 &                                       & Cylindrical                  & Spherical                & Spherical                    & Spherical           & Deformed                      & 3D Cartesian                            &                              &  \\
\textbf{Oscillator Shells}        &                                       & 13                           &                          & 8                            & 10                  & 19                            & 16                                      &                              &  \\
\textbf{Gaussian Basis}           &                                       &                             &                           &                              &                    &                                &                                        &                              &  Spherical      \\
\textbf{Spatial coordinates}      & 3D Cartesian                          &                              &                          &                              &                     &                               &                                         &  3D Cartesian                &  \\
\hline
\textbf{Conserved Parity}         & \checkmark                            & \checkmark                   & \checkmark               & \checkmark                   & \checkmark          & \checkmark                    & \checkmark                              &  \checkmark                  & \checkmark        \\
\textbf{Conserved Signature}      & \checkmark                            &                              & \checkmark               & \checkmark                   & \checkmark          & \checkmark                    &                               & \checkmark                             &  \\
\textbf{Conserved Time reversal}  &                                       &                              & \checkmark               & \checkmark                   & \checkmark          & \checkmark                    &                                         &                              &                   \\
\hline
\textbf{Spherical States}         &                                       &                              & \checkmark               &                              & \checkmark          &                               &                                         &                              &  \\
\textbf{Axial States}             &                                       & \checkmark                   &                          &                              &                     & \checkmark                    & \checkmark                              &                              & \checkmark        \\
\textbf{Triaxial States}\footnote{See the original publications regarding the employed relationship between the conserved signature and triaxial shapes.}          & \checkmark                            &                              &                          & \checkmark                   &                     &                               &                                         &  \checkmark                  &  \\
\hline
\textbf{Reference Frame}          & AMP+PNP                               & Intrinsic                    & Laboratory               & AMP+PNP                      & Laboratory          & Intrinsic                     & AMP                                     & Intrinsic                    & Intrinsic         \\
\hline
\end{tabular}}
\end{center}
\caption{Tabular comparison of the differences of work done by other groups and our own, detailing the methods and symmetries imposed.
}
\label{tab:ProgressionTable}
\end{table*}

Table~\ref{tab:ProgressionTable} gives an overview of the work done by theoretical groups that have previously studied nuclear electromagnetic moments using different versions of nuclear DFT. Let us begin by briefly discussing the essential elements of previous works, which are listed at the top of Table~\ref{tab:ProgressionTable}, specifically, the conserved symmetries. The parity symmetry has been conserved in all applications so far, since the nuclei studied were not in regions of the nuclear chart where static octupole deformation occurs. This area of research lies ahead, but the initial applications are on the way~\cite{(Res25a)}.

The signature-symmetry operator is defined as a 180$^\circ$ rotation around one of the principal axes of the intrinsic shape. For axial nuclei, the signature with respect to the axis of axial symmetry is clearly not an independent one, and the only signature symmetry worth considering is the one with respect to the axis perpendicular to the axis of axial symmetry. Then, conservation of signature means that the angular momentum of an odd nucleus is aligned along that axis. In contrast, the angular momentum aligned along the axis of axial symmetry breaks the signature symmetry, because the signature operator then inverts the direction of the angular momentum.

As it turns out, the direction of the intrinsic angular momentum is a key feature of the intrinsic state (before symmetry restoration) that determines its magnetic properties. Therefore, the conservation or breaking of the signature symmetry is crucial. After the rotational symmetry is restored, the signature symmetry, which commutes with the angular momentum $\bm{I}$, becomes a function of the total angular momentum quantum number $I$ equal to $(-1)^I$ and defines the so-called signature-staggering effects.

The conservation or breaking of time-reversal symmetry is another crucial aspect of describing nuclear magnetic properties. Regardless of the angular momentum's alignment relative to the intrinsic shape, time reversal always inverts its direction. Therefore, before symmetry restoration, time-reversal symmetry cannot be conserved in odd nuclei. In nuclear DFT, it also plays a fundamental role for another reason. Indeed, only when time-reversal symmetry is broken does the time-odd mean-field sector of the functional activate, enabling it to induce a polarized angular momentum and a non-zero magnetic moment in the core.

Below, we summarize previous works, with the columns of Table~\ref{tab:ProgressionTable} arranged by the publication year of the cited references. Studies by Bally {\it et al.}~\cite{(Bal14b),(Bal22a),(Bal23)} employed multi-reference collective mixing of blocked triaxial Hartree-Fock-Bogoliubov (HFB) states that incorporate angular-momentum projection (AMP) and particle-number projection (PNP) for symmetry restoration~\cite{(She21)}, while conserving parity and signature symmetries and breaking time reversal. Similarly, in the paper by Ryssens {\it et al.} ~\cite{(Rys22b)}, the blocked triaxial HFB calculations were carried out in the intrinsic reference frame (without symmetry restoration). Both approaches involved microscopic core polarizations performed in 3D spatial coordinates.

Turning to Bonnaeu {\it et al.}~\cite{(Bon15)}, who considered a range of mass numbers from $A \approx 50$ to $A \approx 236$, their approach involved applying the Hartree-Fock (HF) plus BCS method with Skyrme interactions, incorporating conserved parity while allowing for broken time reversal and signature. They used axial deformations in their calculations. Similar to P{\'e}ru {\it et al.}~\cite{(Per21a)}, they employed effective spin $g$~factors ranging from 0.7 to 0.9 and blocked the lowest aligned $K=I$ states above the even-even core. They did not perform AMP in their work.

The paper by Co' {\it et al.}~\cite{(Co15)} used the HF method to build a spherical s.p.\ basis and imposed the parity, signature, and time-reversal symmetries. Using no effective $g$~factors, they relied on a residual interaction by implementing the random phase approximation. They also added meson-exchange corrections (MEC) to the magnetic dipole moments.

Borrajo and Egido~\cite{(Bor16),(Bor17)} studied Mg isotopes using the HFB approach with the Gogny D1S parameterization. In their calculations, they maintained symmetries such as parity, signature, and time reversal, and they employed triaxial deformations with blocking. They applied collective mixing through symmetry-conserving configuration mixing (SCCM) with AMP and PNP. An interesting hybrid method was used, involving PNP-VAP (variation after projection) followed by AMP-PAV (projection after variation).

In papers using covariant density functional theory (CDFT), see Li and Meng~\cite{(Li18)} and the works cited therein, the authors included first-order (spin polarization effect) and second-order (quadrupole polarization effect) configuration mixing along with MEC. This group limited their applications to spherical symmetry, which enabled them to carry out calculations in the laboratory reference frame that conserved angular momentum.

P{\'e}ru {\it et al.}~\cite{(Per21a)} obtained results for the Hg isotopes and utilized the HFB approach with the Gogny D1M parametrization. This work preserved parity, signature, and time-reversal symmetries while employing axial deformation. This group adopted a spin quenching factor (effective spin $g$~factor) of 0.75 to fit experimental data. They utilized the blocking of neutron states to accurately describe the ground states of the nuclei. Although no AMP was used, they corrected the magnetic dipole moments phenomenologically to account for the effects of symmetry restoration.

As noted in the introduction, our previous work~\cite{(Sas22c)} proposed a novel method for calculating nuclear electromagnetic moments by breaking rotational, signature, and time-reversal symmetries, thereby enabling the alignment of intrinsic angular momenta along the axis of axial symmetry before restoring good angular momentum; these crucial elements have never been implemented together before, as shown in Table~\ref{tab:ProgressionTable}. In our analyses~\cite{(Sas22c),(Bon23c),(Wib25d)}, we highlighted that nuclear DFT methods benefit from the ability to use a sufficiently large s.p.\ phase space and, therefore, should not be supplemented with effective charges or effective $g$~factors.

Recently, Nakada and Iwata~\cite{(Nak26)} calculated the magnetic dipole and electric quadrupole moments of near doubly magic nuclei using the HF method with the functional M3Y-P6 derived from the Yukawa interaction. They studied axial states with broken signature and time reversal, and without symmetry restoration.

\section{Theory}\label{sect:deformed}

\begin{figure}
\begin{center}
\includegraphics[width=0.48\textwidth]{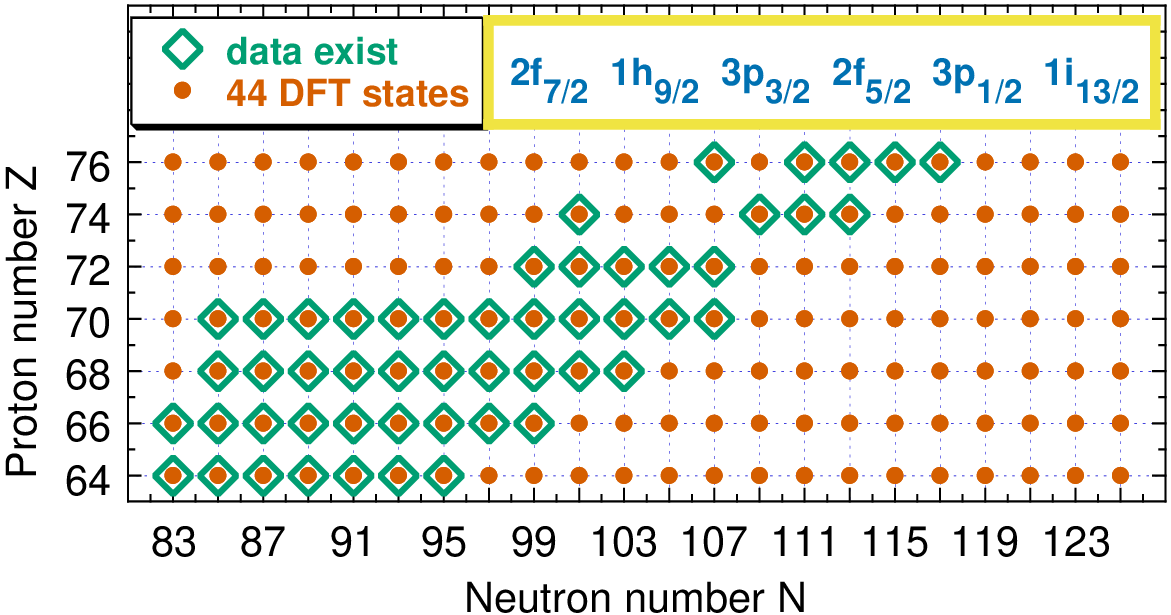}
\end{center}
\caption{Diagram illustrating the set of odd-$N$, even-$Z$ open-shell elements and isotopes considered in this study for which (i) the calculations were performed (dots) and (ii) experimental data exist (diamonds).
\label{Deformed_scheme}
}
\end{figure}

As illustrated in Fig.~\ref{Deformed_scheme}, we determined the electromagnetic moments of odd-$N$ isotopes of even-$Z$ elements, specifically between gadolinium and osmium. This includes 154 nuclei with $83\leq{N}\leq125$ and $64\leq{Z}\leq76$. For each nucleus, we obtained results for 22 prolate and 22 oblate states corresponding to the blocked quasiparticles state within the same neutron major shell of $83\leq{N}\leq125$. We performed all calculations for the Skyrme functional UNEDF1~\cite{(Kor12b)} using the computer code {\sc hfodd} (v3.33b)~\cite{(Dob21f),(Dob26a)}. Details of calculations remained identical to those previously described in Ref.~\cite{(Wib25d)} and will not be repeated here.

\subsection{Methodology}\label{Methodology}

In this work, we follow the methodology for determining electromagnetic moments in heavy deformed odd nuclei that we developed in Ref.~\cite{(Bon23c)} and detailed in Ref.~\cite{(Wib25d)}. Here, we briefly outline this methodology, highlighting its unique or novel aspects. In the Supplemental Material~\cite{supp-GdOs} and the raw data repository~\cite{rep-GdOs}, we provide the results database in both numerical and graphical formats.

The specific steps required to achieve the results outlined below were as follows:

\begin{figure}
\begin{center}
\includegraphics[width=0.48\textwidth]{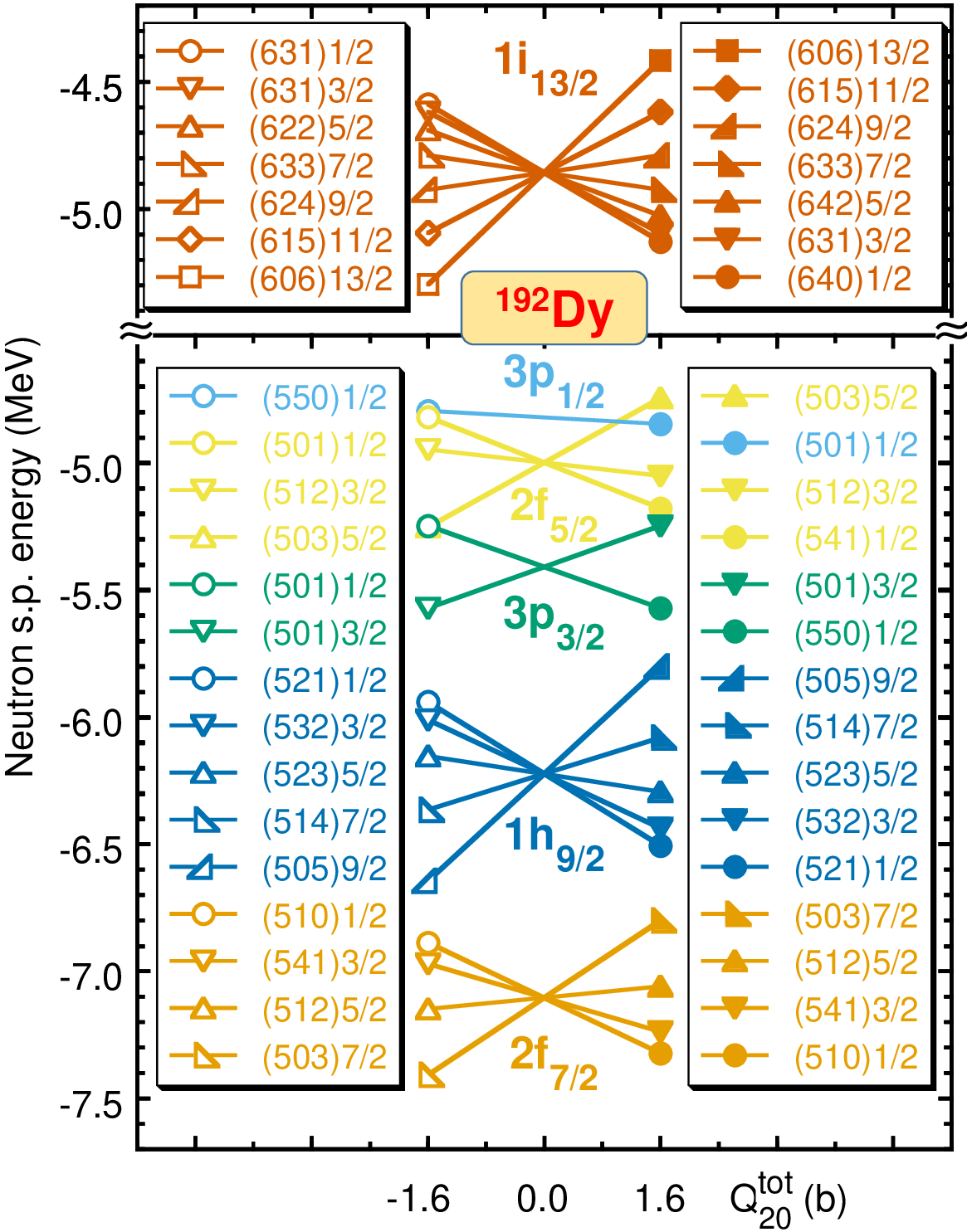}
\end{center}
\caption{(Colorblind-friendly palette~\protect\cite{(Won11)} online) Diagram illustrating the deformation splitting of neutron s.p.\ states in $^{192}$Dy, see the text for the discussion of that choice, which serves as a convenient theoretical starting point for proceeding to calculations of experimentally accessible nuclei. Calculations were carried out using constrained total axial intrinsic quadrupole moments $Q_{20}^{\text{tot}}=\pm0.4$\,b, Eq.~(\protect\ref{Qtot}), and then extrapolated to $\pm1.6$\,b for clear visualization of the splitting.
\label{dy192.spectrum}
}
\end{figure}

\begin{enumerate}
     \item We began by determining the self-consistent neutron s.p.\ energies and wavefunctions in the spherical semi-magic nucleus $^{192}$Dy. This yielded degenerate orbitals characterized by the standard spherical quantum numbers of 2f$_{7/2}$, 1h$_{9/2}$, 3p$_{3/2}$, 2f$_{5/2}$, 3p$_{1/2}$, and  1i$_{13/2}$, as shown by the central points in Fig.~\ref{dy192.spectrum}.

     The choice of $^{192}$Dy was convenient but not essential, even though it is an experimentally inaccessible isotope. Another exotic semi-magic isotope, such as $^{148}$Dy, a semi-magic isotope of a different element, or a state of any even-even nucleus in the region constrained to spherical shape, could have served the purpose just as well. The reason is that in any of those systems, the s.p.\ neutron states are very similar, and only their geometrical structure matters for the tagging mechanism mentioned below.

    \item Next, we determined the deformed and angular-momentum polarized s.p.\ states of $^{192}$Dy by applying small constraints on the total axial intrinsic quadrupole moments $Q_{20}^{\text{tot}}=\pm0.4$\,b,
    \begin{equation}
    Q_{20}^{\text{tot}}=Q^{\text{neu}}_{20}+Q^{\text{prot}}_{20},
    \label{Qtot}
    \end{equation}
    which is the sum of the neutron and proton contributions, along with a small cranking frequency of $\hbar\omega_z=1$\,keV and enforced $z$-axial symmetry. As shown by the side points in Fig.~\ref{dy192.spectrum}, each spherical orbital with total angular momentum $j$ splits into $2j+1$ deformed and aligned orbitals that have well-defined quantized angular momentum projections $\Omega$ along the axis of axial symmetry, ranging from $-j$ to $j$. On the scale of Fig.~\ref{dy192.spectrum}, small energy differences between pairs of states with $\pm{\Omega}$ are not visible. The 44 states with positive projections ($\Omega>0$), of which 22 were on the prolate side (prolate tag states) and 22 on the oblate side (oblate tag states), were recorded and used to define the quasiparticles for blocking. The tagging mechanism was explained thoroughly in Ref.~\cite{(Wib25d)}. Overall, at this stage, analyzing seven elements and 22 isotopes, this large-scale project required $44 \times 7 \times 22 = 6,776$ independent self-consistent calculations.

    We note that the set of tag states used in this work might not be enough to obtain quasiparticle configurations originating from the spherical s.p.\ states at $N<82$ or $N>126$, which, with increasing deformation, could become relevant within the studied nuclei.

    Further discussion of the results requires establishing a proper naming convention. To this end, we used the calculated dominant Nilsson labels of tag states, which represent their largest overlaps with the Nilsson states (the eigenstates of the axially deformed harmonic oscillator~\cite{(Rin80)}), denoted by $(N_0n_z\Lambda)K$, with $K=|\Omega|$. We denote the Nilsson labels of tag states by parentheses and reserve the standard notation of square brackets for the Nilsson labels of the self-consistent configurations, see below.

    The calculated Nilsson labels of the tag states are shown in Fig.~\ref{dy192.spectrum}. As one can see, for fixed projections $\Omega$, the Nilsson labels computed on the prolate and oblate sides often match, indicating that the structure of the related tag states is similar. Therefore, identifying the tag state requires indicating whether it is prolate or oblate. Additionally, in one case, the oblate-tag Nilsson label (501)1/2 appears to dominate two different states. Then, the tag states can only be distinguished by the context. We note that the Nilsson labels assigned to the tag states do not necessarily have to match those assigned to the self-consistent quasiparticle states in individual nuclei, which we discuss below.

    \item For 22 prolate and 22 oblate tag states in each studied nucleus, we performed blocked-quasiparticle calculations by applying constraints on the total axial intrinsic quadrupole moments of $Q_{20}^{\text{tot}}=+10$ and $-$10\,b, respectively, while keeping the constant pairing gaps at 1\,MeV. The goal of this intermediate step was to establish stable starting points for the next step.

    \item The final self-consistent calculations were conducted by removing the constraints from the previous step, which enabled the determination of self-consistent deformations and pairing correlations. At this step, 408 out of 6,776 calculations have not converged; see the discussion in the next point. However, as discussed in Ref.~\cite{(Wib25d)}, a smooth dependence of the electromagnetic moments on the neutron number often allows for a reasonably safe reconstruction of the non-converged points through interpolation.

    \item A more problematic situation arose in 233 other cases when identical solutions were obtained for either two different prolate tag states or two different oblate tag states. Such situations occur when two different tag states have the strongest affinity with the same quasiparticle, leading to duplicated solutions and causing one solution to be missed. To address this issue and find more converged solutions, we performed the second round of calculations with tag states defined not in $^{192}$Dy but in the specified nucleus. In this way, we managed to converge 162 more states and also find 177 new previously missed solutions, leaving only 246 non-converged and 56 missed solutions out of 6,776.

    We observed that the second round of calculations yielded fewer converged solutions and significantly more duplicates, confirming that running the $^{192}$Dy tag states first was the correct approach. We note that using the modified tag states in the second round required manually attributing the newly found solutions to the specific sequences of the $^{192}$Dy tag states used in the first round. We also note that cases where duplicated solutions are found for one oblate tag state and another prolate tag state are entirely normal and occur when, in a given nucleus, only one minimum—either prolate or oblate—exists.\label{point4}

    \item For all self-consistent solutions obtained, we performed angular momentum symmetry restoration (AMP)~\cite{(She21)}, which allowed us to determine the spectroscopic magnetic dipole and electric quadrupole moments, the main results analyzed in this work. At the same time, we determined the Nilsson labels $[N_0n_z\Lambda]K$ of the self-consistent configurations, denoted by the square brackets. These labels provide similar information about the deformed quasiparticle configurations as those determined in various phenomenological models, see, e.g, Ref.~\cite{(Fir99a)}.

\end{enumerate}

\subsection{DFT results for dysprosium isotopes}

In Figs.~\ref{fig:Q2,mu_Dy_prol_2}--\ref{fig:Spect_log_Dy}, we present the complete set of results obtained for dysprosium isotopes. For other elements, the corresponding figures are included in the Supplemental Material~\cite{supp-GdOs}. Additionally, the raw numerical data collected in this work are available in~\cite{rep-GdOs}.

As it turns out, presenting the results organized in lines that connect points corresponding to given tag states is hugely beneficial. In this way, one can easily follow specific structures of quasiparticle states across the entire shell and identify particular points where those structures change, irrespective of their excitation energies over the ground states.

\begin{figure*}
\begin{center}
\includegraphics[width=0.98\textwidth]{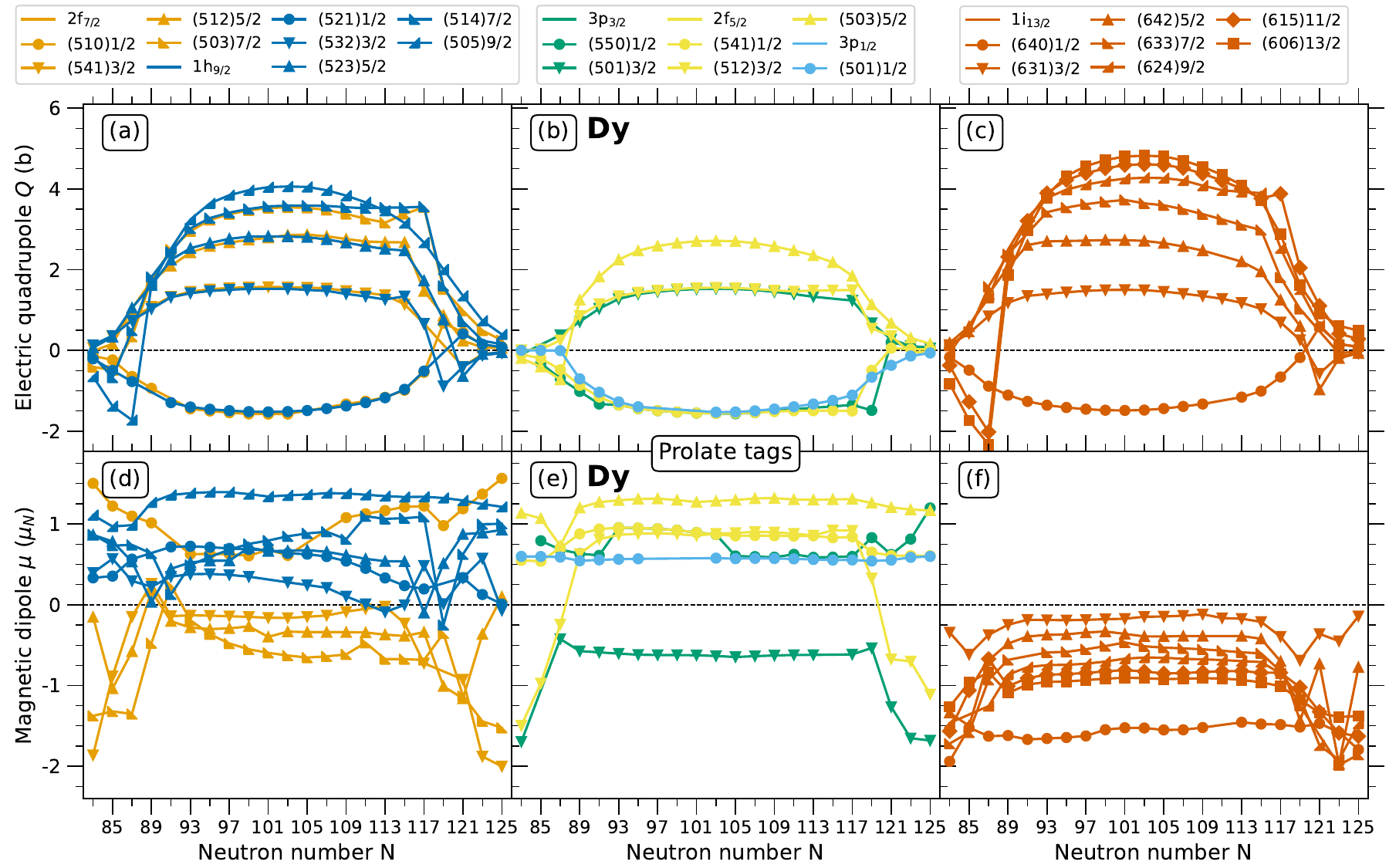}\vspace*{-7mm}
\end{center}
\caption{Upper panels (a), (b), and (c) show the DFT spectroscopic electric quadrupole moments $Q$ (in barn). Lower panels (d), (e), and (f) show the DFT spectroscopic magnetic dipole moments $\mu$ (in $\mu_N$).
Left panels (a) and (d) correspond to tag states originating from the spherical orbitals 2f$_{7/2}$ and 1h$_{9/2}$, middle panels (b) and (e) to those of 3p$_{3/2}$, 2f$_{5/2}$, and 3p$_{1/2}$, and right panels (c) and (f) to those of 1i$_{13/2}$. The figure displays all results determined for prolate tag states in dysprosium isotopes (full symbols). Note that for the $\Omega=1/2$ states, we plotted the values of $Q$ corresponding to the $I=3/2$ members of the rotational bands.
\label{fig:Q2,mu_Dy_prol_2}
}
\end{figure*}
\begin{figure*}
\begin{center}
\includegraphics[width=0.98\textwidth]{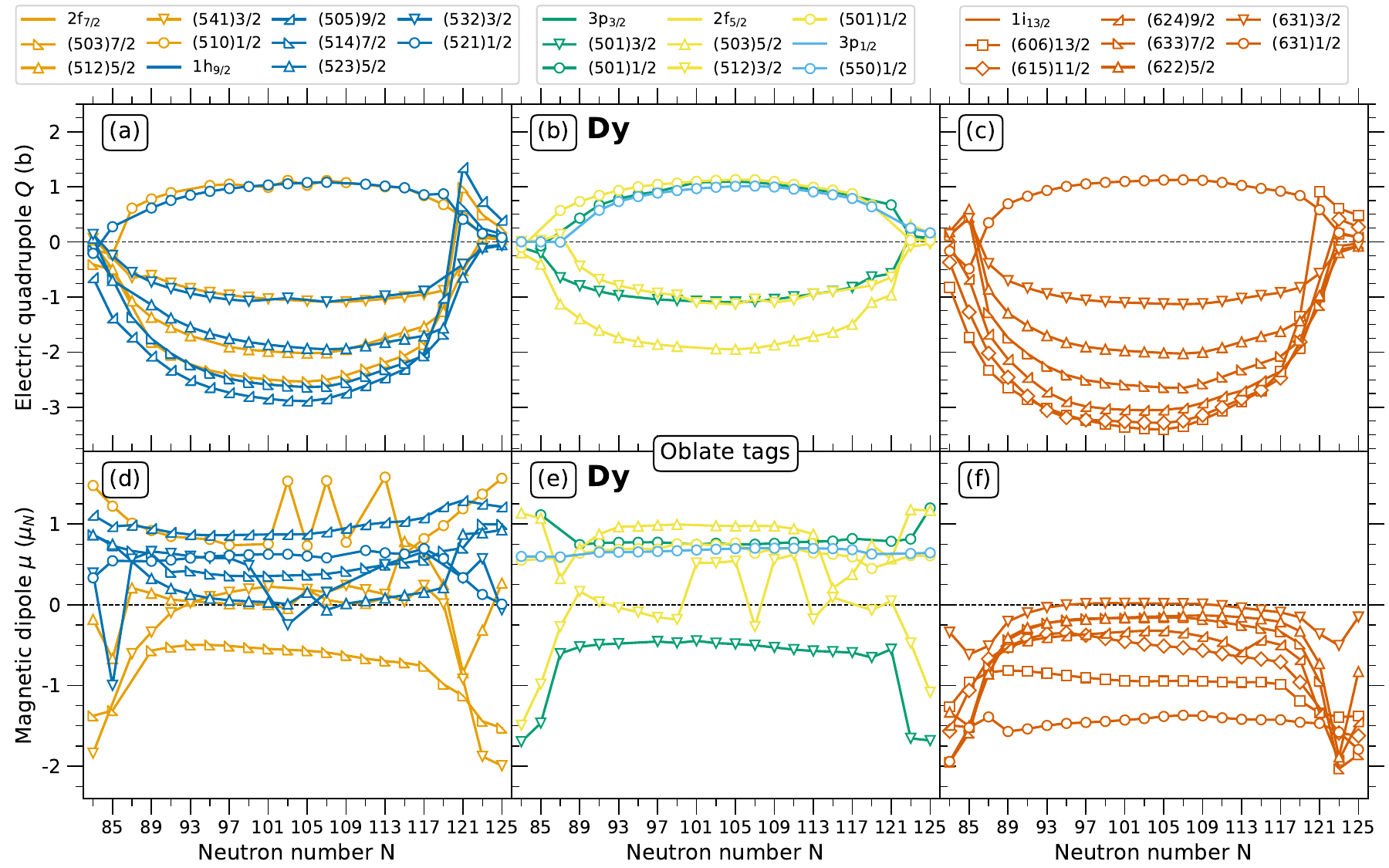}\vspace*{-7mm}
\end{center}
\caption{Same as illustrated in Fig.~\protect\ref{fig:Q2,mu_Dy_prol_2}, but showing the results obtained for oblate tag states, as indicated by the open symbols.
\label{fig:Q2,mu_Dy_obl_2}
}
\end{figure*}

\begin{figure*}
\begin{center}
\includegraphics[width=0.98\textwidth]{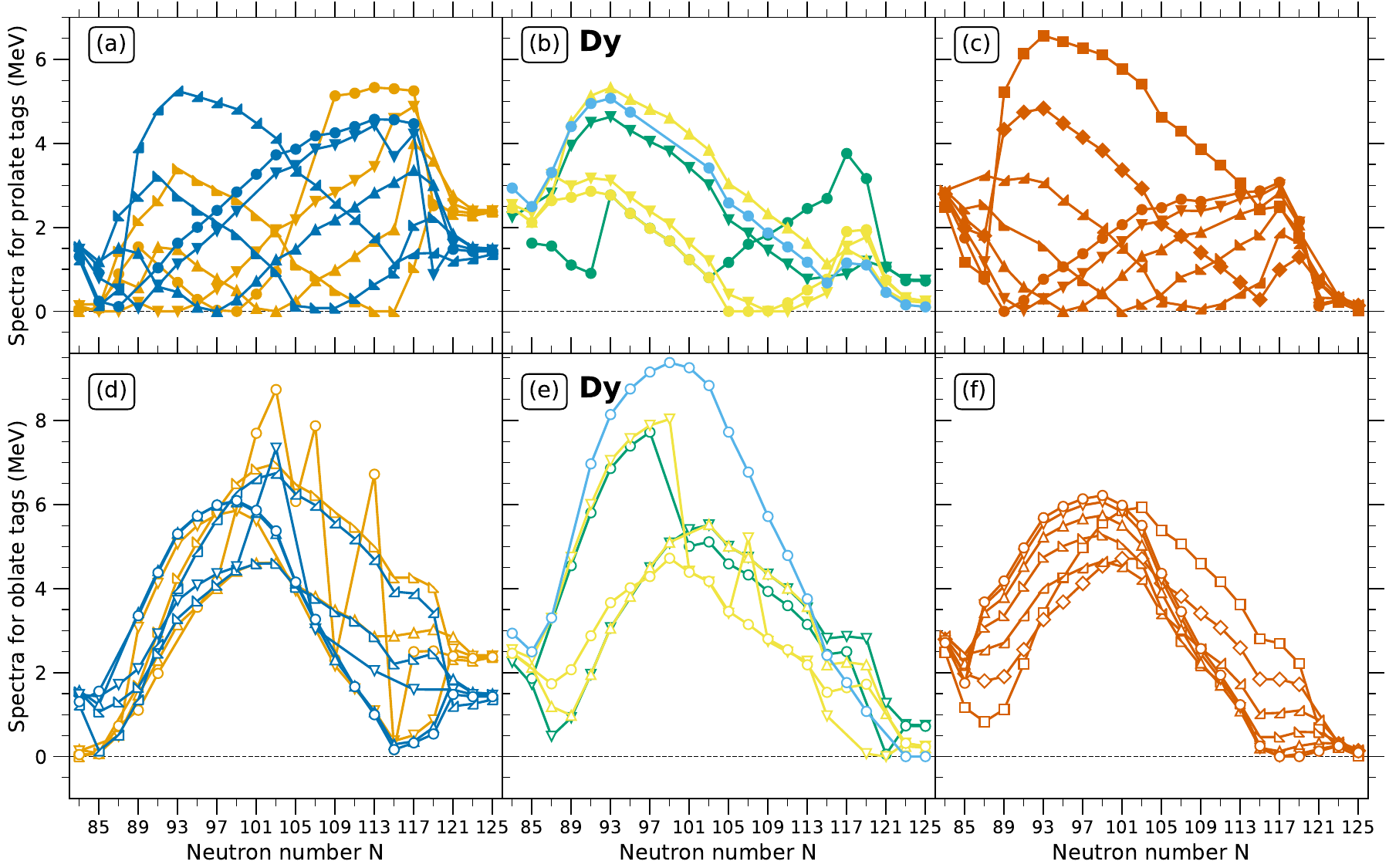}\vspace*{-7mm}
\end{center}
\caption{Excitation energies $E_{\text{exc}}$ of states in dysprosium isotopes calculated for the prolate tag states, upper panels (a), (b), and (c) and oblate tag states, lower panels (c), (d), and (f).
Legends of symbols are shown in Figs.~\protect\ref{fig:Q2,mu_Dy_prol_2} and~\protect\ref{fig:Q2,mu_Dy_obl_2}.
\label{fig:Spect_Dy}
}
\end{figure*}
\begin{figure*}
\begin{center}
\includegraphics[width=0.98\textwidth]{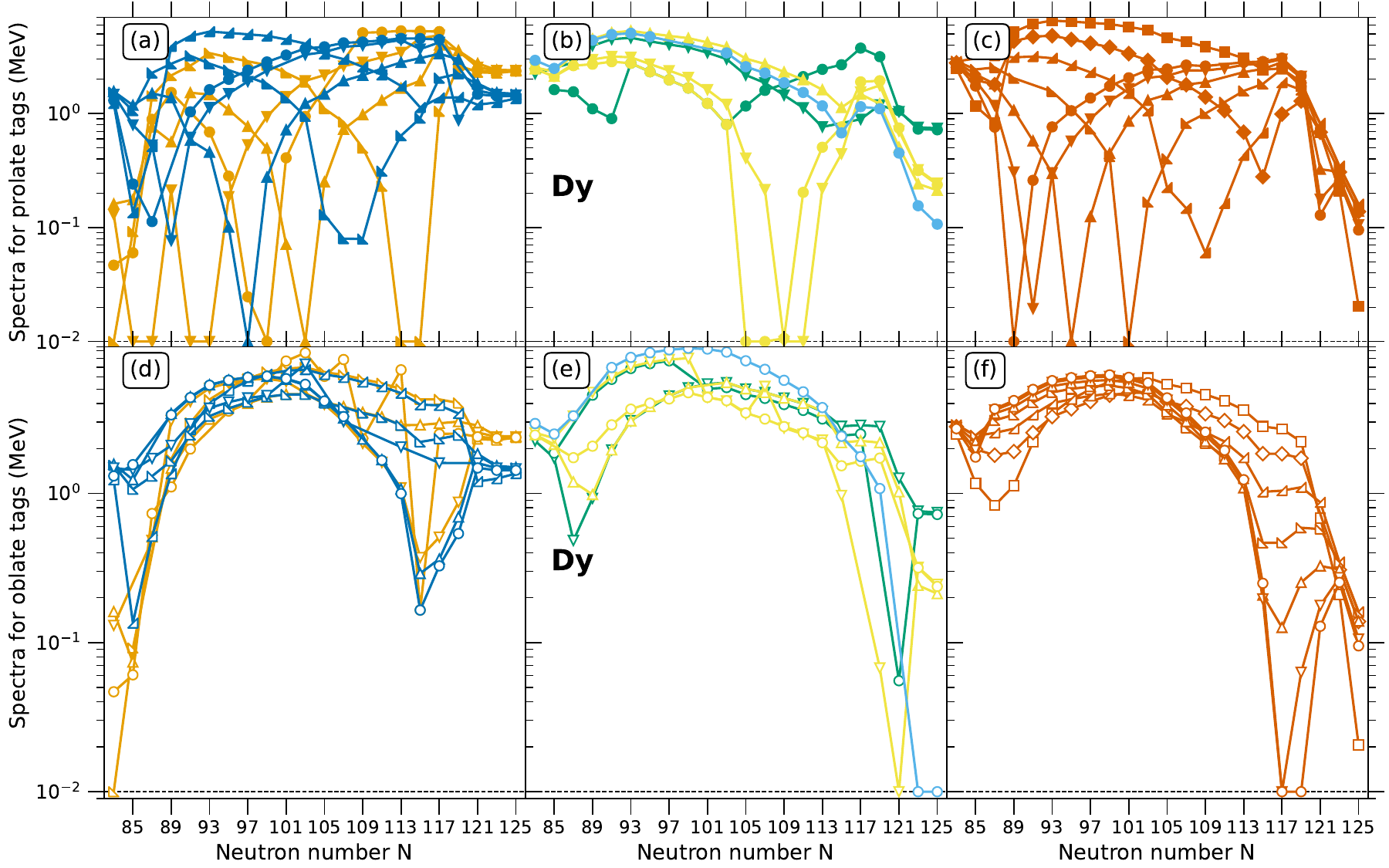}\vspace*{-7mm}
\end{center}
\caption{Same as in Fig.~\protect\ref{fig:Spect_Dy} but plotted in a logarithmic scale with  $E_{\text{exc}}=0$ (ground states) plotted artificially at $E_{\text{exc}}=0.01$\,MeV.
\label{fig:Spect_log_Dy}
}
\end{figure*}

For the prolate tag states, as illustrated in Fig.~\ref{fig:Q2,mu_Dy_prol_2}, we present the calculated spectroscopic electric quadrupole moments (top panels) and magnetic dipole moments (bottom panels). To improve the figure's readability, the results are organized into three columns, each displaying a different group of tag states as indicated in the legend. In addition, seven distinct shapes of symbols correspond to various angular momenta $I=1/2,\ldots,13/2$. In this figure, all symbols are filled, whereas the corresponding open symbols in Fig.~\ref{fig:Q2,mu_Dy_obl_2} represent results obtained for the oblate tag states. The same shapes of filled and open symbols are also displayed in Fig.~\ref{dy192.spectrum}.

The first and most striking observation is that in the majority of cases, the configurations defined by the tag states determined in spherical $^{192}$Dy smoothly extend across the large-deformation region of open-shell isotopes. This indicates that the affinity of deformed quasiparticles with nearly spherical s.p.\ states is quite strong and can indeed be employed to track configurations in seemingly diverse systems.

The missing points along the lines in Figs.~\ref{fig:Q2,mu_Dy_prol_2}-\ref{fig:Spect_log_Dy} correspond to the non-converged cases mentioned in point~\ref{point4}.\ of Sec.~\ref{Methodology}. For example, results regarding the non-converged solutions associated with the prolate tag states (640)1/2 and (642)5/2 at $N=111$ can be easily interpolated from those obtained at $N=109$ and $N=113$. Conversely, those linked to the prolate tag state (510)1/2 at $N=105$ and 107 demonstrate a distinct configuration change and, therefore, cannot be interpolated.

Examples of more problematic situations, discussed in point~\ref{point4}.\ of Sec.~\ref{Methodology}, occur between $N=93$ and $N=103$, where the prolate tag states (550)1/2 and (541)1/2 produced duplicated solutions, Fig.~\ref{fig:Spect_Dy}(b), indicating that another $I=1/2^-$ solution was missed. In those cases, results for heavier elements suggest that the missed states may have low energies near $N=97$ or $N=99$. A thorough search will be necessary to find these or other overlooked solutions, especially when new experimental data demands it. However, in the particular case of the prolate tag states (550)1/2 and (541)1/2, one can see that the missing low-energy solution between $N=93$ and $N=103$ corresponds to the tag state (510)1/2 visible in Fig.~\ref{fig:Spect_Dy}(a).

The excitation energies shown in Fig.~\ref{fig:Spect_Dy} (linear scale) or in Fig.~\ref{fig:Spect_log_Dy} (logarithmic scale) are essential for identifying configurations that may become ground states or low-lying excited states in an isotope. To this end, either all six panels of the figures should be examined, or the databases in the raw data repository~\cite{rep-GdOs} described in the Supplemental Material~\cite{supp-GdOs} can be reviewed. The logarithmic scale used in Fig.~\ref{fig:Spect_log_Dy} helps reveal details of the low-energy spectra.

Configurations associated with various tag states exhibit a clear, distinct pattern of evolution as neutron numbers increase. For example, the $I=5/2^-$ configuration of the prolate tag state (512)5/2 begins at $N=83$ within the nearly degenerate group of the ground-state particle 2f$_{7/2}$ states, Figs.~\ref{fig:Spect_Dy}(a) and~\ref{fig:Spect_log_Dy}(a). Then, the energy splitting due to deformation raises it above the neutron Fermi energy, leading it to transition to higher excitation energies and become a particle-like quasiparticle. As the neutron number increases, the rising neutron Fermi energy reaches this configuration. At $N=103$, it appears as the ground state, accompanied by the low-lying $I=7/2^+$ prolate tag state (633)7/2 configuration at $E_{\text{exc}}=167$\,keV, Figs.~\ref{fig:Spect_Dy}(c) and~\ref{fig:Spect_log_Dy}(c). At even larger neutron numbers and higher neutron Fermi energies, the prolate tag state (512)5/2 configuration appears below the neutron Fermi energy. Then it transitions again to higher excitation energies, this time as a hole-like quasiparticle state. Ultimately, its evolution ends at $N=125$ within the nearly degenerate group of the excited hole 2f$_{7/2}$ states.

Figures~\ref{fig:Spect_Dy} and~\ref{fig:Spect_log_Dy} clearly show the reasons for frequent disagreement between the measured and calculated excitation-energy sequences in odd nuclei, see, for instance, Tables~\ref{yyxxx-ttt-v67-170-HFT-N16-siq-UDF1.print1} and~\ref{yyxxx-ttt-v67-170-HFT-N16-siq-UDF1.print2}. Indeed, the spherical s.p.\ spectra that characterize different functionals differ, and also differ from the evaluated experimental data~\cite{(Tar14b),(Tar14a)}. This uncertainty of the theoretical description translates into different deformation dependencies of the s.p.\ energies, such as shown in the Nilsson diagram of Fig.~\ref{Nilsson}. It is evident that even minor shifts of spherical s.p.\ energies, which result in minor shifts of deformed s.p.\ energies, may induce significant differences in the evolution of excitation energies of odd nuclei as functions of particle numbers, Figures~\ref{fig:Spect_Dy} and~\ref{fig:Spect_log_Dy}. In particular, even if a given energy level of an odd nucleus correctly appears in a given isotope at low energy, it may not appear as its ground state. In contrast, the particle-number dependencies of electromagnetic moments, Figs.~\ref{fig:Q2,mu_Dy_prol_2} and~\ref{fig:Q2,mu_Dy_obl_2}, do not directly depend on the s.p.\ energies but rather on s.p.\ wave functions, and therefore are probably more robust, as discussed in Ref.~\cite{(Sas22c)}.


\subsection{DFT results for $^{161}$Dy\label{161}}

Using the results obtained for $^{161}$Dy as an example, here we discuss how electromagnetic moments depend on angular momentum and deformation.

\subsubsection{Dependence on angular momentum\label{angular}}

\begin{figure*}
\begin{center}
\includegraphics[width=0.98\textwidth]{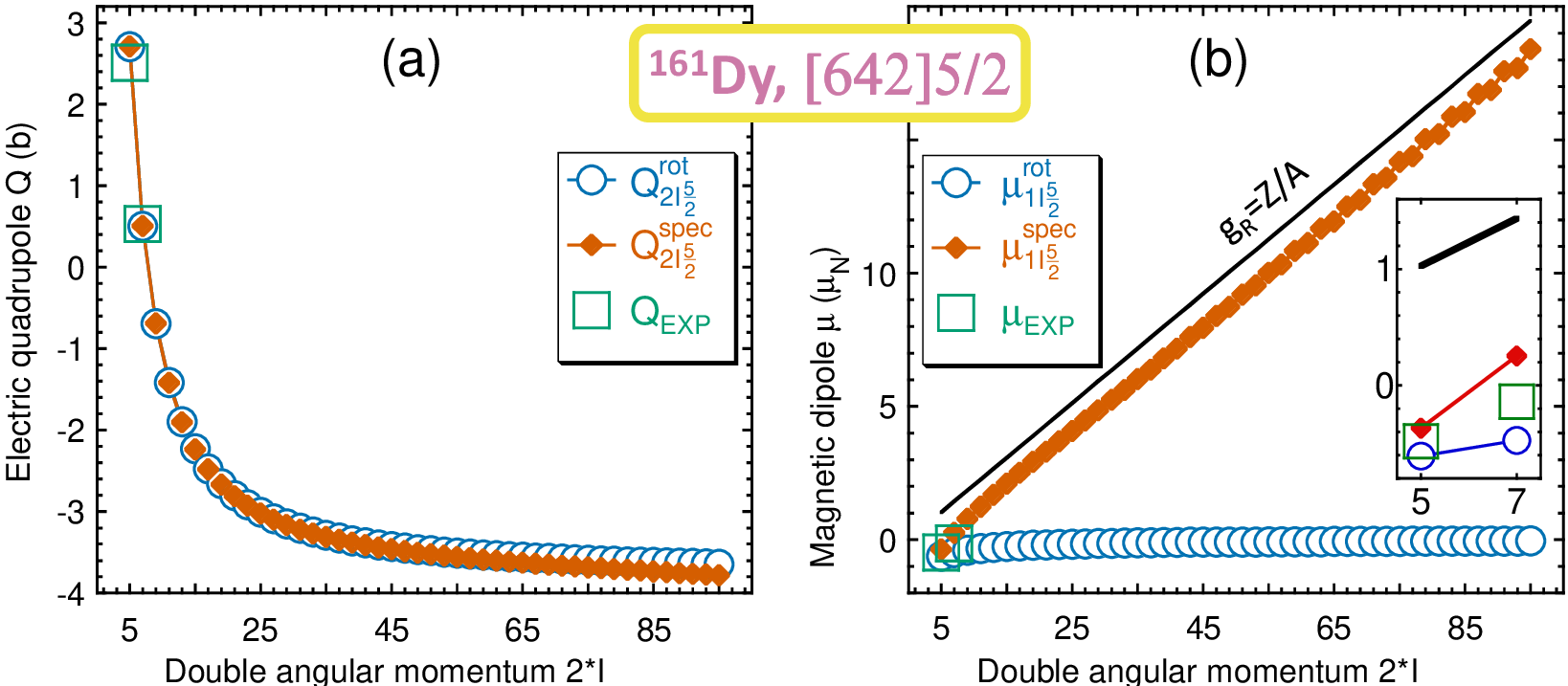}
\end{center}
\caption{Electric quadrupole (a) and magnetic dipole (b) moments determined in this work for the $\Omega=5/2$ [642]5/2 rotational band, $I^\pi=5/2^+\ldots95/2^+$, in $^{161}$Dy. The panels illustrate the comparison between the microscopically calculated (spec, diamonds) and large-axial-deformation approximate (rot, circles) spectroscopic moments. Squares display the known experimental data~\cite{(Syl73),(Fer74)}. The thick line depicts the magnetic dipole moment corresponding to the phenomenological estimate of the rotational collective $g$~factor, $g_R\simeq{}Z/A$. The inset presents the results for $I^\pi=5/2^+$ and $7/2^+$ plotted in an extended scale.
}
\label{dy161_2_5+_rotational_band}
\end{figure*}
In this section, we compare the DFT spectroscopic electromagnetic moments with their large-axial-deformation (rotational) approximations~\footnote{See Ref.~\protect\cite{(Rin80)}, Eq.~(11.143).}. As for the other calculations performed in this work, the AMP results were obtained by restoring the rotational symmetry of the axially deformed intrinsic state. Here we focus on the [642]5/2 intrinsic configuration of $^{161}$Dy, which is characterized by the projection $\Omega=+5/2$ of angular momentum along the axis of axial symmetry. This intrinsic state was projected onto good total angular momenta between the band head of $I=5/2$ and $I=95/2$.

The integral over the Euler angle $\beta$ was evaluated using 100 nodes of the Gauss-Legendre quadrature. We note in passing that this band is experimentally known up to $I=47/2$~\cite{(Jun03)}. However, we must remember that the AMP of a non-rotating intrinsic state ignores structural changes due to rotation, which can only be accounted for using the cranking method~\cite{(Rin80)}. The application of this method will be covered in a future publication.

In Figs.~\ref{dy161_2_5+_rotational_band}(a) and (b), we presented the AMP results obtained for the electric quadrupole and magnetic dipole moments, respectively, in the [642]5/2 $^{161}$Dy rotational band, see the Supplemental Material~\cite{supp-GdOs}. They demonstrate that our AMP calculations achieve exceptionally high numerical precision, enabling us to restore rotational symmetry at very high angular momenta. This is mainly because the singularities and self-interactions~\cite{(She21)} that trouble most density functionals currently in use have no impact on the one-body observables, such as the electromagnetic moments studied in this work. Additionally, calculations performed in the spherical HO basis, implemented with highly precise Gauss-Hermit quadratures~\cite{(Dob97b)}, yield results with significant digits practically limited only by the CPU's precision. Furthermore, the use of highly precise evaluations of the Wigner functions $D^I_{MK}$ is essential~\cite{(Fen15)}. Overall, these aspects of our implementation enable us to correctly identify the AMP components of the $^{161}$Dy intrinsic state, starting from a small norm of 3.44\% at $I=5/2$, reaching a maximum norm of 6.78\% at $I=17/2$, and dropping to a very small norm of 0.0001\% at $I=95/2$ (cited percentage values correspond to the case illustrated in Fig.~\ref{dy161_2_5+_rotational_band}).

The large-axial-deformation approximation links the calculated spectroscopic moments, ${O}^{\text{spec}}_{\lambda{I\Omega}}=\langle{II\Omega}|\hat{O}_{\lambda0}|{II\Omega}\rangle$, to the calculated intrinsic moments, ${O}^{\text{intr}}_{\lambda{\Omega}}=\langle{\Omega}|\hat{O}_{\lambda0}|{\Omega}\rangle$, of generic spherical-tensor observables $\hat{O}_{\lambda0}$, where $|IM\Omega\rangle={\cal{N}}_{I\Omega}\hat{P}^I_{M\Omega}|\Omega\rangle$ are normalized AMP states~\cite{(She21)}. This approximation depends on the assumption that the overlaps between the axial $z$-aligned broken-symmetry self-consistent intrinsic states $|\Omega\rangle$, rotated by an angle $\beta$ around the perpendicular $y$-axis, are positive and sharply peaked at $\beta=0$.

The large-axial-deformation approximation can be used in two flavors. First, it can serve us to define an approximation $Q^{\text{rot}}_{\lambda{I\Omega}}$ of the spectroscopic moment ${O}^{\text{spec}}_{\lambda{I\Omega}}$ in terms of the calculated intrinsic moment ${O}^{\text{intr}}_{\lambda\Omega}$, that is,
\begin{equation}
\label{Qrot1}
{O}^{\text{spec}}_{\lambda{I\Omega}}\simeq{}Q^{\text{rot}}_{\lambda{I\Omega}}\equiv{O}^{\text{intr}}_{\lambda\Omega}\times C^{II}_{II,\lambda0}\times C^{I\Omega}_{I\Omega,\lambda0},
\end{equation}
where $C^{IM}_{IM,\lambda0}$ are the Clebsch-Gordan coefficients~\cite{(Var88)}.

Second, it can be employed to define the effective intrinsic moments ${O}^{\text{intr}}_{\text{eff}}({\lambda{I\Omega}})$ as useful parameterizations of the calculated or measured spectroscopic moment ${O}^{\text{spec}}_{\lambda{I\Omega}}$, that is,
\begin{eqnarray}
\label{Qrot2eff}
{O}^{\text{spec}}_{\lambda{I\Omega}}={O}^{\text{intr}}_{\text{eff}}({\lambda{I\Omega}})\times C^{II}_{II,\lambda0}\times C^{I\Omega}_{I\Omega,\lambda0} .
\end{eqnarray}
We note that for $I>K=|\Omega|$, the definition of the experimental effective intrinsic moment ${O}^{\text{intr}}_{\text{eff}}({\lambda{IK}})$ requires independently assigning the $K$ quantum number to the experimental rotational band. Usually, $K$ is associated with the band-head spin, $K\equiv{}I_{\text{min}}$.

The effective intrinsic moments ${O}^{\text{intr}}_{\text{eff}}({\lambda{IK}})$ are (in the spirit of the large-axial-deformation approximation) useful indicators of the intrinsic shapes. Since the shapes are not observables, their values are vital for the interpretation of the experimental moments and are routinely calculated in experimental analyses, see Sect.~\ref{sect:deformed-cf-experiment} below.

On the one hand, theoretical implementations capable of calculating both the spectroscopic and intrinsic moments, like the one in this work, can test the accuracy of the large-axial-deformation approximation ${O}^{\text{spec}}_{\lambda{I\Omega}}\simeq{}Q^{\text{rot}}_{\lambda{I\Omega}}$, Eq.~(\ref{Qrot1}). The tests performed in nuclei between tin and gadolinium~\cite{(Wib25d)} indicate that such an approximation is accurate within 1-2\% in well-deformed nuclei but can decline to 30\% near the closed shells.

On the other hand, parameterization in Eq.~(\ref{Qrot2eff}) does not have an approximative character. The comparison between theory and experiment can be performed equally well by comparing the spectroscopic moments ${O}^{\text{spec}}_{\lambda{I\Omega}}$ directly, or their corresponding effective intrinsic moments ${O}^{\text{intr}}_{\text{eff}}({\lambda{I\Omega}})$.

Specifically, the large-axial-deformation approximations of the electric quadrupole and magnetic dipole operators read as follows,
\begin{eqnarray}
\label{Qrot}
{Q}^{\text{spec}}_{2{I\Omega}}\simeq{Q}^{\text{rot}}_{2{I\Omega}}&=&{Q}^{\text{intr}}_{2\Omega}
\frac{3\Omega^2-I(I+1)}{(2I+3)(I+1)}, \\
\label{miurot}
{\mu}^{\text{spec}}_{1{I\Omega}}\simeq{\mu}^{\text{rot}}_{1{I\Omega}}&=&{\mu}^{\text{intr}}_{1\Omega}
\frac{\Omega}{I+1}.
\end{eqnarray}

As illustrated in Fig.~\ref{dy161_2_5+_rotational_band}(a), for the [642]5/2 $^{161}$Dy rotational band, the large-axial-deformation approximation $Q^{\text{spec}}_{2I\frac{5}{2}}\simeq{}Q^{\text{rot}}_{2I\frac{5}{2}}$ of Eq.~(\ref{Qrot}), evaluated for the calculated intrinsic electric quadrupole moment of $Q^{\text{intr}}_{2\frac{5}{2}}=7.58$\,b, accurately represents the AMP spectroscopic electric quadrupole moments for all $I$. Additionally, the experimental values for the two lowest states are well reproduced.

Analogous conclusions do not apply to the magnetic dipole moments shown in Fig.~\ref{dy161_2_5+_rotational_band}(b). Using the calculated intrinsic magnetic dipole moment of $\mu^{\text{intr}}_{1\frac{5}{2}}=-0.85$\,$\mu_N$, the large-axial-deformation approximation $\mu^{\text{spec}}_{1I\frac{5}{2}}\simeq{}\mu^{\text{rot}}_{1I\frac{5}{2}}$ of Eq.~(\ref{miurot}) does not reproduce the AMP spectroscopic magnetic dipole moments and entirely overlooks the rotational collective growth associated with the standard rotational $g$~factor, $g_R\simeq{}Z/A$, see Ref.~\cite{(Boh75)}, Sec.~4-3c. The reason for this discrepancy is that the magnetic dipole moments are primarily given by the s.p.\ magnetic moments of the odd nucleon; thus, the assumptions of the large-axial-deformation approximation are not met. As a result, the magnetic dipole moments and transitions in rotational bands are typically described within the particle-rotor model, with values of $g_R$ adjusted to match the data. Here, we observe in Fig.~\ref{dy161_2_5+_rotational_band}(b) that the AMP results consistently reproduce the rotational collective trend of $g_R\simeq{}Z/A$ without any adjustments.

\subsubsection{Dependence on deformation\label{deformation}}

\begin{figure}
\begin{center}
\includegraphics[width=0.48\textwidth]{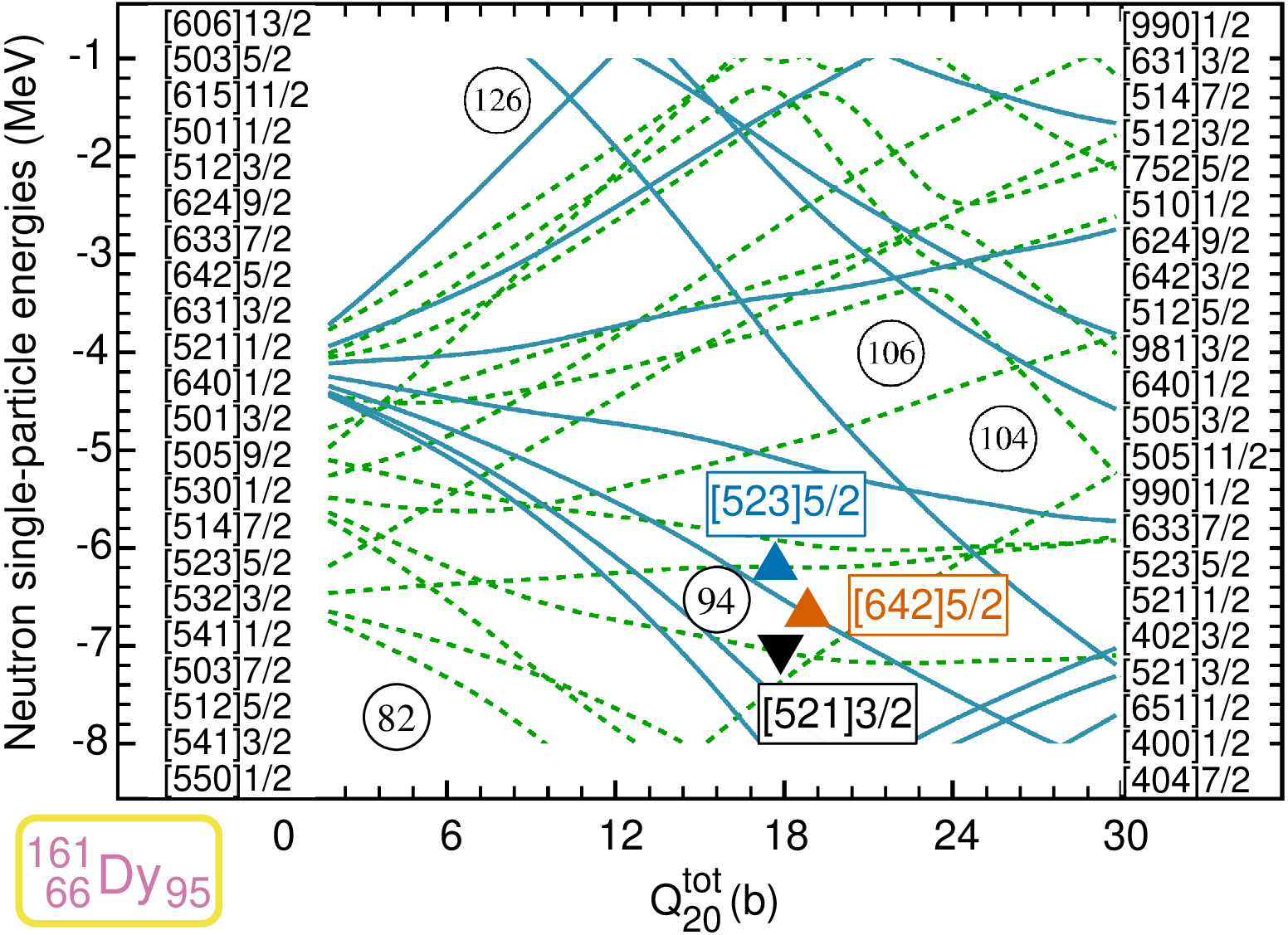}
\end{center}
\caption{The Nilsson diagram of the neutron s.p.\ energies in the false-vacuum configuration of $^{161}$Dy, see text.
}
\label{Nilsson}
\end{figure}

The so-called Nilsson diagram, that is, a plot of the s.p.\ energies as functions of deformation, constitutes an invaluable tool to analyze nuclear deformation properties. In the phenomenological mean-field models, two generic diagrams --one for neutrons and one for protons-- illustrate the properties of a large set of nuclei, specifically those for which the parameters of the mean-field potentials are fixed. In contrast, in the self-consistent models, the diagrams obtained for different configurations in different nuclei differ. Rather than discussing 6,776 possible pairs of Nilsson diagrams that could have been generated in this work, in Fig.~\ref{Nilsson} we show only one.

The solid and dashed lines represent the positive- and negative-parity neutron s.p.\ energies, respectively, calculated for the so-called false-vacuum (no blocking) state of $^{161}$Dy. To generate it, we performed calculations by constraining the total axial intrinsic quadrupole moments $Q_{20}^{\text{tot}}$, Eq.~(\ref{Qtot}), of $^{161}$Dy to values ranging from 0\,b to 30\,b in steps of 1\,b, without blocking any quasiparticle. Such self-consistent solutions are called false vacuums. Although they do not correspond to any physical state in this nucleus, they enable the analysis of a generic Nilsson diagram, which is approximately valid for any blocked quasiparticle discussed below. The left (right) sets of self-consistent Nilsson labels were computed at deformations corresponding to the left (right) endpoints of the lines. They are printed from bottom to top in the same order as those endpoints.

We observe that the three experimentally identified configurations~\cite{(Jun03)}, two particle-type ones, [642]5/2 and [523]5/2, are correctly located just above a small deformed shell gap of $N=94$, and one hole-type, [521]3/2, just below it. We also observe that the states [523]5/2 and [521]3/2 near sphericity appear as [512]5/2 and [532]3/2, respectively. Only near the self-consistent minima at $Q_{20}^{\text{tot}}\simeq18$\,b, they revert to their final configurations; see also the discussion below.

\begin{figure*}
\begin{center}
\includegraphics[width=0.98\textwidth]{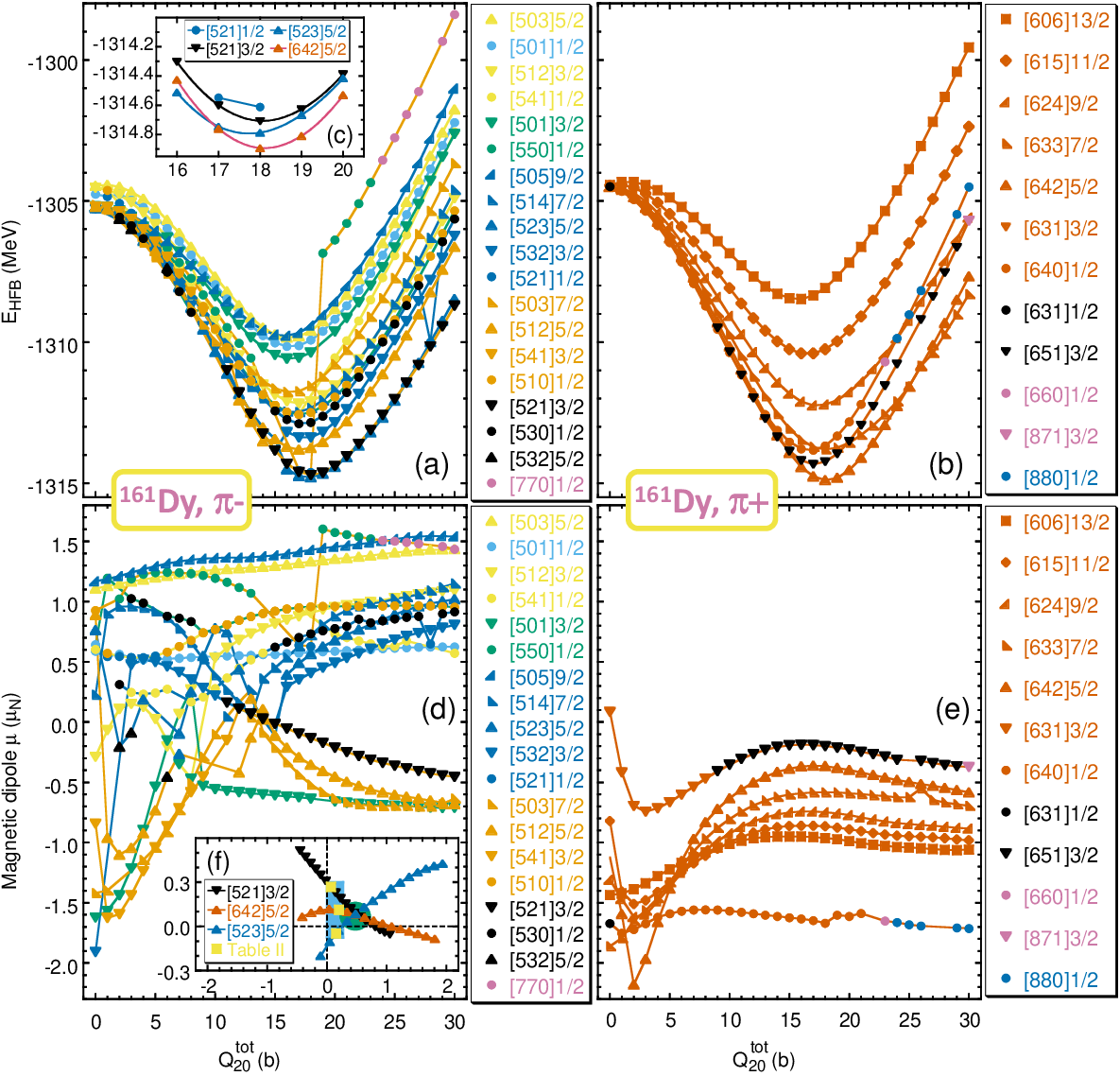}
\end{center}
\caption{The HFB energies, (a) and (b), and magnetic dipole moments, (c) and (d), of $^{161}$Dy plotted as functions of the total axial intrinsic quadrupole moment $Q_{20}^{\text{tot}}$, Eq.~(\protect\ref{Qtot}). The left and right panels display the results for the negative ($\pi-$) and positive ($\pi+$) parities, respectively. The inset (c) presents the lowest HFB energies for both parities plotted in an extended scale, cf.~Fig.~\protect\ref{Nilsson}. The inset (f) presents the residuals of the magnetic dipole moment plotted against those of the electric quadrupole moment; see the text.
}
\label{dy161-PES}
\end{figure*}

Continuing the example of $^{161}$Dy, we performed similar constrained calculations by blocking at each deformation one of the 22 states corresponding to the prolate tags. This required $31\times22=682$ independent calculations, of which 35 have not converged. We note that the results constrained to $Q_{20}^{\text{tot}}=0$\,b do not correspond to spherical shapes. In fact, these points represent states with nonzero quadrupole moments of the odd neutron that, by construction, cancel the nonzero quadrupole moments of the even-even core. The top and bottom panels of Fig.~\ref{dy161-PES} show how the obtained total intrinsic HFB energies $E_{\text{HFB}}$ and spectroscopic magnetic dipole moments, respectively, depend on quadrupole deformations.

The results obtained for fifteen negative-parity (seven positive-parity) prolate tag states are shown in the left (right) panels of Fig.~\ref{dy161-PES} in function of $Q_{20}^{\text{tot}}$, Eq.~(\ref{Qtot}). The lines drawn in the figure connect points corresponding to individual prolate tag states. The Nilsson labels of those tag states are listed in the right column of Fig.~\ref{dy192.spectrum}. However, the symbols and legends shown in Fig.~\ref{dy161-PES} correspond to the calculated self-consistent Nilsson labels. In this way, Fig.~\ref{dy161-PES} illustrates connections between the Nilsson labels of tag states on one hand and the self-consistent Nilsson labels and configurations on the other, with the latter varying with deformations. For fine details presented in Fig.~\ref{dy161-PES}, where curves and symbols overlap, the reader is invited to review the databases described in the Supplemental Material~\cite{supp-GdOs}.

Close inspection of Fig.~\ref{dy161-PES} shows that, for all configurations with $I\geq7/2$, the self-consistent Nilsson labels consistently match those of the tag states. It is again remarkable to see that the $I\geq7/2$ states, when deformed up to $Q_{20}^{\text{tot}}=30$\,b, which corresponds to the Bohr deformation parameter of $\beta\simeq0.5$, still maintain their strong affinity with the tag states determined near sphericity.

\begin{figure*}
\begin{center}
\includegraphics[width=0.98\textwidth]{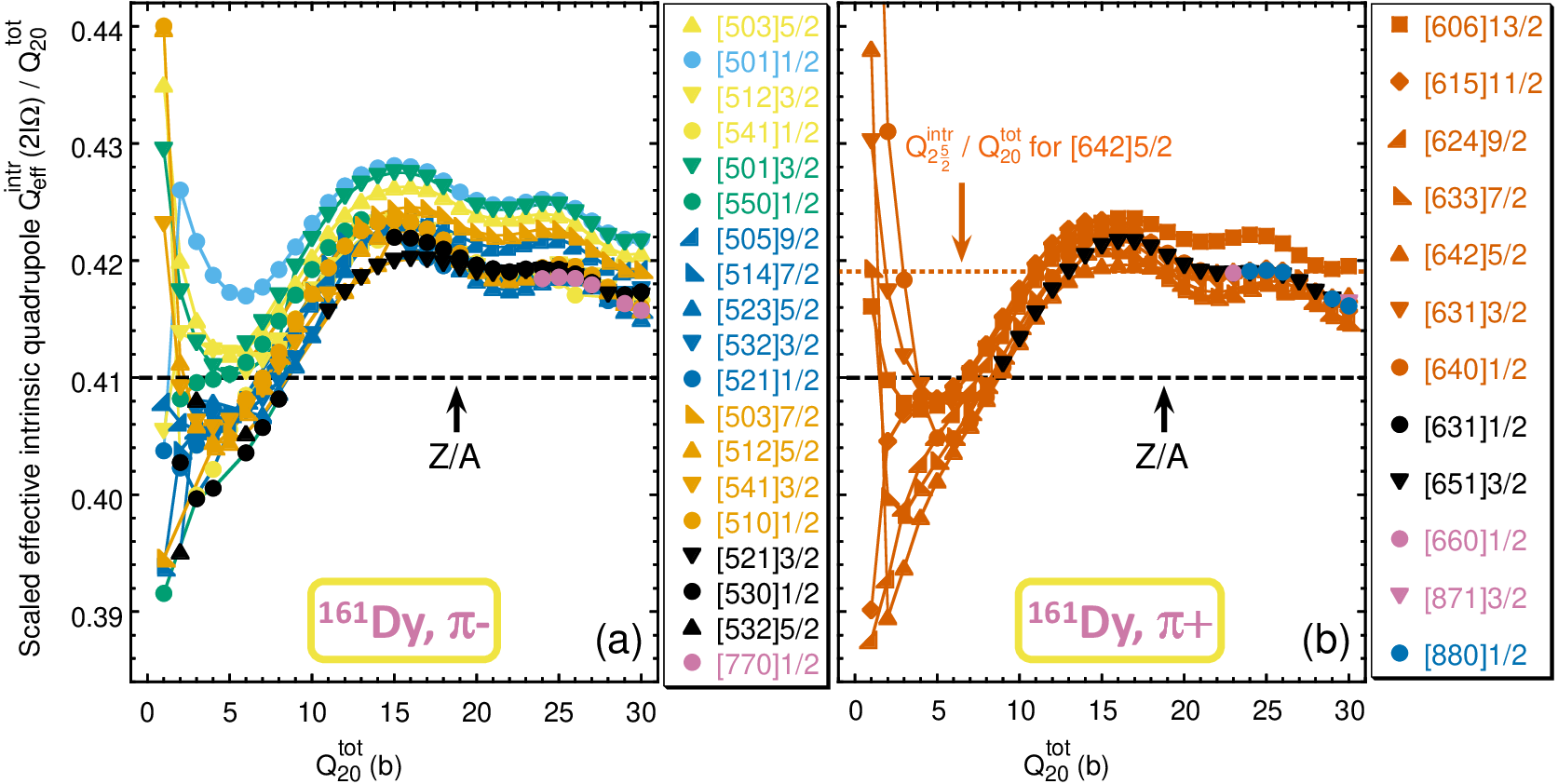}
\end{center}
\caption{Effective intrinsic electric quadrupole moments ${Q}^{\text{intr}}_{\text{eff}}({2{I\Omega}})$ (\protect\ref{Qrot2eff}) in $^{161}$Dy scaled by the total axial intrinsic quadrupole moment $Q_{20}^{\text{tot}}$, Eq.~(\protect\ref{Qtot}), see text. The left (a) and right (b) panels display the results for the negative ($\pi-$) and positive ($\pi+$) parities, respectively. Note again that for the $\Omega=1/2$ states, we plotted the values of $Q$ corresponding to the $I=3/2$ members of the rotational bands.
The point of [640]1/2 at $Q_{20}^{\text{tot}}=1$\,b, ${Q}^{\text{intr}}_{\text{eff}}({2{\frac{3}{2}\frac{1}{2}}})/Q_{20}^{\text{tot}}=$0.476, is beyond the scale of the figure.
}
\label{dy161-quad}
\end{figure*}

Configurations with $I<7/2$ exhibit various patterns of deformation dependence. In general, the total HFB energies $E_{\text{HFB}}$, Figs.~\ref{dy161-PES}(a) and (b), behave very smoothly and do not reflect changes in the internal odd-neutron's structure. A unique, conspicuous example to the contrary is that of the sequence of configurations affine to the (510)1/2 tag state. Only at $Q_{20}^{\text{tot}}=0$\,b the corresponding self-consistent Nilsson label is also [510]1/2. However, already at $Q_{20}^{\text{tot}}=3$\,b and up to 13\,b, it continues as the configuration [550]1/2. Then, after a sequence of non-converged points, at 17 and 18\,b it jumps to the configuration [521]1/2, where it appears at very low energy, see Fig.~\ref{dy161-PES}(c). Later, at 19\,b, it reverts to [550]1/2 and jumps to a very high energy, without continuing its earlier section. Finally, from 24\,b on, it smoothly changes to [770]1/2.

Another example of changing configurations, this time without any visible impact on the energies or magnetic dipole moments, is the evolution of configurations affine to the (640)1/2 tag state. This one at $Q_{20}^{\text{tot}}=0$\,b starts as [631]1/2 but from 1 to 21\,b continues as [640]1/2, then changing to [660]1/2 and from 24\,b on continuing as [880]1/2. We observe that even the self-consistent states corresponding to the Nilsson labels from two major shells above may not significantly affect the results.

Contrary to the smooth deformation dependence of the total HFB energies $E_{\text{HFB}}$, a relatively complicated pattern of dependence is obtained for the magnetic dipole moments, Figs.~\ref{dy161-PES}(d) and (e). In some cases, we observe that the magnetic dipole moments can exhibit rapid changes along the sequences of states corresponding to the same tag states and self-consistent Nilsson labels. In other cases, changes of the self-consistent Nilsson labels do not induce jumps of the magnetic dipole moments. We can conclude that a correct simultaneous description of the dipole and quadrupole moments is nontrivial.

In this context, in Fig.~\ref{dy161-PES}(f) we present the deformation dependence of the calculated dipole and quadrupole moments of the three lowest states, cf.~Figs.~\ref{Nilsson} and~\ref{dy161-PES}(c), where the experimental data are available. The figure represents the residuals, see Table~\ref{yyxxx-ttt-v67-170-HFT-N16-siq-UDF1.print1}, of spectroscopic magnetic dipole moments, $\mu_{\text{R}}\equiv\mu_{\text{DFT}}-\mu_{\text{EXP}}$, plotted against those of the spectroscopic electric quadrupole moments, $Q_{\text{R}}\equiv{}Q_{\text{DFT}}-Q_{\text{EXP}}$ as functions of $Q_{20}^{\text{tot}}$, Eq.~(\protect\ref{Qtot}).

We observe that the unconstrained self-consistent calculations (squares) reproduce the experimental electric quadrupole moments very well, up to 0.1\,b, but fail to reproduce the magnetic dipole moments (rectangle) up to 0.3\,$\mu_N$. However, we also observe that at slightly higher deformations, all three magnetic dipole moments cluster in a single point (circle), which is 0.1\,$\mu_N$ away from the data. This example suggests that some fine-tuning of the DFT functionals, which could shift the curves in Fig.~\ref{dy161-PES}(f) upwards or downwards, might still improve the agreement with data without needing to introduce new elements to the approach. Since magnetic dipole moments are primarily sensitive to the time-odd mean-field sector of the functional, improvements can be made there.

We also observe that at moderate deformations, complex patterns of the magnetic dipole moments develop. This suggests that the transition between the weak and strong-coupling schemes, discussed in the introduction, may not be achievable by either of the two. In contrast, it is accessible via the approach presented in this work.

Spectroscopic electric quadrupole moments increase nearly linearly with $Q_{20}^{\text{tot}}$. To display them meaningfully, in Fig.~\ref{dy161-quad}, we plotted the corresponding effective intrinsic moments ${Q}^{\text{intr}}_{\text{eff}}({2{I\Omega}})$ (\ref{Qrot2eff}) scaled by the total axial intrinsic quadrupole moments $Q_{20}^{\text{tot}}$. We observe that at large deformations, the scaled values remain nearly constant, supporting the validity of the large-axial-deformation approximation. At smaller deformations, notable differences emerge, which nonetheless do not exceed $\pm6\%$.

To put the scaled values in the right perspective, in Fig.~\ref{dy161-quad} we added two horizontal lines. The dashed line represents the value of $Z/A=0.4099$, which corresponds to the ratio of the proton and mass quadrupole moments equal to that of particle numbers. Similarly, the dotted line represents the value of
$Q^{\text{intr}}_{2\frac{5}{2}}/Q_{20}^{\text{tot}}=7.58/18.10=0.4188$ corresponding to the self-consistent [642]5/2 solution in $^{161}$Dy.


\section{Comparison with experimental data}\label{sect:deformed-cf-experiment}

\newcommand{\ph}{~~\,}
\begin{table*}
\begin{center}
\renewcommand{\arraystretch}{1.25}
$\begin{array}{|rrrr|rrl|lrrr|rr|}
\cline{5-13}
\multicolumn{4}{r|}{} & \multicolumn{3}{c|}{\mbox{Experiment}} &  \multicolumn{4}{c|}{\mbox{DFT}} & \multicolumn{2}{c|}{\mbox{Residuals}} \\
\hline
\mbox{No.}& \mbox{Nuclide} & N      & I^\pi~~  &E~\mbox{(keV)}&Q\,\mbox{(barn)}&~~\mu\,(\mu_N)&\mbox{Nilsson}&E~\mbox{(keV)}&Q\,\mbox{(barn)}&\mu\,(\mu_N)~~& Q_{\text{R}}\,\mbox{(barn)} & \mu_{\text{R}}\,(\mu_N) \\
\hline
 1    & ^{147}\mathrm{Gd} & 83     & {7/2} ^- & 0     &             &\ph1.02(9)    & [503]7/2       & 0      & -0.380 & -1.390(69) &        & -0.370  \\
 2    & ^{147}\mathrm{Gd} & 83     & {13/2}^+ & 997   & -0.70(8)    &  +0.49(2)    & [606]13/2      & 2575   & -0.759 & -1.289(86) & -0.059 & -1.779  \\
 3    & ^{149}\mathrm{Gd} & 85     & {7/2} ^- & 0     &             &\ph0.88(4)    & [503]7/2       & 135    & -0.400 & -1.359(73) &        & -0.479  \\
 4    & ^{149}\mathrm{Gd} & 85     & {5/2} ^- & 165   &             &  -0.9(2)     & [512]5/2       & 113    & -0.436 & -0.716(40) &        & +0.184  \\
 5    & ^{151}\mathrm{Gd} & 87     & {7/2} ^- & 0     &             &\ph0.77(6)    & [503]7/2       & 1064   & -1.293 & -0.806(68) &        & -0.036  \\
 6    & ^{151}\mathrm{Gd} & 87     & {5/2} ^- & 108   &             &  -1.08(13)   & [532]5/2       & 647    & -0.944 & +0.046(12) &        & +1.126  \\
 7    & ^{151}\mathrm{Gd} & 87     & {3/2} ^- & 395   &             &  -2.5(8)     & [541]3/2       & 0      & +0.783 & -0.209(26) &        & +2.291  \\
 8    & ^{153}\mathrm{Gd} & 89     & {3/2} ^- & 0     &             &\ph0.38(8)    & [501]3/2       & 1196   & -0.766 & -0.540(39) &        & -0.160  \\
 9    & ^{153}\mathrm{Gd} & 89     & {5/2} ^- & 110   &             &  +0.40(15)   & [523]5/2       & 607    & +1.643 & +0.053(12) &        & -0.347  \\
10    & ^{153}\mathrm{Gd} & 89     & {3/2} ^- & 129   &             &  +0.37(7)    & [532]3/2       & 120    & +0.999 & +0.264(12) &        & -0.107  \\
11    & ^{155}\mathrm{Gd} & 91     & {3/2} ^- & 0     & +1.27(3)    &  -0.2591(4)  & [521]3/2       & 58     & +1.287 & -0.176(25) & +0.017 & +0.083  \\
12    & ^{155}\mathrm{Gd} & 91     & {5/2} ^+ & 87 ^* & +0.110(8)   &  -0.525(2)   & [651]3/2       &        & -0.467 & +0.259(23) & -0.577 & +0.784  \\
13    & ^{155}\mathrm{Gd} & 91     & {3/2} ^+ & 105   & +1.27(5)    &  +0.143(5)   & [651]3/2       & 0      & +1.308 & -0.191(32) & +0.038 & -0.334  \\
14    & ^{157}\mathrm{Gd} & 93     & {3/2} ^- & 0     & +1.35(3)    &  -0.3398(6)  & [521]3/2       & 0      & +1.412 & -0.125(21) & +0.062 & +0.215  \\
15    & ^{157}\mathrm{Gd} & 93     & {5/2} ^+ & 64    & +2.43(7)    &  -0.464(11)  & [642]5/2       & 341    & +2.626 & -0.389(50) & +0.196 & +0.075  \\
16    & ^{159}\mathrm{Gd} & 95     & {3/2} ^- & 0     &             &  -0.44(3)    & [521]3/2       & 190    & +1.478 & -0.140(22) &        & +0.300  \\
17    & ^{149}\mathrm{Dy} & 83     & {7/2} ^- & 0     & -0.62(5)    &              & [503]7/2       & 0      & -0.407 & -1.375(70) & +0.213 &         \\
18    & ^{151}\mathrm{Dy} & 85     & {7/2} ^- & 0     & -0.30(5)    &  -0.945(7)   & [503]7/2       & 92     & -0.492 & -1.320(74) & -0.192 & -0.375  \\
19    & ^{153}\mathrm{Dy} & 87     & {7/2} ^- & 0     & -0.15(9)    &  -0.712(6)   & [503]7/2       & 1428   & +0.341 & -1.352(71) & +0.491 & -0.640  \\
20    & ^{155}\mathrm{Dy} & 89     & {3/2} ^- & 0     & +0.96(2)    &  -0.337(3)   & [521]3/2       & 77     & +1.028 & +0.224(15) & +0.068 & +0.561  \\
21    & ^{157}\mathrm{Dy} & 91     & {3/2} ^- & 0     & +1.29(2)    &  -0.301(2)   & [521]3/2       & 0      & +1.333 & -0.140(25) & +0.043 & +0.161  \\
22    & ^{159}\mathrm{Dy} & 93     & {3/2} ^- & 0     & +1.37(2)    &  -0.354(3)   & [521]3/2       & 0      & +1.455 & -0.131(23) & +0.085 & +0.223  \\
23    & ^{161}\mathrm{Dy} & 95     & {5/2} ^+ & 0     & +2.51(2)    &  -0.479(3)   & [642]5/2       & 0      & +2.711 & -0.367(48) & +0.201 & +0.112  \\
24    & ^{161}\mathrm{Dy} & 95     & {5/2} ^- & 26    & +2.51(2)    &  +0.594(3)   & [523]5/2       & 101    & +2.660 & +0.546(20) & +0.150 & -0.048  \\
25    & ^{161}\mathrm{Dy} & 95     & {7/2} ^+ & 44 ^* & +0.53(13)  &  -0.140(5)   & [642]5/2       &        & +0.506 & +0.254(38) & -0.024 & +0.394  \\
26    & ^{161}\mathrm{Dy} & 95     & {3/2} ^- & 75    & +1.45(6)    &  -0.403(4)   & [521]3/2       & 187    & +1.515 & -0.135(24) & +0.065 & +0.268  \\
27    & ^{163}\mathrm{Dy} & 97     & {5/2} ^- & 0     & +2.65(2)    &  +0.671(4)   & [523]5/2       & 0      & +2.766 & +0.546(22) & +0.116 & -0.125  \\
28    & ^{165}\mathrm{Dy} & 99     & {7/2} ^+ & 0     & +3.48(7)    &  -0.518(6)   & [633]7/2       & 420    & +3.683 & -0.512(64) & +0.203 & +0.006  \\
29    & ^{153}\mathrm{Er} & 85     &({7/2} ^-)& 0     & -0.42(2)    &  -0.932(7)   & [503]7/2       & 107    & -0.708 & -1.205(72) & -0.288 & -0.273  \\
30    & ^{155}\mathrm{Er} & 87     & {7/2} ^- & 0     & -0.27(2)    &  -0.666(5)   & [503]7/2       & 886    & -1.550 & -0.668(39) & -1.280 & -0.002  \\
31    & ^{155}\mathrm{Er} & 87     & {13/2}^+ & 563   &             &  -0.55(3)    & [606]13/2      & 818    & -2.445 & -0.805(93) &        & -0.255  \\
32    & ^{157}\mathrm{Er} & 89     & {3/2} ^- & 0     & +0.920(10)  &  -0.411(3)   & [532]3/2       & 185    & +1.025 & +0.220(02) & +0.105 & +0.631  \\
33    & ^{159}\mathrm{Er} & 91     & {3/2} ^- & 0     & +1.170(10)  &  -0.303(2)   & [521]3/2       & 0      & +1.249 & -0.112(23) & +0.079 & +0.191  \\
34    & ^{161}\mathrm{Er} & 93     & {3/2} ^- & 0     & +1.363(8)   &  -0.364(3)   & [521]3/2       & 0      & +1.443 & -0.134(23) & +0.080 & +0.230  \\
35    & ^{163}\mathrm{Er} & 95     & {5/2} ^- & 0     & +2.56(2)    &  +0.556(4)   & [523]5/2       & 13     & +2.692 & +0.560(22) & +0.132 & +0.004  \\
36    & ^{165}\mathrm{Er} & 97     & {5/2} ^- & 0     & +2.71(3)    &  +0.641(4)   & [523]5/2       & 13     & +2.831 & +0.556(24) & +0.121 & -0.085  \\
37    & ^{167}\mathrm{Er} & 99     & {7/2} ^+ & 0     & +3.57(3)    &  -0.5623(4)  & [633]7/2       & 393    & +3.766 & -0.521(63) & +0.196 & +0.041  \\
38    & ^{169}\mathrm{Er} & 101    & {1/2} ^- & 0     &             &  +0.4828(4)  & [521]1/2       & 412    &        & +0.642(06) &        & +0.159  \\
39    & ^{171}\mathrm{Er} & 103    & {5/2} ^- & 0     & +2.86(9)    &\ph0.657(10)  & [512]5/2       & 0      & +2.925 & -0.377(41) & +0.065 & +0.280  \\
\hline
\end{array}$
\end{center}
\caption{Electric quadrupole moments $Q$ (in barn) and magnetic dipole moments $\mu$ (in $\mu_N$) of Gd, Dy, and Er isotopes determined in this work within nuclear DFT and compared with experimental data~\cite{(Sto19a),(Sto20),(Sto21)}. Both for experiment and DFT, energies $E$ (in keV) denote excitation energies relative to the ground states, which are indicated by $E=0$. The last two columns display the residuals between the theoretical results and experiment, that is, $Q_{\text{R}}\equiv{}Q_{\text{DFT}}-Q_{\text{EXP}}$ and $\mu_{\text{R}}\equiv\mu_{\text{DFT}}-\mu_{\text{EXP}}$. Asterisks ($*$) mark higher-spin members of the rotational bands. Values without signs indicate experimental results where the signs were not measured; then the residuals were determined with the DFT signs assigned to the experimental values. The DFT values and residuals are given up to three decimal points, with uncertainties of the DFT magnetic dipole moments given up to two last digits, see Sec.~\ref{sect:deformed-cf-experiment}.}
\label{yyxxx-ttt-v67-170-HFT-N16-siq-UDF1.print1}
\end{table*}
\begin{table*}
\begin{center}
\renewcommand{\arraystretch}{1.25}
$\begin{array}{|rrrr|rrl|lrrr|rr|}
\cline{5-13}
\multicolumn{4}{r|}{} & \multicolumn{3}{c|}{\mbox{Experiment}} &  \multicolumn{4}{c|}{\mbox{DFT}} & \multicolumn{2}{c|}{\mbox{Residuals}} \\
\hline
\mbox{No.}& \mbox{Nuclide} & N      & I^\pi~~  &E~\mbox{(keV)}&Q\,\mbox{(barn)}&~~\mu\,(\mu_N)&\mbox{Nilsson}&E~\mbox{(keV)}&Q\,\mbox{(barn)}&\mu\,(\mu_N)~~& Q_{\text{R}}\,\mbox{(barn)} & \mu_{\text{R}}\,(\mu_N) \\
\hline
40    & ^{155}\mathrm{Yb} & 85     &({7/2} ^-)& 0     & -0.5(3)     &  -0.91(2)    & [503]7/2       & 115    & -0.829 & -1.138(70) & -0.329 & -0.228  \\
41    & ^{157}\mathrm{Yb} & 87     & {7/2} ^- & 0     &             &  -0.637(8)   & [503]7/2       & 1007   & -1.495 & -0.743(62) &        & -0.106  \\
42    & ^{157}\mathrm{Yb} & 87     & {13/2}^+ & 529   &             &  -0.75(8)    & [606]13/2      & 1012   & -2.451 & -0.809(95) &        & -0.059  \\
43    & ^{159}\mathrm{Yb} & 89     & {5/2}^-  & 0     & -0.22(2)    &  -0.365(8)   & [523]5/2       & 254    & +1.647 & +0.217(02) & +1.867 & +0.582  \\
44    & ^{161}\mathrm{Yb} & 91     & {3/2} ^- & 0     & +1.03(2)    &  -0.326(8)   & [521]3/2       & 0      & +1.162 & -0.096(22) & +0.132 & +0.230  \\
45    & ^{163}\mathrm{Yb} & 93     & {3/2} ^- & 0     & +1.24(2)    &  -0.373(8)   & [521]3/2       & 0      & +1.367 & -0.132(23) & +0.127 & +0.241  \\
46    & ^{165}\mathrm{Yb} & 95     & {5/2} ^- & 0     & +2.48(4)    &  +0.477(8)   & [523]5/2       & 0      & +2.649 & +0.565(23) & +0.169 & +0.088  \\
47    & ^{167}\mathrm{Yb} & 97     & {5/2} ^- & 0     & +2.70(4)    &  +0.621(8)   & [523]5/2       & 31     & +2.943 & +0.609(28) & +0.243 & -0.012  \\
48    & ^{169}\mathrm{Yb} & 99     & {7/2} ^+ & 0     & +3.54(6)    &  -0.633(8)   & [633]7/2       & 361    & +3.926 & -0.517(62) & +0.386 & +0.116  \\
49    & ^{169}\mathrm{Yb} & 99     & {1/2} ^- & 24    &             &  +0.506(8)   & [521]1/2       & 0      &        & +0.586(13) &        & +0.080  \\
50    & ^{171}\mathrm{Yb} & 101    & {1/2} ^- & 0     &             &  +0.4923(4)  & [521]1/2       & 418    &        & +0.613(09) &        & +0.121  \\
51    & ^{171}\mathrm{Yb} & 101    & {3/2} ^- & 67 ^* & -1.64(8)    &  +0.350(2)   & [521]1/2       &        & -1.665 & +0.364(15) & -0.025 & +0.014  \\
52    & ^{171}\mathrm{Yb} & 101    & {5/2} ^- & 76 ^* & -2.22(7)    &  +1.015(5)   & [521]1/2       &        & -2.378 & +1.420(01) & -0.158 & +0.405  \\
53    & ^{173}\mathrm{Yb} & 103    & {5/2} ^- & 0     & +2.80(4)    &  -0.6780(6)  & [512]5/2       & 0      & +3.012 & -0.402(44) & +0.212 & +0.276  \\
54    & ^{173}\mathrm{Yb} & 103    & {7/2} ^- & 79 ^* &             &  -0.20(7)    & [512]5/2       &        & +0.562 & +0.267(34) &        & +0.467  \\
55    & ^{175}\mathrm{Yb} & 105    & {7/2} ^- & 0     & +3.52(5)    &\ph0.766(8)   & [514]7/2       & 0      & +3.791 & +1.040(53) & +0.271 & +0.272  \\
56    & ^{177}\mathrm{Yb} & 107    & {9/2} ^+ & 0     & +4.03(6)    &  -0.695(15)  & [624]9/2       & 30     & +4.372 & -0.626(75) & +0.342 & +0.069  \\
57    & ^{177}\mathrm{Yb} & 107    & {1/2} ^- & 332   &             &  +0.151(15)  & [510]1/2       & 0      &        & +0.867(16) &        & +0.716  \\
58    & ^{171}\mathrm{Hf} & 99     & {7/2} ^+ & 0     & +3.46(3)    &  -0.674(12)  & [633]7/2       & 385    & +4.337 & -0.523(60) & +0.877 & +0.151  \\
59    & ^{171}\mathrm{Hf} & 99     & {1/2} ^- & 22    &             &  +0.526(16)  & [521]1/2       & 0      &        & +0.581(14) &        & +0.055  \\
60    & ^{173}\mathrm{Hf} & 101    & {1/2} ^- & 0     &             &  +0.502(7)   & [521]1/2       & 245    &        & +0.597(12) &        & +0.095  \\
61    & ^{175}\mathrm{Hf} & 103    & {5/2} ^- & 0     & +2.72(2)    &  -0.677(9)   & [512]5/2       & 0      & +3.294 & -0.467(45) & +0.574 & +0.210  \\
62    & ^{177}\mathrm{Hf} & 105    & {7/2} ^- & 0     & +3.37(3)    &  +0.7910(9)  & [514]7/2       & 0      & +3.750 & +1.007(52) & +0.380 & +0.216  \\
63    & ^{177}\mathrm{Hf} & 105    & {9/2} ^- & 113^* & +1.30(2)    &  +1.03(3)    & [514]7/2       &        & +1.460 & +1.406(43) & +0.160 & +0.376  \\
64    & ^{177}\mathrm{Hf} & 105    & {11/2}^- & 250^* &             &  +1.5(5)     & [514]7/2       &        & +0.087 & +1.791(36) &        & +0.291  \\
65    & ^{177}\mathrm{Hf} & 105    & {9/2} ^+ & 321   &             &  -0.73(9)    & [624]9/2       & 157    & +4.597 & -0.673(72) &        & +0.057  \\
66    & ^{179}\mathrm{Hf} & 107    & {9/2} ^+ & 0     & +3.79(3)    &  -0.6389(14) & [624]9/2       & 0      & +4.123 & -0.661(74) & +0.333 & -0.023  \\
67    & ^{179}\mathrm{Hf} & 107    & {11/2}^+ & 123^* & +1.88(3)    &              & [624]9/2       &        & +2.075 & +0.027(63) & +0.195 &         \\
68    & ^{175}\mathrm{W}  & 101    & {7/2} ^+ & 235   &             &  -0.65(2)    & [633]7/2       & 0      & +4.312 & -0.520(61) &        & +0.130  \\
69    & ^{183}\mathrm{W}  & 109    & {1/2} ^- & 0     &             &  +0.11739(11)& [510]1/2       & 0      &        & +0.837(16) &        & +0.719  \\
70    & ^{183}\mathrm{W}  & 109    & {3/2} ^- & 47 ^* & -1.8(4)     &  -0.10(10)   & [510]1/2       &        & -1.354 & -0.986(17) & +0.446 & -0.886  \\
71    & ^{183}\mathrm{W}  & 109    & {9/2} ^- & 309^* &             &\ph1.53(14)   & [510]1/2       &        & -2.462 & +2.951(01) &        & +1.421  \\
72    & ^{183}\mathrm{W}  & 109    & {9/2} ^- & 551^* &             &\ph2.2(9)     & [512]3/2       &        & -1.843 & +1.760(17) &        & -0.440  \\
73    & ^{185}\mathrm{W}  & 111    & {3/2} ^- & 0     &             &  +0.543(14)  & [512]3/2       & 0      & +1.304 & +0.826(37) &        & +0.283  \\
74    & ^{187}\mathrm{W}  & 113    & {3/2} ^- & 0     &             &\ph0.621(15)  & [512]3/2       & 332    & +1.179 & +0.803(39) &        & +0.182  \\
75    & ^{183}\mathrm{Os} & 107    & {9/2} ^+ & 0     & +3.1(3)     &\ph0.794(14)  & [624]9/2       & 0      & +3.895 & -0.682(72) & +0.795 & +0.112  \\
76    & ^{187}\mathrm{Os} & 111    & {1/2} ^- & 0     &             &  +0.06442(7) & [510]1/2       & 55     &        & +0.831(13) &        & +0.766  \\
77    & ^{189}\mathrm{Os} & 113    & {3/2} ^- & 0     & +0.86(3)    &  +0.6576(7)  & [512]3/2       & 200    & +1.082 & +0.785(35) & +0.222 & +0.128  \\
78    & ^{189}\mathrm{Os} & 113    & {1/2} ^- & 36    &             &  +0.23(3)    & [510]1/2       & 272    &        & +0.830(13) &        & +0.600  \\
79    & ^{189}\mathrm{Os} & 113    & {5/2} ^- & 70 ^* & -0.63(2)    &  +0.981(9)   & [512]3/2       &        & -0.387 & +1.113(25) & +0.243 & +0.132  \\
80    & ^{189}\mathrm{Os} & 113    & {3/2} ^- & 95 ^* &             &  -0.32(5)    & [510]1/2       &        & -1.078 & -0.864(18) &        & -0.544  \\
81    & ^{191}\mathrm{Os} & 115    & {9/2} ^- & 0     & +2.53(16)   &  +0.96(3)    & [505]9/2       & 218    & +2.651 & +1.241(71) & +0.121 & +0.281  \\
82    & ^{193}\mathrm{Os} & 117    & {3/2} ^- & 0     & +0.48(6)    &  +0.730(2)   & [512]3/2       & 1036   & +0.812 & +0.727(28) & +0.332 & -0.003  \\
\hline
\end{array}$
\end{center}
\caption{Same as in Table~\protect\ref{yyxxx-ttt-v67-170-HFT-N16-siq-UDF1.print1} but for the isotopes of Yb, Hf, W, and Os. Experimental data taken from Refs.~\cite{(Sto19a),(Sto20),(Sto21),Pli71}}
\label{yyxxx-ttt-v67-170-HFT-N16-siq-UDF1.print2}
\end{table*}

In this section, we compare theory and experiment, as shown in Tables~\ref{yyxxx-ttt-v67-170-HFT-N16-siq-UDF1.print1} and~\ref{yyxxx-ttt-v67-170-HFT-N16-siq-UDF1.print2}. Theoretical uncertainties of the electric quadrupole moments were not evaluated. However, based on the results in Ref.~\cite{(Sas22c)} for various Skyrme functionals, the estimated deviations are negligible compared to the overall deviations from the data. Theoretical uncertainties of the magnetic dipole moments were evaluated by varying the isovector Landau parameter, $g_0'=1.7(4)$, as adjusted in Ref.~\cite{(Sas22c)}, to $g_0'=1.3$ and 2.1, and by taking the averages of deviations from those determined at $g_0'=1.7$.

We begin with a survey of results for each element ($Z$ value) and then discuss trends for various Nilsson orbits. Where relevant, some remarks are made on the choice or evaluation of experimental data.

We systematically associate every experimental data point with the lowest converged configuration having the same spin and parity.

In general, the description of the electric quadrupole moments is excellent across the whole range of isotopes. In contrast, the prediction of the magnetic dipole moments is less reliable, and they will be the primary focus of the following discussion.

To assist the discussion of magnetic dipole moments and relate them to the traditional Nilsson (particle plus rotor) model descriptions, we estimate the Nilsson quadrupole deformation parameter $\epsilon$ using the following relation,
\begin{equation}
Q_0 \simeq\frac{4}{5}ZR^2 \epsilon(1+\frac{\epsilon}{2}),
\end{equation}
where $R = 1.2A^{1/3}$
is a suitable parameterization of the nuclear radius in fm, and $Q_0\equiv{Q}^{\text{intr}}_{\text{eff}}({2{IK}})$, cf.~Eq.~(\ref{Qrot2eff}) for $K=|\Omega|$, is an experimental estimate of the intrinsic electric quadrupole moment,
\begin{equation}
\label{Q0}
Q(I)=Q_0 \frac{3K^2-I(I+1)}{(2I+3)(I+1)},
\end{equation}
in terms of the measured spectroscopic electric quadrupole moment $Q(I)\equiv{Q}^{\text{spec}}_{\lambda{IK}}$.

Some of the spins and/or parities are given in parentheses in the compilations \cite{(Sto19a),(Sto21),ensdf} because they have not been directly measured. In Tables~\ref{yyxxx-ttt-v67-170-HFT-N16-siq-UDF1.print1}  and~\ref{yyxxx-ttt-v67-170-HFT-N16-siq-UDF1.print2}, we have retained the parentheses on the $(7/2^-)$ ground states of $^{153}$Er and $^{155}$Yb as neither the spin nor the parity has been measured. However, as these nuclei are near spherical and the Fermi surface is in the $\nu f_{7/2}$ orbit, there can be little doubt about the $7/2^-$ spin-parity assignment, and we show the theory only for $I^{\pi}=7/2^-$.

The parities are listed as uncertain in ENSDF for the ground-states of $^{171}$Hf and $^{175}$Hf, and for the $K=1/2$ state at an excitation energy of 22 keV in $^{171}$Hf. In these two nuclei, there are well-characterized rotational bands built on the states in question.
Hence, the ground state of $^{171}$Hf can confidently be associated with the $[633]7/2^+$ Nilsson orbit~\footnote{To increase the readability of presentation, in Sect.~\protect\ref{sect:deformed-cf-experiment}, we explicitly added the parities $\pi=(-1)^{N_0}$ to the Nilsson labels $[N_0n_z\Lambda]K^\pi$.} and assigned $I^{\pi}=7/2^+$. Likewise, the ground-state band of $^{175}$Hf can be assigned the Nilsson orbit $[512]5/2^-$ and hence $I^{\pi}=5/2^-$. Similarly, the band built on the state at an excitation energy of 22 keV in $^{171}$Hf is clearly a $K=1/2$ band and its $\gamma$-ray decay properties allow a firm Nilsson assignment; both signatures of the $[521]1/2^-$ band are observed to high spin affirming $I^{\pi}=1/2^-$ for the band head.

The ground-state parity of $^{159}$Yb is tentatively assigned as negative in ENSDF. The spin $I=5/2$ is from laser spectroscopy \cite{Neu83,Sch91c}, whereas the tentative negative parity is from systematics. Given the near spherical shape implied by the quadrupole moment, and the likely position of the Fermi surface at $N=89$, we consider the positive parity highly unlikely and have adopted
$I=5/2^-$.

\subsection{Gadolinium} \label{sect:aesGd}

\begin{figure*}
\begin{center}
\includegraphics[width=\textwidth]{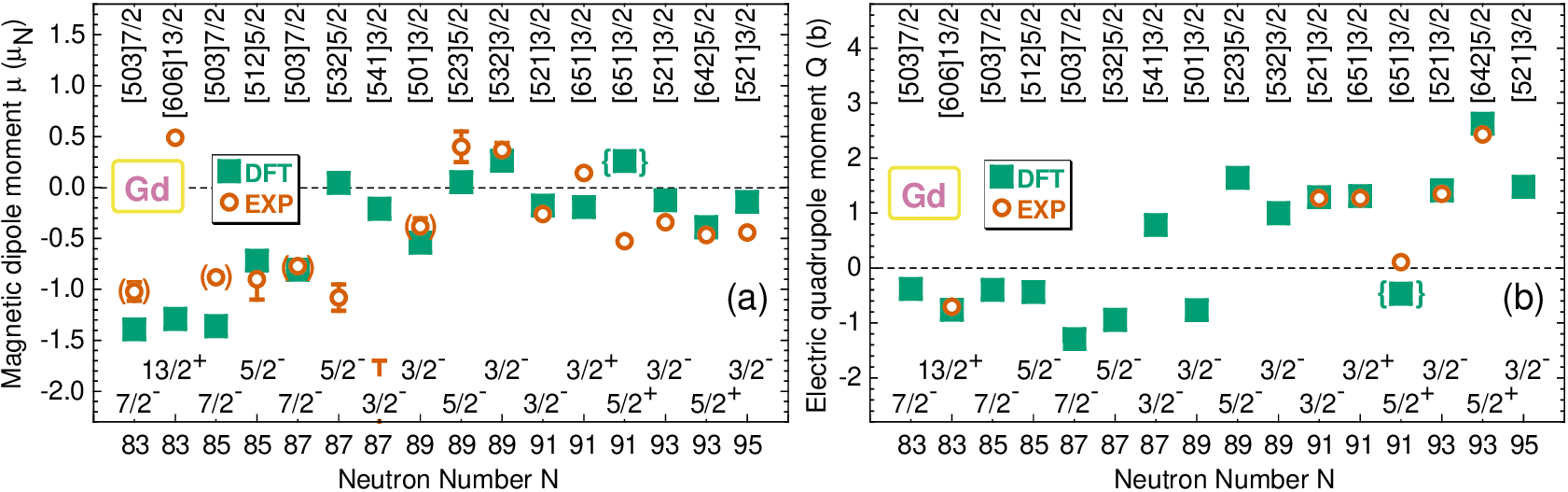}
\end{center}
\caption{Experimental and theoretical magnetic dipole (a) and electric quadrupole (b) moments in gadolinium isotopes ($Z=64$). Braces denote the values obtained for the higher-spin members of the rotational bands. Parentheses denote the signs of the calculated magnetic dipole moments assigned to the experimental values with unmeasured signs. In most cases, the experimental and theoretical error bars are smaller than the sizes of symbols. The experimental value of $\mu=-2.5(8)\,\mu_N$ for the 3/2$^-$ state in $^{151}$Gd ($N=87$) is outside the scale of the figure.
}
\label{gdxxx-exp-the}
\end{figure*}

The moments for the gadolinium isotopes are shown in Fig.~\ref{gdxxx-exp-the}. The first case of $^{147}$Gd has $N=83$ and a deformation inferred from the quadrupole deformation of $\epsilon \simeq -0.05$, slightly oblate in shape. The deformation increases to $\epsilon \simeq +0.28$ at $^{159}$Gd, $N=95$. Clear band structures have developed in $^{153}$Gd, $N=89$, where $\epsilon \approx 0.2$, and are observed for the heavier isotopes, making it possible to identify the states with a specific Nilsson orbit. For the lighter isotopes, we cannot rely on band structures to identify the Nilsson orbit.

Whereas the  detailed agreement of the magnetic dipole moments with experiment is not very good, the trend is correct in that the moments start negative with values of around $-1\,\mu_N$ near $N=82$, tend to become positive around $N=90$ and then become negative again, but with smaller absolute values, near $N=96$. The strongest disagreement is for the $N=83$, [606]13/2$^+$ state, and for the $N=87$, [532]5/2$^-$ and [541]3/2$^-$ states.

For the 13/2$^+$ state in $^{147}$Gd, the measurement of Dafni {\it et al.}~\cite{Daf87}, which shows high quality data, is adopted in preference to the earlier measurement~\cite{Hausser1979}. The reason for the present discrepancy between theory and experiment is likely related to octupole-vibration mixing between the 7/2$^-$ and 13/2$^+$ states (see~\cite{Daf87} and references therein), which may not be captured in the DFT. More specifically, the first excited state of $^{146}$Gd is a 3$^-$ state at an excitation energy of 1.579 MeV which decays by a strong 37(5) W.u. $E3$ transition to the ground state. The 13/2$^+$ first-excited state of $^{147}$Gd at 0.997 MeV likewise decays with a similar $E3$ transition strength of 44.9$^{+2.4}_{-2.2}$ W.u. to the 7/2$^-$ ground state. Dafni {\it et al.}~\cite{Daf87} suggest that the structure of the 13/2$^+$ state is predominantly $[f_{7/2^-} \otimes 3^-]_{13/2^+}$. The measured magnetic dipole moment and transition strength are consistent with this scenario.

While there could be a $[i_{13/2^+} \otimes 3^-]_{7/2^-}$ contribution to the ground state moment, the energy separation between this octupole-coupled configuration and the $f_{7/2}$ single-particle orbit is considerable and mixing seems to be very small; the magnetic dipole moment of such an octupole-coupled configuration is $\mu \approx -2.6\,\mu_N$, which is roughly twice the experimental value in magnitude and would tend to increase the discrepancy between theory and experiment.

The discrepancies for the $N=87$, $^{151}$Gd case are more difficult to explain. The deformation of the nucleus and the magnetic dipole moments are consistent with the three observed moments being associated with a weak-coupling scenario between the $f_{7/2}$ neutron and the quadrupole excitation of the core: $0^+ \otimes f_{7/2}$ for the ground state and $2^+ \otimes f_{7/2}$ for the excited 3/2$^-$ and 5/2$^-$ states. However, the long lifetimes of the excited states do not support this interpretation. These moments remain a puzzle.

We also observe that at $N=89$, the configuration [521]3/2$^-$ has not converged. Whether the high-energy [501]3/2$^-$ configuration reproduces the measured magnetic dipole moment is not known because the experimental sign was not measured.

Finally, we note that experimental values are given in the Evaluated Nuclear Structure Data File (ENSDF)~\cite{ensdf} for the magnetic moments of the 5/2$^-$ states in the [521]3/2$^-$ bands of $^{155}$Gd and $^{157}$Gd, $N=91$ and $N=93$, respectively. These values are excluded from the most recent compilation~\cite{(Sto21)} because they are not measurements but model estimates used in the analysis of the muonic x-ray measurements on the 3/2$^-$ ground states~\cite{Lau83}.


\subsection{Dysprosium} \label{sect:aesDy}

\begin{figure*}
\begin{center}
\includegraphics[width=\textwidth]{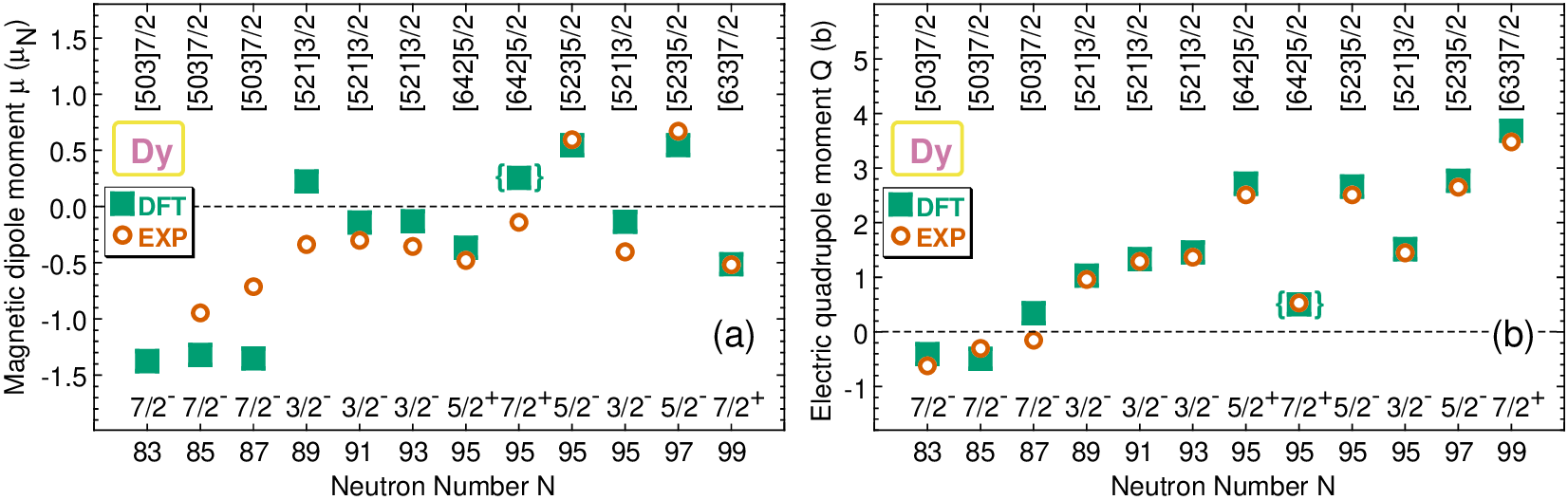}
\end{center}
\caption{Same as in Fig.~\ref{gdxxx-exp-the} but for the dysprosium isotopes ($Z=66$).
}
\label{dyxxx-exp-the}
\end{figure*}

\begin{figure}
\begin{center}
\includegraphics[width=0.50\textwidth]{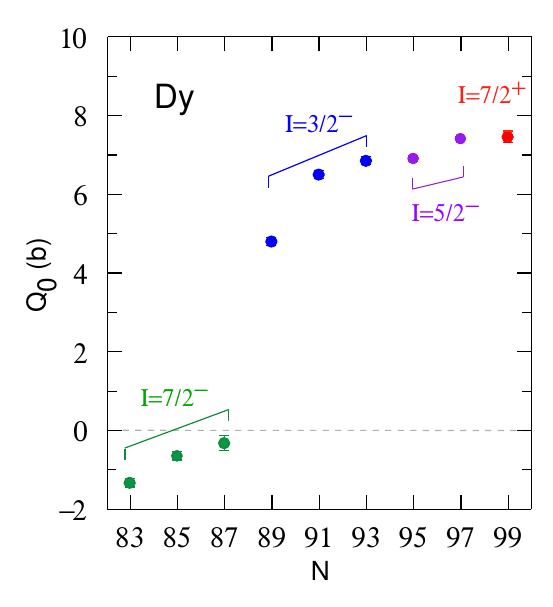}
\end{center}
\caption{Effective intrinsic electric quadrupole moments~(\protect\ref{Q0}) in the Dy isotopes versus neutron number.
}
\label{dy-Q0}
\end{figure}

The moments for the dysprosium isotopes are shown in Fig.~\ref{dyxxx-exp-the}. The calculated electric quadrupole moments are in excellent agreement with experiment. The intrinsic electric quadrupole moments are displayed versus neutron number in Fig.~\ref{dy-Q0}. Near $N=82$, the isotopes have weak oblate deformation $\epsilon \approx -0.06$, becoming prolate rotors with $\epsilon \approx +0.29$ toward $N=100$.
Similar to the Gd isotopes, band structures have developed at $N=89$, $^{155}$Dy, where again $\epsilon \approx 0.2$.

Overall agreement between theory and experiment for the magnetic dipole moments is better than for the Gd isotopes. The overall trend as the Fermi energy moves through the neutron shell is reproduced.
Note that four moments are calculated in $^{161}$Dy, $N=95$, and that the experimental trends are well reproduced. These include the [642]5/2$^+$ band head and its $7/2^+$ excited state.
Like in the Gd isotopes, discrepancies again occur for the $7/2^-$ ground states near $N=83$.

The evaluated data~\cite{ensdf} include a magnetic dipole moment value of $-0.119\,\mu_N$ for the 7/2$^-$ ground state of $^{149}$Dy, which we have excluded from the comparison of theory and experiment.
This value originates from a private communication to the 1989 tabulation of nuclear moments~\cite{Rag89}.
There are several other measurements included in the same private communication which agree with independent measurements and fit in with systematics of the the 7/2$^-$ states in neighboring nuclei, but this point for $^{149}$Dy is so very different that we suspect a typographical error.

\subsection{Erbium} \label{sect:aesEr}

\begin{figure*}
\begin{center}
\includegraphics[width=\textwidth]{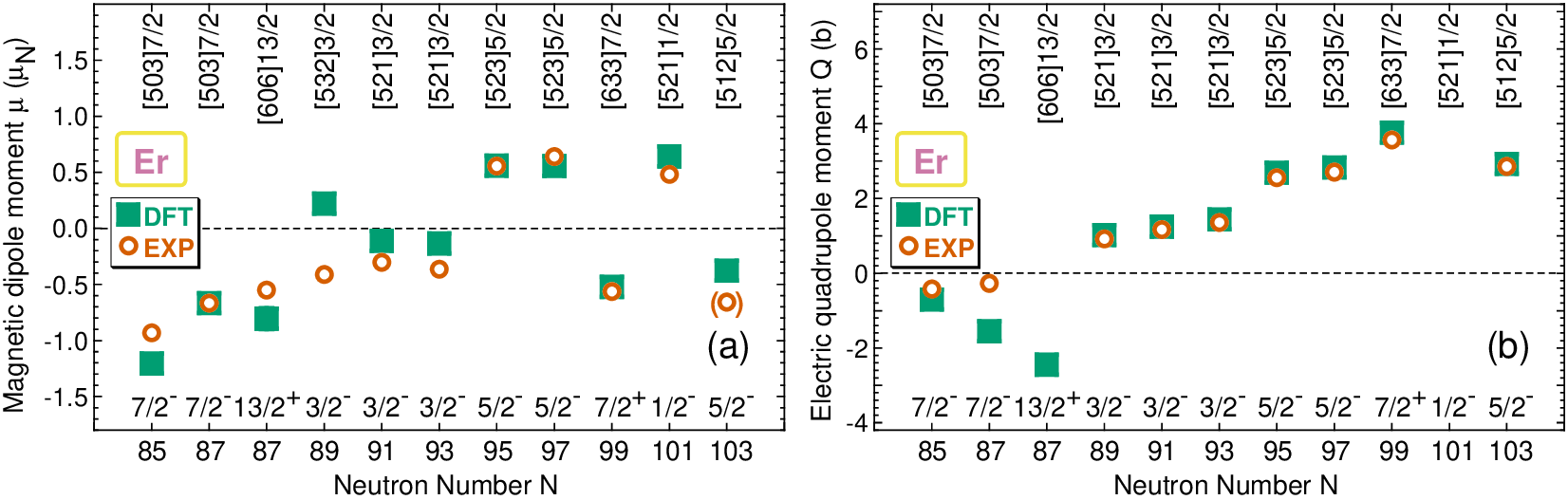}
\end{center}
\caption{Same as in Fig.~\ref{gdxxx-exp-the} but for the erbium isotopes ($Z=68$).
}
\label{erxxx-exp-the}
\end{figure*}

Experimental and theoretical moments for the erbium isotopes are shown in Fig.~\ref{erxxx-exp-the}.
The data now begin with $N=85$, however the trends in both the magnetic dipole moments and electric quadrupole moments are similar to those in the dysprosium isotopes. The description of the electric quadrupole moments is again excellent, except for the [503]7/2$^-$ state in $^{155}$Er, $N=87$. Ironically, the magnetic dipole moment for this state is in near perfect agreement with experiment. More generally, the magnetic dipole moments are quite well described, and those of the 7/2$^-$ and 13/2$^+$ states in the $N=85,87$ isotopes, which were poorly described by theory in the Gd and Dy sequences, are now better described, possibly related to a weakening influence of octupole correlations.

There are several cases where the same Nilsson state is considered at the same neutron number in both the Er and Dy isotopes. In these cases, the comparison of experimental and theoretical magnetic dipole moments is seen to be very similar in both isotopic sequences.

Similar to the Gd and Dy isotopes, band structures have developed at $N=89$, $^{157}$Er, where again $\epsilon \approx 0.2$.

On the theoretical side, we observe that at $N=89$, the configuration [521]3/2$^-$ has not converged, and the lowest converged one [532]3/2$^-$ does not reproduce the measured magnetic dipole moment. Inspection of a Nilsson level scheme reveals that these two orbits exhibit a weak avoided crossing near $\epsilon \approx 0.2$, which may be related to the convergence challenges encountered in the DFT. This conjecture requires further investigation. There may be some room for experimental uncertainty on the Nilsson assignment for this state as well. The adopted spin is $I=3/2$ from atomic-beam magnetic resonance \cite{EKSTROM196}. The assignment of the configuration and hence the parity was then based on the proximity of $K=3/2$ Nilsson orbits to the Fermi surface. Thus [521]3/2$^-$ was assigned, although [651]3/2$^+$ was also listed as a possibility \cite{EKSTROM196}. Unfortunately, the data on excited band members built on the ground state are not altogether transparent. The signature $\alpha=+1/2$ band beginning at 5/2$^-$ is connected by an unobserved low-energy transition to the ground state, while the signature $\alpha=-1/2$ sequence (of which the ground state is a member) is identified only at higher spins (15/2$^-$ to 27/2$^-$). Members of the [651]3/2$^+$ band have been assigned above 17/2$^+$. No band structure associated with [532]3/2$^-$ is reported in ENSDF. For this reason, despite the shortcomings of the evidence favoring the [521]3/2$^-$ assignment, there is no experimental basis for a [532]3/2$^-$ assignment. This conclusion accounts for the disagreement between the experimental moment and the DFT calculation, as the lowest converged configuration appears not to be that of the experimental ground state.

\subsection{Ytterbium} \label{sect:aesYb}

\begin{figure*}
\begin{center}
\includegraphics[width=\textwidth]{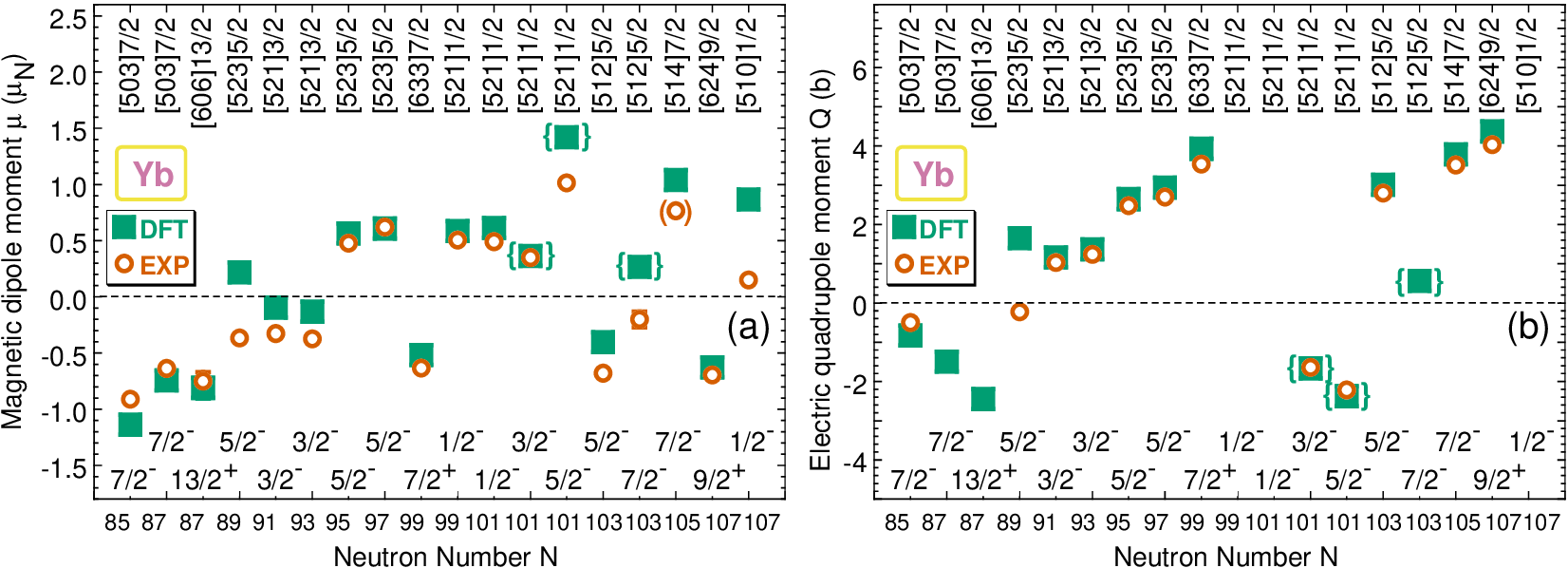}
\end{center}
\caption{Same as in Fig.~\ref{gdxxx-exp-the} but for the ytterbium isotopes ($Z=70$).
}
\label{ybxxx-exp-the}
\end{figure*}

\begin{figure}
\begin{center}
\includegraphics[width=0.50\textwidth]{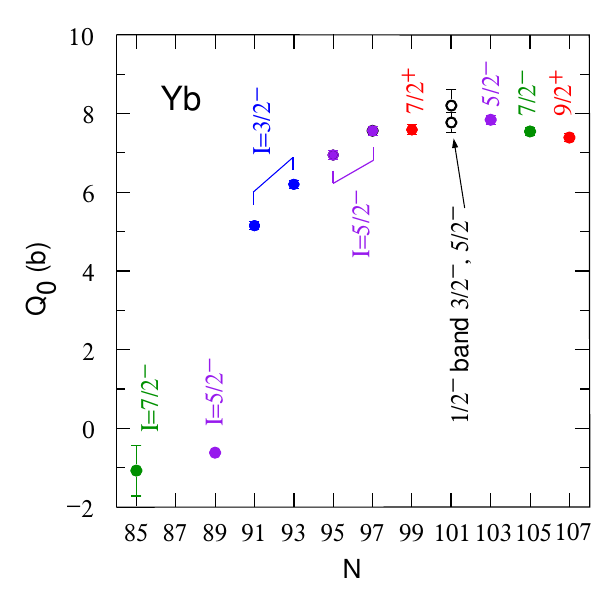}
\end{center}
\caption{Effective intrinsic electric quadrupole moments~(\protect\ref{Q0}) in the Yb isotopes versus neutron number.
}
\label{yb-Q0}
\end{figure}

Experimental and theoretical moments for the ytterbium isotopes are shown in Fig.~\ref{ybxxx-exp-the}. The intrinsic electric quadrupole moments are displayed versus neutron number in Fig.~\ref{yb-Q0}.

Note that there is an error in the value of the electric quadrupole moment of the $3/2^-$ state in $^{171}$Yb as reported by the Evaluated Nuclear Structure Data File (ENSDF)~\cite{Bag18}, taken from~\cite{Sto16} and still present in~\cite{(Sto21)}. The correct value is $Q(3/2^-)=-1.64(8)$\,b. The  data are from Plingen, Wolbeck, Schroder~\cite{Pli71} who measured the ratio $Q(3/2^-)/Q(5/2^-)$. The value of $Q(5/2^-) = -2.22(7)$\,b is taken as the reference. It appears that in \cite{Sto16} the {\em spectroscopic} electric quadrupole moments were accidentally scaled using the ratio applicable to the {\em intrinsic} electric quadrupole moments.


The description of the electric quadrupole moments of the Yb isotopes is again excellent, except for the [523]5/2$^-$ state in $^{159}$Yb, $N=89$, for which the experimental and theoretical magnetic dipole moments disagree. In contrast, the experimental and theoretical moments (both quadrupole and dipole) agree very well for the [523]5/2$^-$ states in $^{163}$Er and $^{165}$Er, $N=95,97$. The difference is that the states in the Er isotopes are prolate deformed, whereas the nominally equivalent state in $^{159}$Yb is weakly oblate deformed. It is also relevant to note that Neugart {\it et al.}~\cite{Neu83} pointed out that their measured moments in $^{159}$Yb have no straightforward explanation, which may suggest the need for a new measurement.

Apart from this case in $^{159}$Yb, the agreement between theory and experiment for the magnetic dipole moments of the Yb isotopes is rather good. For example, the moments of the  7/2$^-$ and 13/2$^+$ states in the $N=85,87$ isotopes, which were poorly described by theory in the Gd and Dy sequences, are now well described.

A band-like structure built on the $\nu 2f_{7/2}$ ground state is observed in $^{157}$Yb, $N=87$. The electric quadrupole moment is not measured, but the electric quadrupole moment of the $5/2^-$ ground state of $^{159}$Yb, $N=89$, implies a very weakly deformed oblate shape, $\epsilon \approx -0.03$. There is limited spectroscopic data on $^{159}$Yb; no band built on the ground state has been observed, but bands built on  $\nu 1i_{13/2}$ are observed to high spin. More complete spectroscopy on $^{159}$Yb would help resolve the disparity between the theoretical and experimental moments.

As shown in Fig.~\ref{yb-Q0}, the intrinsic electric quadrupole moments
of the Yb isotopes begin with small negative values (weakly oblate shapes) and then jump to positive values (prolate shapes) and increase smoothly to $\epsilon \approx 0.3$ with neutron number above $N=90$, independent of the spin and configuration of the state. There is a slight decrease in deformation beyond $N=101$ (midshell is at $N=104$).

Among the odd-$A$ ytterbium isotopes, we have cases where $K=1/2$, for which the so-called magnetic decoupling effect can have a strong effect on the magnetic dipole moments. In the case of [521]1/2$^-$ in $^{171}$Yb, the magnetic decoupling effect is reasonably well reproduced by theory for the lowest three states of the band. In contrast, the magnetic dipole moment of the [510]1/2$^-$ band head in $^{177}$Yb is not well described by the DFT calculation. It will be seen below that there are similar discrepancies between theory and experiment for this orbit in the W and Os isotopes, where it is closer to the Fermi surface. It will also be shown in section \ref{sect:510band} that the magnetic dipole moment of this orbit is not accounted for by the Nilsson model, which is unusual.

\subsection{Hafnium} \label{sect:aesHf}

\begin{figure*}
\begin{center}
\includegraphics[width=\textwidth]{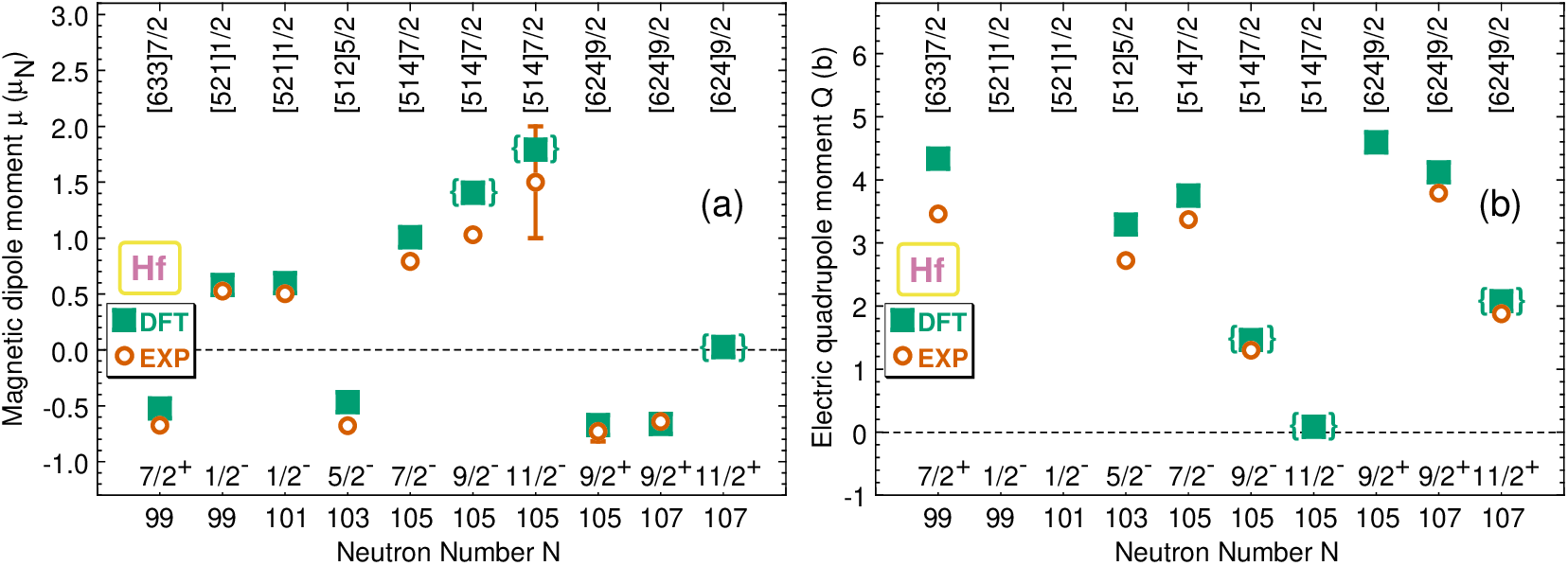}
\end{center}
\caption{Same as in Fig.~\ref{gdxxx-exp-the} but for the hafnium isotopes ($Z=72$).
}
\label{hfxxx-exp-the}
\end{figure*}

Moments for the hafnium isotopes, beginning with $^{171}$Hf at $N=99$ and concluding with $^{179}$Hf at $N=107$, are shown in Fig.~\ref{hfxxx-exp-the}.
All of the nuclides in this isotopic sequence have rotational bands observed above the states for which the moments have been measured. The quadrupole deformation is almost constant; the experimental spectroscopic electric quadrupole moments all correspond to an intrinsic electric quadrupole moment of $Q_0 \simeq 7$\,b and a deformation of $\epsilon \approx 0.25$.

Overall, the electric quadrupole moments are not as well described as in the lower-$Z$ cases; however, the trends are accurately described. The agreement between theory and experiment for the magnetic dipole moments is very good.

A feature of the Hf sequence is the measurement and calculation of four magnetic dipole moments in $^{177}$Hf ($N=105$), three of them in the [514]7/2$^-$ band. The DFT magnetic dipole moments in the [514]7/2$^-$ band all exceed the experimental values by a few tenths of a nuclear magneton. It is instructive to analyze these data in terms of the particle-rotor model, beginning with the expression for a pure Nilsson band with $K \neq 1/2$ :
\begin{equation}\label{eq:gNils}
    \mu(I)=g_R I + (g_K-g_R)\frac{K^2}{I+1},
\end{equation}
where the rotational $g$~factor, $g_R$, and the projection of the s.p.\ magnetic moment on the symmetry axis of the nucleus, $g_K$, are treated as parameters.
In Table~\ref{tab:Hf177gKgR} the experimental $g$~factors are compared with the DFT calculation as well as the Nilsson model (i.e. particle-rotor model with no Coriolis mixing) and a full particle-rotor (PR) calculation~\cite{Rag88} including (unattenuated) Coriolis mixing. The effective values of $g_K$ and $g_R$ for each case are shown in the last two rows. The Nilsson model and particle-rotor calculations assumed $\epsilon =0.25$ and quenched the spin $g$~factor of the neutron to 0.7 times the free nucleon $g$~factor. These calculations (Nilsson and PR)  also set $g_R=0.315(30)$, which is the experimental $g(2^+_1)$ value of $^{176}$Hf~\cite{Alfter1996}. The theoretical $g$~factors from the Nilsson and PR models are given in Table~\ref{tab:Hf177gKgR} with an estimate of the uncertainty arising from the uncertainty in the adopted $g_R$ value, which scales as $1-K^2/I(I+1)$.

In principle, this analysis separates the contributions to the magnetic dipole moment from the core and the odd neutron. There is close agreement between the DFT calculation and the Nilsson model for $g_K$. Whereas it is usually claimed that $g_R$ in the odd-$N$ nucleus is smaller than $g(2^+_1)$ in the even-even neighbor (see~\cite{(Boh75)}, pp. 256 and 303), the DFT predicts the opposite trend here. The PR calculation suggests that Coriolis interactions effect a small reduction in $g_R$, but the observed effective $g_R$ is smaller than that predicted by any of the models. The empirical $g_K$ is also smaller than that given (or adopted) by the models. The difference between the DFT and empirical effective $g$~factors is about 20\% for $g_K$ and about 50\% for $g_R$.

Although this analysis may suggest that the difference between the DFT and experiment in the [514]7/2$^-$ band of $^{177}$Hf is primarily due to the core contribution, it would be premature to draw a firm conclusion. These observations do, however, suggest a direction for further investigation: they indicate the value of precise data on excited states in rotational bands to resolve the origins of discrepancies between theory and experiment.

\begin{table}[htbp]
\begin{center}
\caption{Comparison of theory and experiment for members of the [514]7/2$^-$ band in $^{177}$Hf. The last two rows give $g_K$ and $g_R$ derived from Eq.~(\ref{eq:gNils}) as described in the text.}
\label{tab:Hf177gKgR}
\begin{tabular}{cllll}
\hline
\multicolumn{1}{c}{$I^{\pi}$} &
\multicolumn{4}{c}{$g$~factor} \\
\cline{2-5}
 & Exp. & DFT & Nilsson & PR\\
 \hline
 7/2$^-$  & $+0.2267(2)$ & $+0.288$ & $+0.276(7)$  & $+0.267(7)$ \\
 9/2$^-$  & $+0.229(7)$  & $+0.312$ & $+0.290(15)$ & $+0.279(15)$ \\
 11/2$^-$ & $+0.27(9)$   & $+0.326$ & $+0.298(20)$ & $+0.285(20)$ \\
\hline
$g_K$: & $+0.225$  & $+0.268$ & $+0.265$ & $+0.258$ \\
$g_R$: & $+0.233$  & $+0.355$ & $+0.315(30)$ \tablenotemark[1] & $+0.299$ \\
\hline
\end{tabular}
\tablenotetext{This is the experimental value of $g(2^+_1)$ in $^{176}$Hf~\cite{Alfter1996}.}
\end{center}
\end{table}

Finally, a remark on the experimental data for $^{175}$Hf. We have adopted the moments for the [512]5/2$^-$ ground state from Stone~\cite{Sto16}, taken from a laser spectroscopy measurement published in 2002~\cite{Nie02}. ENSDF (with cut off date in 2005)~\cite{ensdf} has adopted older values from NMR measurements. The quadrupole moments differ little, although the laser measurement is much more precise. The magnetic moments, however, differ somewhat, namely $-0.54(3)$~\cite{ensdf} cf. $-0.677(9)$~\cite{Sto16,Nie02}. While the difference between theory and experiment is greater for the laser spectroscopy measurement, it fits well with the trend observed for other [512]5/2$^-$ band heads in neighboring nuclei. See section \ref{sect:aesBands} below for further discussion.

\subsection{Tungsten} \label{sect:aesW}

\begin{figure}
\begin{center}

\includegraphics[width=0.48\textwidth]{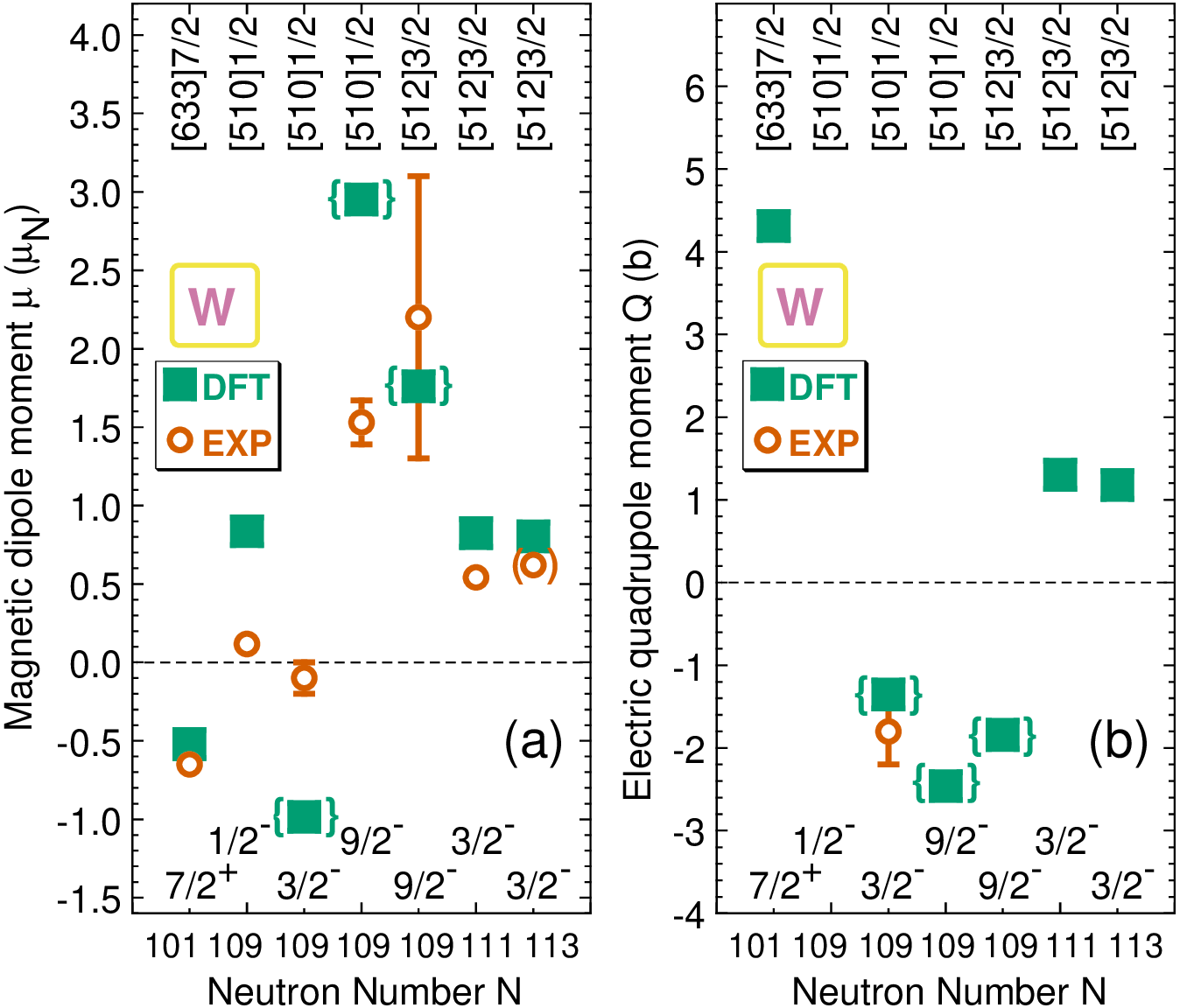}
\end{center}
\caption{Same as in Fig.~\ref{gdxxx-exp-the} but for the tungsten isotopes ($Z=74$).
}
\label{w_xxx-exp-the}
\end{figure}

Moments for the tungsten isotopes, beginning with $^{175}$W at $N=101$ and then jumping to  $^{183,185,187}$W  ($N=109, 111, 113$), are shown in Fig.~\ref{w_xxx-exp-the}. The single measured electric quadrupole moment, for the 3/2$^-$ member of the [510]1/2$^-$ ground-state band, is in agreement with theory. Deformations in the vicinity of $\epsilon \approx 0.20 - 0.25$ are expected for these isotopes~\cite{Naz90}.

Turning to the magnetic dipole moments, there is good agreement for the [633]7/2$^+$ band head in $^{175}$W, and satisfactory agreement for the heads of the [512]3/2$^-$ bands in   $^{183,185,187}$W. However, the description of the 1/2$^-$, 3/2$^-$, and 9/2$^-$ members of the [510]1/2$^-$ ground-state band in $^{183}$W is poor. A similarly poor description of this band is found in $^{189}$Os. This case will be discussed in more detail below in section \ref{sect:510band}.


\subsection{Osmium} \label{sect:aesOs}

\begin{figure}
\begin{center}
\includegraphics[width=0.48\textwidth]{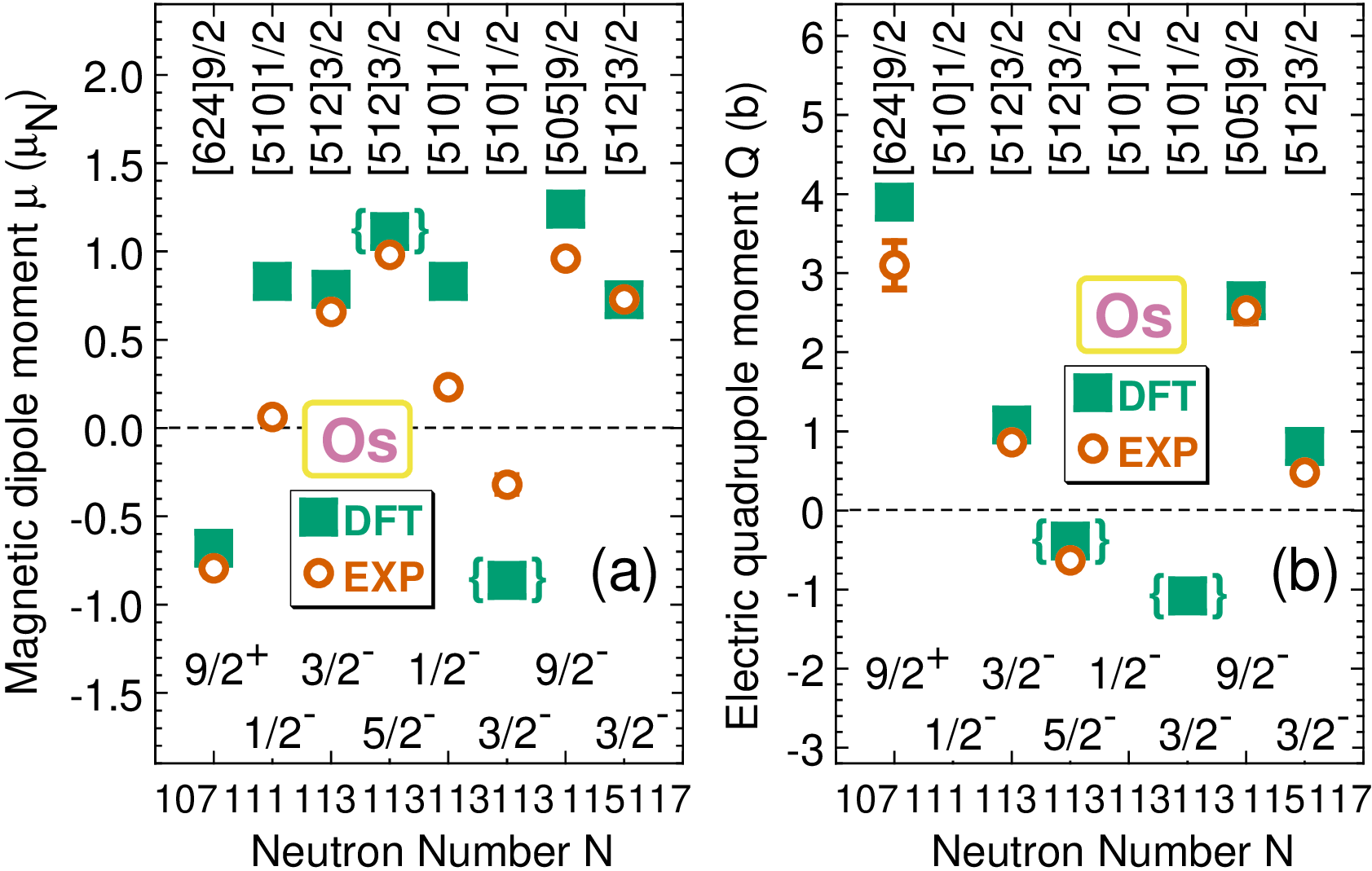}
\end{center}
\caption{Same as in Fig.~\ref{gdxxx-exp-the} but for the osmium isotopes ($Z=76$).
}
\label{osxxx-exp-the}
\end{figure}

Moments for osmium isotopes with neutron numbers between $N=107$ and $N=117$, including $^{183}$Os, $^{187}$Os, $^{189}$Os, $^{191}$Os, and $^{193}$Os, are shown in Fig.~\ref{osxxx-exp-the}. The electric quadrupole moments are well described despite a tendency to sit above the experimental data (regardless of whether the moment is positive or negative).

Concerning the magnetic dipole moments, those associated with the [510]1/2$^-$ band are poorly described. Otherwise, the agreement is good, although there is a tendency for the theoretical values to sit above the data. The case of the [510]1/2$^-$ band is discussed in section  \ref{sect:510band} below.

\subsection{[521]3/2$^-$, [523]5/2$^-$, and [512]5/2$^-$ bands}
\label{sect:aesBands}

The band heads and some excited states in the [521]3/2$^-$, [523]5/2$^-$, and [512]5/2$^-$ bands have known moments across several isotopes in several isotopic sequences. The trends and behavior as $Z$ varies are discussed here.

The [521]3/2$^-$ band has extensive data, specifically in $^{155,157,159}$Gd ($N=91,93,95$), $^{155,157,159,161}$Dy ($N=89,91,93,95$), $^{159,161}$Er ($N=91,93$), and in $^{161,163}$Yb ($N=91,93$). We exclude $^{157}$Er ($N=89$) from the discussion here due to the DFT convergence issue discussed in section \ref{sect:aesEr}. From an experimental perspective, it could be included in this group.

The DFT theory for [521]3/2$^-$ virtually always sits above experiment. It is striking that this trend is independent of both $N$ and $Z$. The discrepancy in the case of the [521]3/2$^-$ bands may be related to Coriolis interactions, which mix bands with $\Delta K=1$. $K=3/2$ bands are special in that they interact directly with $K=1/2$ bands for which both Coriolis interactions and the $M1$ operator connect states with $\Omega=1/2$ and $\Omega =-1/2$, giving rise to the so-called decoupling or signature-splitting effects.

An early examination of energy spacings in [521]3/2$^-$ bands across the rare earth region and their Coriolis interactions with $K=1/2$ states was presented by Joshi and Sood~\cite{Jos74}. Concerning the magnetic dipole moments, the [521]3/2$^-$ band in $^{155}$Gd was studied in some detail by Stuchbery, Lampard, and Bolotin~\cite{Stu98}. Experimental moments were compared with particle-rotor calculations. It was found that Coriolis interactions can increase the magnitude of the calculated ground-state magnetic dipole moment by a factor of three, from about $-0.35\,\mu_N$ to about $-1.05\,\mu_N$. Thus, a possible source of the difference between theory and experiment for the [521]3/2$^-$ bands
might be related to an underestimation of Coriolis-like mixing.

A similar pattern of the theoretical magnetic moments sitting above the data is observed for the [512]5/2$^-$ band, which has been measured at $N=85$ in oblate weakly-deformed $^{149}$Gd, and also at $N=103$ in prolate-deformed $^{171}$Er, $^{173}$Yb and $^{175}$Hf. In this case, there is only indirect Coriolis mixing with $K=1/2$ bands; there has to be a $K=3/2$ intermediary. It may be difficult to unravel the role of Coriolis interactions in the absence of moment data on excited states in the bands.

Moments associated with [523]5/2$^-$ have been measured in
$^{153}$Gd ($N=89$),
$^{161}$Dy ($N=95$),
$^{163,165}$Er ($N=95,97$) and
$^{159,165,167}$Yb ($N=89,95,97$). The band is observed in weakly deformed oblate isotopes at $N=89$ and then in prolate deformed nuclei at $N=95,97$. Apart from the problematic case of $^{159}$Yb discussed above in section~\ref{sect:aesYb}, the theoretical and experimental magnetic moments are in good agreement.

The cases discussed in the section suggest that the level of agreement between theory and experiment is more correlated with the particular Nilsson orbit associated with the state in question than it is with the atomic or mass numbers of the nucleus in which the state appears.

\subsection{[510]1/2$^-$ band} \label{sect:510band}

Bands associated with [510]1/2$^-$ have measured magnetic dipole moments in $^{177}$Yb, $^{183}$W and $^{187,189}$Os. In all cases, the agreement between theory and experiment is poor. The measured moments are small $\lesssim +0.2\,\mu_N$, whereas the calculated moments are $\approx +0.8$\,$\mu_N$.

This systematic trend in theory versus experiment is similar in standard particle-rotor calculations for this Nilsson orbit. For example, calculations based on the Woods-Saxon potential for $^{183}$W presented in~\cite{Stu02a} give a ground-state moment of $\mu =+0.53$\,$\mu_N$. The calculation includes a hexadecapole  deformation ($\beta_2 \approx +0.23$, $\beta_4 \approx -0.06$; see~\cite{Naz90})  and Coriolis mixing. To date, no reasonable particle-rotor parameter set has been found that gives a ground-state moment near the experimental value of $\mu = +0.117$\,$\mu_N$. The same situation applies to the equivalent states in $^{187}$Os and $^{189}$Os. It is hardly credible that all of the measurements are incorrect, so there appears to be some significant physical effect missing from both the present DFT and standard particle-rotor descriptions of these states.

\subsection{Residuals}

\begin{figure}
\begin{center}
\includegraphics[width=0.48\textwidth]{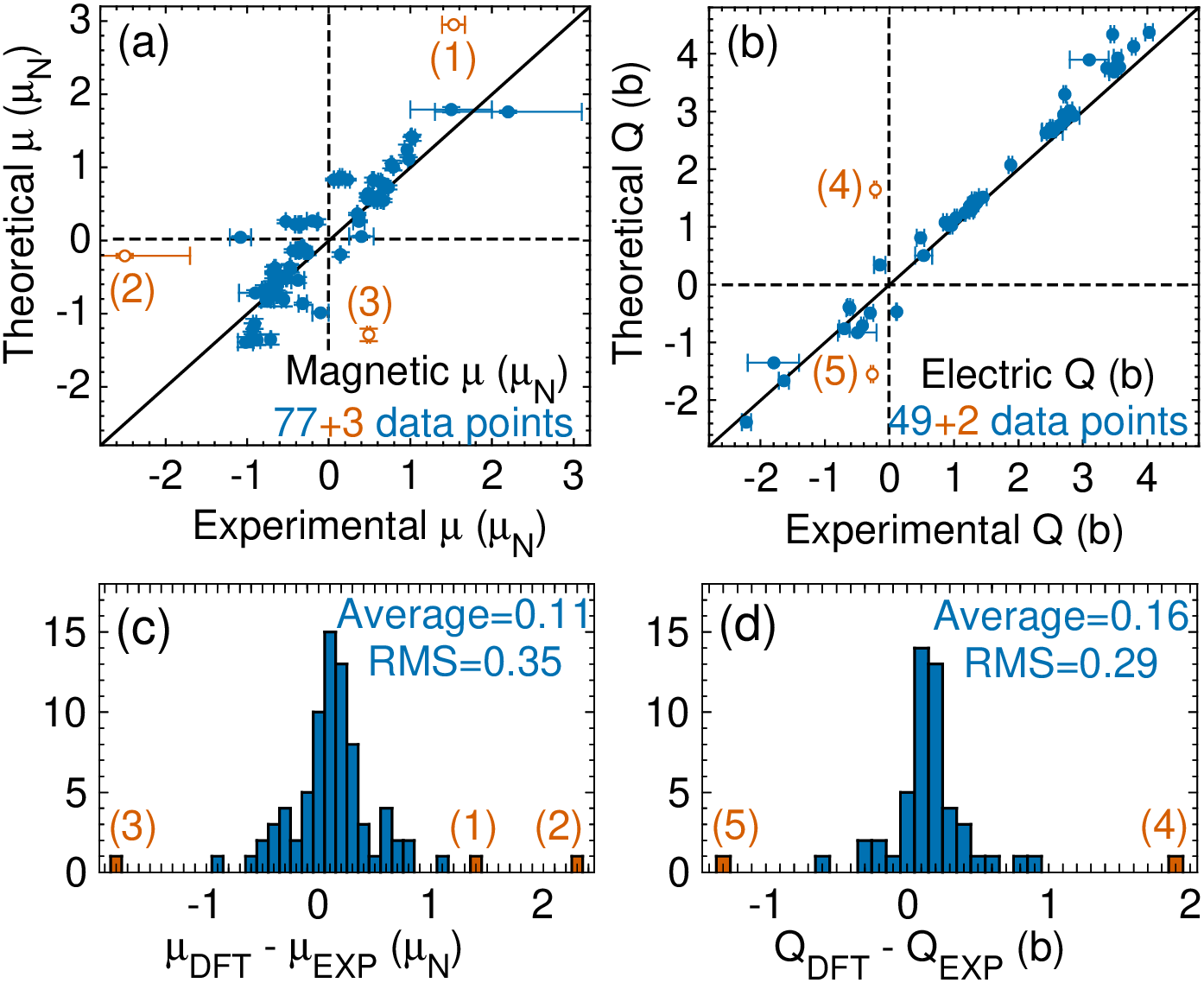}
\end{center}
\caption{Summary comparison of the experimental and theoretical DFT magnetic dipole moments $\mu$ (left panels) and electric quadrupole moments $Q$ (right panels) in odd-$N$ nuclei between gadolinium ($Z=64$) and osmium  ($Z=76$). Open symbols correspond to the outliers identified in this work, see text.
\label{yyxxx-ttt-v67-170-HFT-N16-siq-UDF1_318j}
}
\end{figure}

\begin{figure*}
\begin{center}
\includegraphics[width=\textwidth]{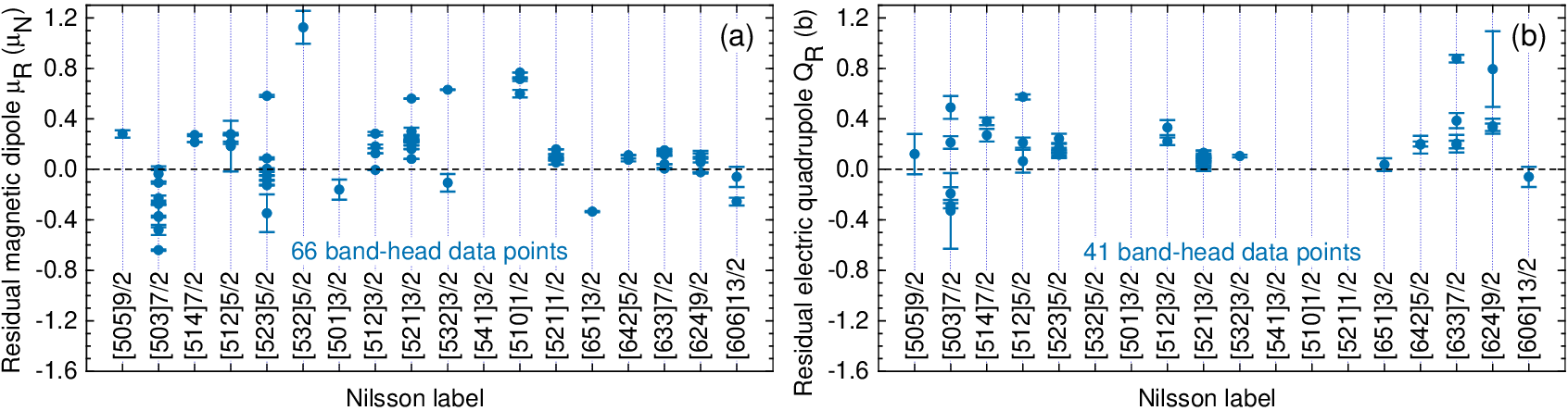}
\end{center}
\caption{Residuals of the magnetic dipole (a) and electric quadrupole (b) moments, see Figs.~\protect\ref{yyxxx-ttt-v67-170-HFT-N16-siq-UDF1_318j}(c) and (d), respectively, plotted for the band-head configurations $I=K$ identified by the self-consistent Nilsson labels, with outliers excluded.
}
\label{yyxxx-ttt-v67-170-HFT-N16-siq-UDF1_318j-25-ResX-Nil}
\end{figure*}

The comparison of theory and experiment above has been focused and detailed. We now present a more global view.

In Figs.~\ref{yyxxx-ttt-v67-170-HFT-N16-siq-UDF1_318j}(a) and (b), we present the theoretical vs.\ experimental results obtained in this work for the magnetic dipole $\mu$ and electric quadrupole $Q$ moments, respectively. Similarly, Figs.~\ref{yyxxx-ttt-v67-170-HFT-N16-siq-UDF1_318j}(c) and (d), display the histograms of residuals $\mu_{\text{R}}\equiv\mu_{\text{DFT}}-\mu_{\text{EXP}}$ and $Q_{\text{R}}\equiv{}Q_{\text{DFT}}-Q_{\text{EXP}}$ listed in Tables~\ref{yyxxx-ttt-v67-170-HFT-N16-siq-UDF1.print1}  and~\ref{yyxxx-ttt-v67-170-HFT-N16-siq-UDF1.print2}.

In Fig.~\ref{yyxxx-ttt-v67-170-HFT-N16-siq-UDF1_318j}, open symbols correspond to the outliers identified in this work for the magnetic dipole moments,
\begin{itemize}
    \item[(1)] $^{183}$W, [510]1/2$^-$, $I=1/2^-$,
    \item[(2)] $^{151}$Gd, [541]3/2$^-$, $I=3/2^-$,
    \item[(3)] $^{147}$Gd, [606]13/2$^+$, $I=13/2^+$,
\end{itemize}
and for the electric quadrupole moments,
\begin{itemize}
    \item[(4)] $^{159}$Yb, [523]5/2$^-$, $I=5/2^-$,
    \item[(5)] $^{155}$Er, [503]7/2$^-$, $I=7/2^-$.
\end{itemize}
Each of these outliers has been discussed in the detailed comparison of theory and experiment in the preceding subsections.
Note that the residuals' distributions shown here do not result from any specific statistical property of the results, so the outlier definitions are somewhat arbitrary. Our choice, therefore, reflects an unusual value that may have either a theoretical or experimental origin, which has yet to be firmly identified; see, however, the discussion above. The calculated average and RMS deviations between theory and experiment, which are given in Figs.~\ref{yyxxx-ttt-v67-170-HFT-N16-siq-UDF1_318j}(c) and (d), serve to illustrate the overall agreement and are almost unaffected by including or excluding the listed outlier values.

We observe that the relative quality of describing electric quadrupole moments is higher than that of the magnetic dipole moments. Indeed, for the former, the RMS deviation of 0.29\,b is about 15\% and 7\% of the maximum oblate and prolate values. For the latter, the RMS deviation of 0.35\,$\mu_N$ is about 30\% of the corresponding maximum values. In addition, both moments are, on average, systematically overestimated by 0.16\,b and 0.11\,$\mu_N$, respectively. These generic features may become the subject of future analyses aiming at a slight decrease in the electric quadrupole and magnetic dipole polarizability strengths of the underlying DFT functional.

However, as illustrated in Fig.~\ref{yyxxx-ttt-v67-170-HFT-N16-siq-UDF1_318j-25-ResX-Nil} and remarked upon in sections \ref{sect:aesBands} and \ref{sect:510band}, the pattern of agreement with data also depends on the type of configuration involved. It thus may require addressing individual properties of different states. One example is the magnetic moment of the [510]1/2$^-$, $I=1/2^-$ state, which appears in several nuclei, and is in all cases poorly described by the DFT calculation, as well as by established particle-rotor models. In contrast, the [503]7/2$^-$ orbit (see Fig.~\ref{yyxxx-ttt-v67-170-HFT-N16-siq-UDF1_318j-25-ResX-Nil}) shows a range of (dis)agreement between theory and experiment.
The implication is that the discrepancies between theory and experiment may have different structural origins, which needs to be explored in future work.

\section{Conlusions} \label{sect:conclusions}

In this work, we presented a comprehensive analysis of the electromagnetic properties of odd-$N$, even-$Z$ open-shell isotopes ranging from gadolinium to osmium. We performed nuclear DFT calculations for 154 nuclei in that region, determining 22 oblate and 22 prolate self-consistent quasiparticle configurations for each, and compared that vast database of theoretical results with experimental data available for 82 states. For this study, no parameters were adjusted.

We demonstrated that tagging blocked quasiparticle states with those identified at the border of the region in $^{192}$Dy enables the tracking of specific single-particle structures from one nearly semi-magic isotope to the next, across the region of large deformations. In these sequences, nuclear moments either remain noticeably similar or undergo well-defined changes related to specific quasiparticle crossings as a function of deformations that vary with neutron numbers. Simultaneously, the excitation energies change rapidly, with a given structure appearing near the ground states in a very narrow range of neutron numbers.

Comparison with data speaks volumes. As our no-parameter study shows for the first time, the main features of the magnetic dipole and electric quadrupole moments seem to be captured; therefore, the key focus is on understanding the details. On one hand, the overall comparison reveals that the particular functional used here, UNEDF1, with the Landau parameter $g_0'$ adjusted near doubly magic nuclei, slightly overestimates both moments, indicating the need to include these data in future fine-tuned versions. On the other hand, specific analyses of individual nuclei suggest that different structural effects may play unique roles in various isotopes. It seems there is no single solution that will fix all issues at once, but there's no harm in continuing to look for one.

The methodology applied here is still worth improving in several aspects, such as missing terms or deficiencies in the interactions or functionals, triaxial and/or octupole deformability and collectivity, contributions from two-body meson-exchange currents~\cite{(Miy24),(Wib25c)}, configuration interaction, $K$-mixing, cranking, or possibly even more unknown ones. The primary strategy must nevertheless be based on a holistic approach, with different effects not being invoked in particular nuclei but tested against a large body of calculations and data, so that the improved agreement in one place does not induce worsening agreement where the theory already performs well.

\begin{acknowledgments}
We acknowledge interesting discussions with Witek Nazarewicz.
This work was partially supported by the STFC Grant Nos.~ST/P003885/1 and~ST/V001035/1, by the Australian Research Council Discovery Grant No. DP250100400, by the Academy of Finland under the Academy Project No. 339243 and
by a Leverhulme Trust Research Project Grant.
We thank the CSC-IT Center for Science Ltd., Finland, and the IFT Computer Center at the University of Warsaw, Poland, for the allocation of computational resources.
This project was partly undertaken on the Viking Cluster,
which is a high performance compute facility provided by the
University of York. We are grateful for computational support
from the University of York High Performance Computing
service, Viking and the Research Computing team.
We thank Grammarly\textsuperscript{\textregistered} for its support with English writing.
\end{acknowledgments}


%

\label{LastBibItem}

\clearpage

\title{Supplemental material for: \protect\mytitle}

\date{\today}

\begin{abstract}
\myabstract
\end{abstract}

\maketitle

\begin{figure*}
\begin{center}
\includegraphics[width=0.93\textwidth]{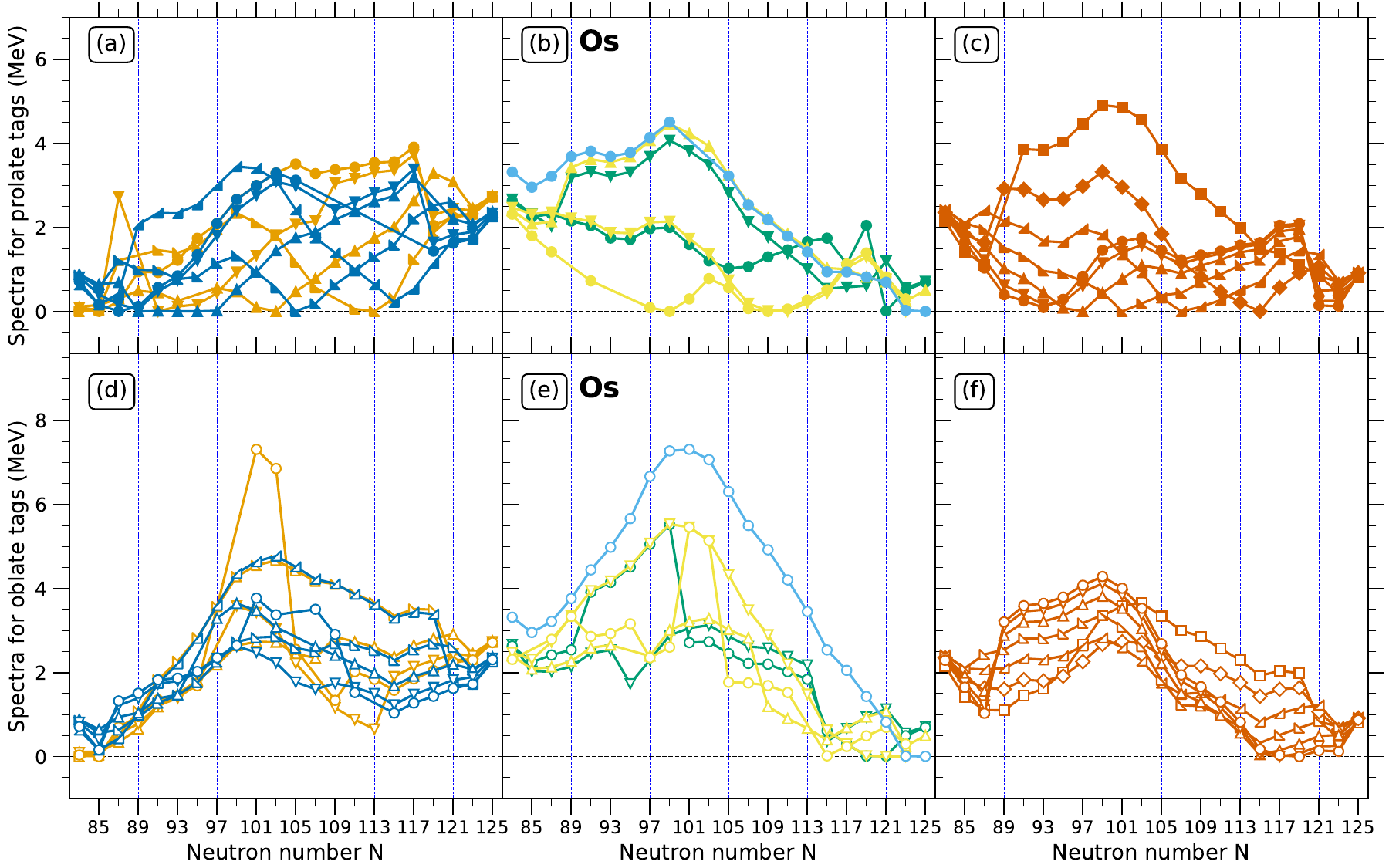}\vspace*{-5mm}
\end{center}
\caption{Same as in Fig.~\protect\ref{fig:Spect_Dy} but for the osmium isotopes.
}
\label{fig:Spect_Os}
\begin{center}
\includegraphics[width=0.98\textwidth]{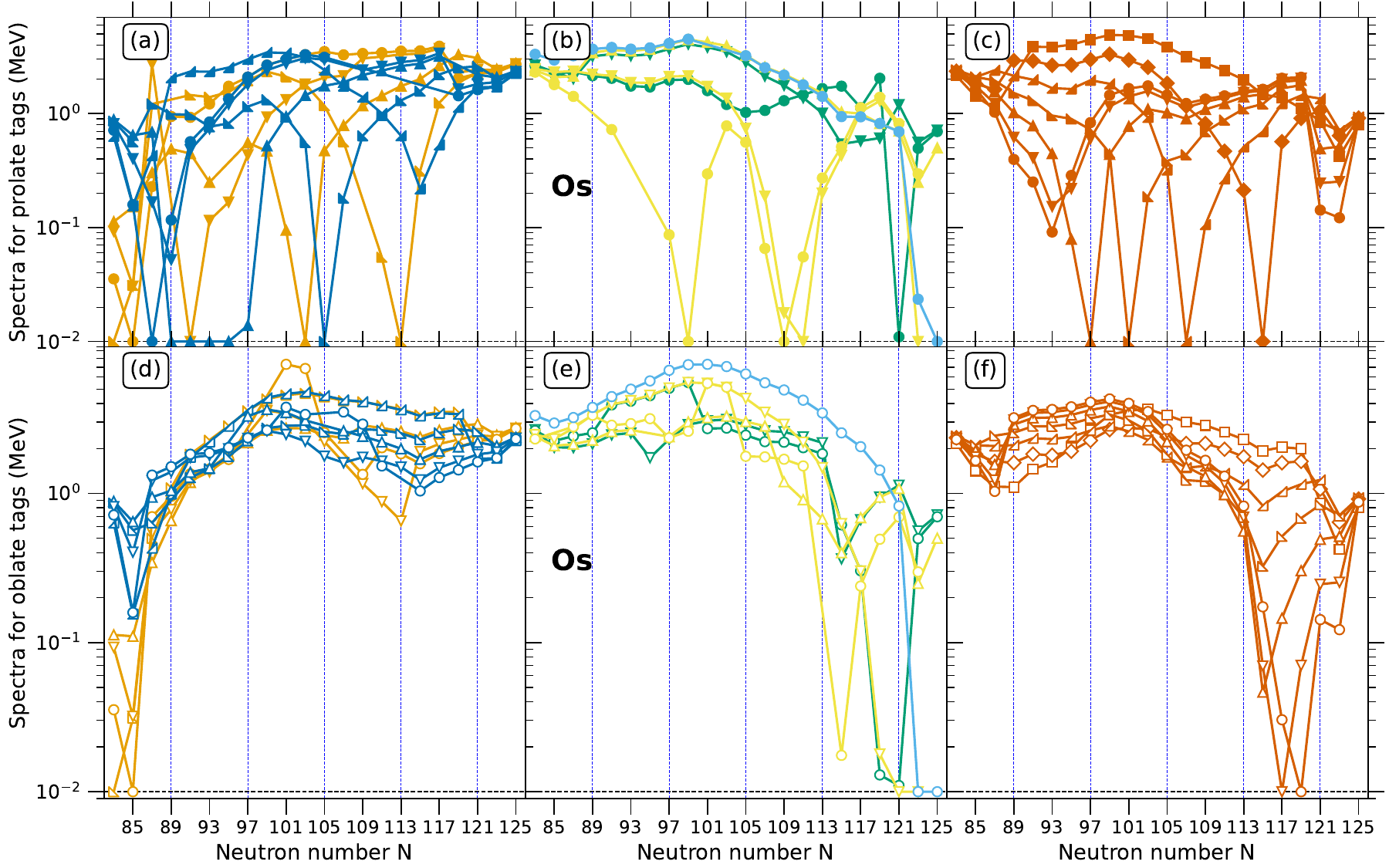}\vspace*{-5mm}
\end{center}
\caption{Same as in Fig.~\protect\ref{fig:Spect_Os} but plotted in a logarithmic scale with  $E_{\text{exc}}=0$ (ground states) plotted artificially at $E_{\text{exc}}=0.01$\,MeV.
\label{fig:Spect_log_Os}
}
\end{figure*}

\begin{widetext}

\section{Databases and files}

The computer files and Excel databases related to the present project are available in~\cite{rep-GdOs}. The computer files are described in the file {\tt Readme.txt} there.

\subsection{Databases with self-consistent HFB results}

In this section, we describe the contents and structure of the seven Excel databases named as {\tt yyxxx-ttt-vws-170-HFT-AMP-N16-siq-UDF1.3uub.23d.version16.xls}, where the first two characters denote element names, that is, {\tt yy = gd, dy, er, yb, hf, w\_, or os}.

\begin{figure*}[t]
\begin{center}`
\includegraphics[width=\textwidth]{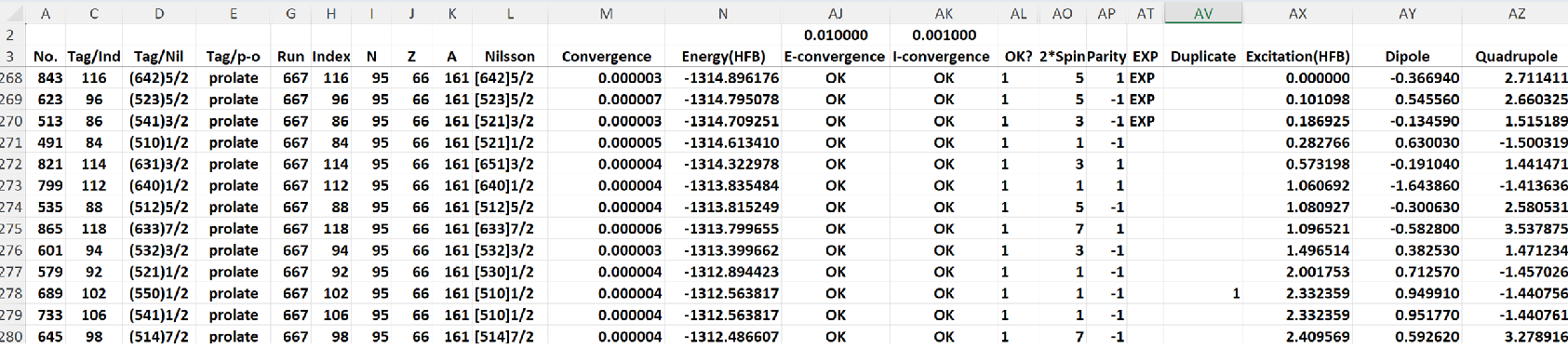}
\end{center}
\caption{Example of the records of Excel databases with self-consistent HFB results, available in the raw data repository~\cite{rep-GdOs}, see text.
}
\label{gdxxx-ttt-vws-170-HFT-AMP-N16-siq-UDF1.3uub.23d}
\end{figure*}

In Fig.~\ref{gdxxx-ttt-vws-170-HFT-AMP-N16-siq-UDF1.3uub.23d}, we present 13 records corresponding to the lowest quasiparticle configurations identified in $^{161}$Dy, selected from 968 records listed in the file {\tt{dyxxx-ttt-vws-170-HFT-AMP-N16-siq-UDF1.3uub.23d.version16.xls}}.

Each Excel database contains 968 records (rows), representing results for 22 oblate and 22 prolate tag states across 22 isotopes, $N=83,\ldots,125$, which means $2\times 22\times 22=968$ calculations per element. Each record includes 22 visible columns, listed on page~\pageref{page1} below, along with several auxiliary invisible columns that are not recommended for viewing.

Records in each Excel database are ordered first by the number of neutrons {\bf N} and next by the calculated HFB excitation energy {\bf Excitation(HFB)}. However, the ordering of records can be changed arbitrarily. For example, by ordering the databases by the record numbers {\bf No.}\label{page-no.}, they become ordered by the oblate and prolate tag states, next by the number of neutrons {\bf N}, and finally by the tag indices {\bf Tag/Ind}. This particular ordering is suitable for producing plots of results such as in Figs.~\ref{fig:Q2,mu_Dy_prol_2}--\ref{fig:Spect_log_Dy} and \ref{fig:Q2,mu_Gd_prol_2}--\ref{fig:Spect_log_Os}.
Another useful ordering is first by the number of neutrons {\bf N}, next by the {\bf Parity}, next by the doubled spin {\bf 2*Spin}, and finally by the HFB excitation energy {\bf Excitation(HFB)}. This particular ordering is suitable for identifying the lowest states for a given parity and spin.

The Excel databases do not include information on the members of rotational bands, cf.\ results marked by asterisks (*) in Tables~\ref{yyxxx-ttt-v67-170-HFT-N16-siq-UDF1.print1} and~\ref{yyxxx-ttt-v67-170-HFT-N16-siq-UDF1.print2}. To find that, one must review individual output files available in the raw data repository~\cite{rep-GdOs}. For example, in Fig.~\ref{dy161-116-667-170-HFT-AMP-N16-siq-UDF1.318j.out.xxxx-01}, we display fragments of the output file {\tt dy161-116-667-170-HFT-AMP-N16-siq-UDF1.318j.out} showing the spectroscopic electric quadrupole (a) and magnetic dipole (b) moments calculated for the $\Omega=5/2$ [642]5/2 rotational band, $I^\pi=5/2^+\ldots95/2^+$, in $^{161}$Dy. This output file can be identified by the part of its name ({\tt{116-667}}) that corresponds to columns {\bf Tag/Ind} and {\bf Run} listed in record 843 of the Excel database {\tt{dyxxx-ttt-vws-170-HFT-AMP-N16-siq-UDF1.3uub.23d.version16.xls}} displayed in Fig.~\ref{gdxxx-ttt-vws-170-HFT-AMP-N16-siq-UDF1.3uub.23d}.

\noindent The headers of visible columns in the databases containing self-consistent HFB results are as follows.\label{page1}

\vspace*{2mm}\hspace*{20mm}
\begin{minipage}{15.5cm}\begin{itemize}
\item[{\bf No.             }] Predefined consecutive number of the record, see description on page~\pageref{page-no.}.\vspace*{-2mm}
\item[{\bf Tag/Ind         }] Indices of tag states, every second from 084 to 126, that define a specific numbering to identify the computer files stored in~\cite{rep-GdOs}. The indices are in one-to-one correspondence with the Nilsson labels of tag states given in column {\bf Tag/Nil}.\vspace*{-2mm}
\item[{\bf Tag/Nil         }] Nilsson labels $(N_0n_z\Lambda)K$ of tag states.\vspace*{-2mm}
\item[{\bf Tag/p-o         }] Denomination of prolate and oblate tag states.\vspace*{-2mm}
\item[{\bf Run             }] Denomination of the performed run of the code to identify the computer files stored in~\cite{rep-GdOs}.\vspace*{-2mm}
\item[{\bf Index           }] Index of the blocked state to identify the computer files stored in~\cite{rep-GdOs}.\vspace*{-2mm}
\item[{\bf N               }] Number of neutrons $N$.\vspace*{-2mm}
\item[{\bf Z               }] Number of protons $Z$.\vspace*{-2mm}
\item[{\bf A               }] Number of nucleons $A=N+Z$.\vspace*{-2mm}
\item[{\bf Nilsson         }] Self-consistent Nilsson labels $[N_0n_z\Lambda]K$ of blocked quasiparticle states.\vspace*{-2mm}
\item[{\bf Convergence     }] Convergence indicator $E_{\text{conv}}$ (in MeV) equal to the stability energy defined in Eq.~(37) of Ref.~\cite{(Dob97b)}.\vspace*{-2mm}
\item[{\bf Energy(HFB)     }] The total HFB energy (in MeV).\vspace*{-2mm}
\item[{\bf E-convergence   }] Value ``OK'' denotes energy-converged calculations defined by $|E_{\text{conv}}|<0.01$\,MeV.\vspace*{-2mm}
\item[{\bf I-convergence   }] Value ``OK'' denotes spin-converged calculations defined by $|\langle{}\hat{I}_z\rangle-I|<0.001\,\hbar$.\vspace*{-2mm}
\item[{\bf OK?             }] Value of ``1'' denotes energy- and spin-converged calculations.\vspace*{-2mm}
\item[{\bf 2*Spin          }] Doubled spin $I$.\vspace*{-2mm}
\item[{\bf Parity          }] Values of $+1$ and $-1$ denote configurations of positive and negative parity, respectively.\vspace*{-2mm}
\item[{\bf EXP             }] Value ``EXP'' denotes the calculated configuration associated with the experimental result listed in Tables~\ref{yyxxx-ttt-v67-170-HFT-N16-siq-UDF1.print1} and~\ref{yyxxx-ttt-v67-170-HFT-N16-siq-UDF1.print2}.\vspace*{-2mm}
\item[{\bf Duplicate       }] Value of ``1'' denotes duplicated configuration.\vspace*{-2mm}
\item[{\bf Excitation(HFB) }] Excitation energy (in MeV) defined as the difference between the total HFB energy {\bf Energy(HFB)} of the given converged configuration ({\bf OK?}=1) and that of the lowest one calculated in a given isotope for any tag state.\vspace*{-2mm}
\item[{\bf Dipole          }] Spectroscopic magnetic dipole moment $\mu$ (in $\mu_N$).\vspace*{-2mm}
\item[{\bf Quadrupole      }] Spectroscopic electric quadrupole moment $Q$ (in barn). We remind the reader that for $\Omega=1/2$ band heads, the values listed in the Excel databases pertain to the $I=3/2$ members of the corresponding rotational bands.\vspace*{-2mm}
\end{itemize}\end{minipage}

\begin{figure*}
\begin{center}
\includegraphics[width=0.45\textwidth]{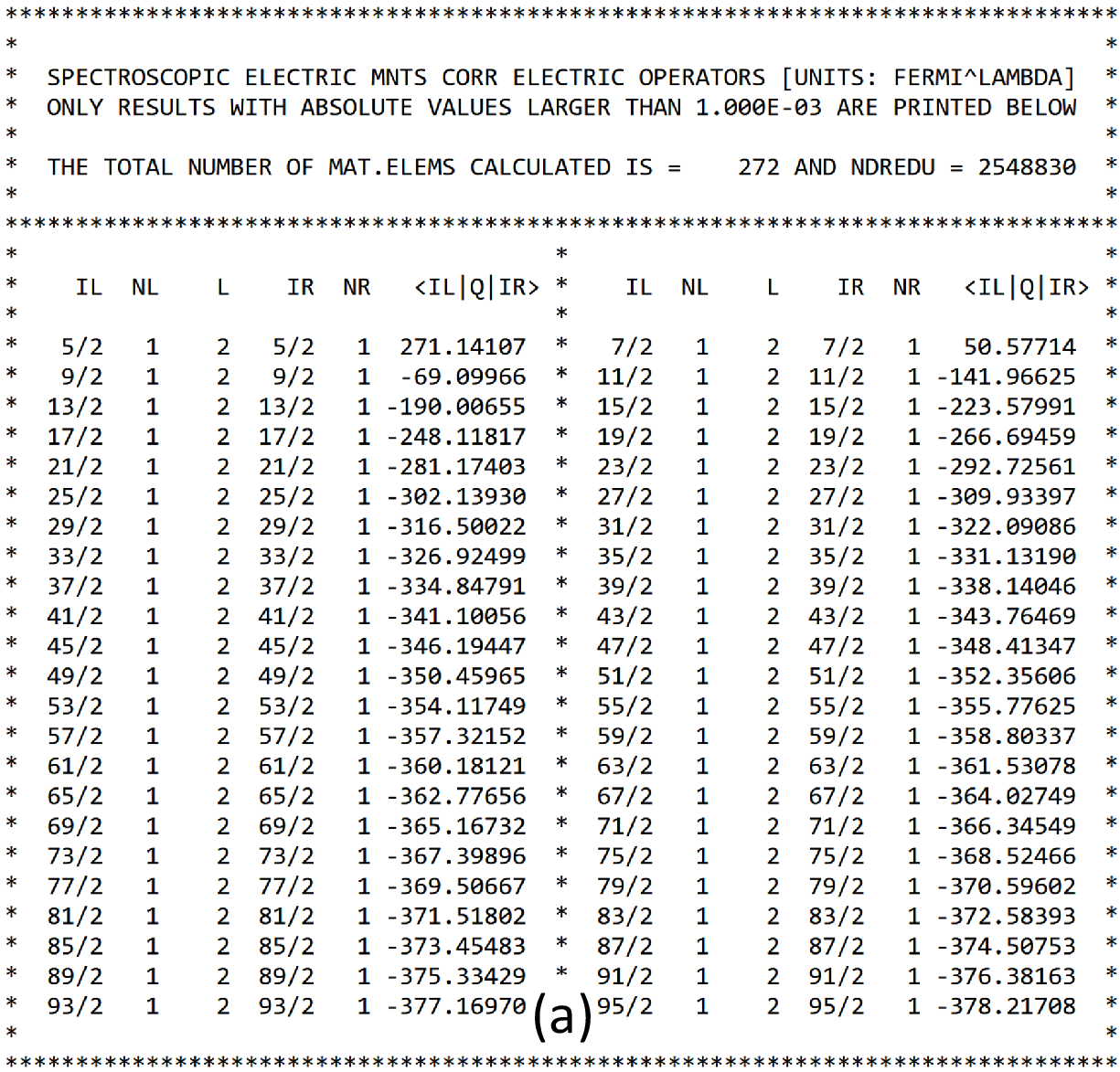}\hspace*{5mm}
\includegraphics[width=0.45\textwidth]{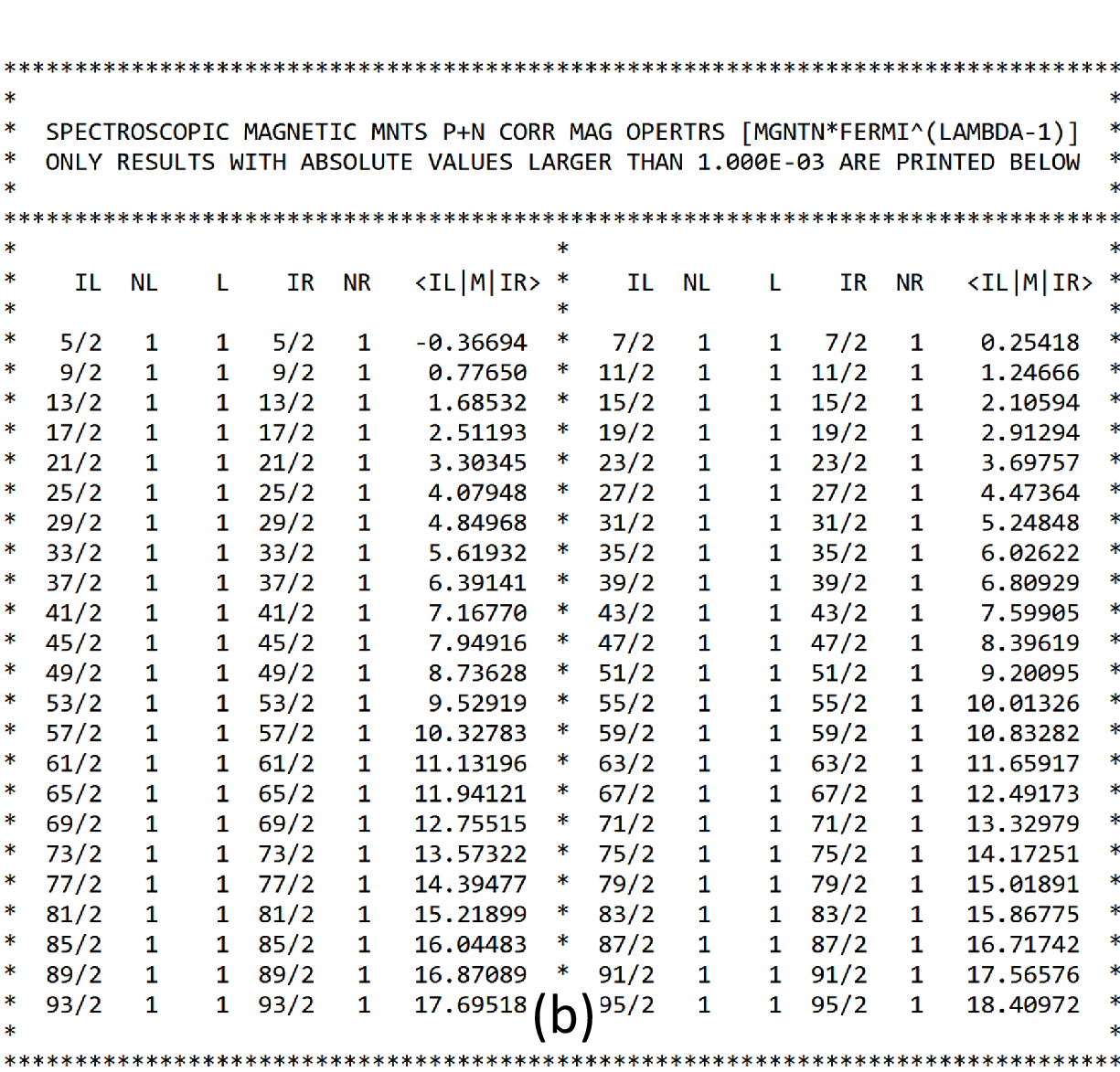}
\end{center}
\caption{Fragments of the output file {\tt dy161-116-667-170-HFT-AMP-N16-siq-UDF1.318j.out} showing the spectroscopic electric quadrupole (a) and magnetic dipole (b) moments calculated for the
$\Omega=5/2$ [642]5/2 rotational band, $I^\pi=5/2^+\ldots95/2^+$, in $^{161}$Dy}.
\label{dy161-116-667-170-HFT-AMP-N16-siq-UDF1.318j.out.xxxx-01}
\end{figure*}

\newpage
\subsection{Databases with the deformation-dependence results}

In this section, we describe the contents and structure of the six Excel databases named as {\tt dy161-ttt-671-170-k+zz0-HFT-AMP-N16-siq-UDF1.333b.OBSERVABLE.NILSSON.version04.xls}, where:
\begin{itemize}
\item
'OBSERVABLE' equals 'energy' (total intrinsic HFB energy $E_{\text{HFB}}$), 'magnetic' (spectroscopic magnetic dipole moment $\mu$), or 'quadrupole' (spectroscopic electric quadrupole moment $Q$),
\item
'NILSSON' equals 'tag' (the database columns refer to the Nilsson labels of the prolate tag states) or 'def' (the database columns refer to the Nilsson labels of the self-consistent states).
\end{itemize}

In Fig.~\ref{dy161-ttt-671-170-k+zz0-HFT-AMP-N16-siq-UDF1.333b.magnetic.tag}, we show the first 31 records out of 682 listed in the file {\tt{dy161-ttt-671-170-k+zz0-HFT-AMP-N16-siq-UDF1.333b.magnetic.tag.version04.xls}}, which correspond to the proton tag state (510)1/2 in $^{161}$Dy.

Each Excel database contains 682 records corresponding to the results obtained for 31 deformations and 22 prolate tag states, performed for $^{161}$Dy. The contents of columns in each database are listed on page~\pageref{page2}. Records in each Excel database are sorted by the record number  {\bf No.}\label{page-no.2}, which aligns with ordered prolate-state-tag indices (see {\bf Tag/Ind} on page~\pageref{page1}), followed by sorted deformations {\bf Q20tot(b)}.

\begin{figure*}[t]
\begin{center}`
\includegraphics[width=\textwidth]{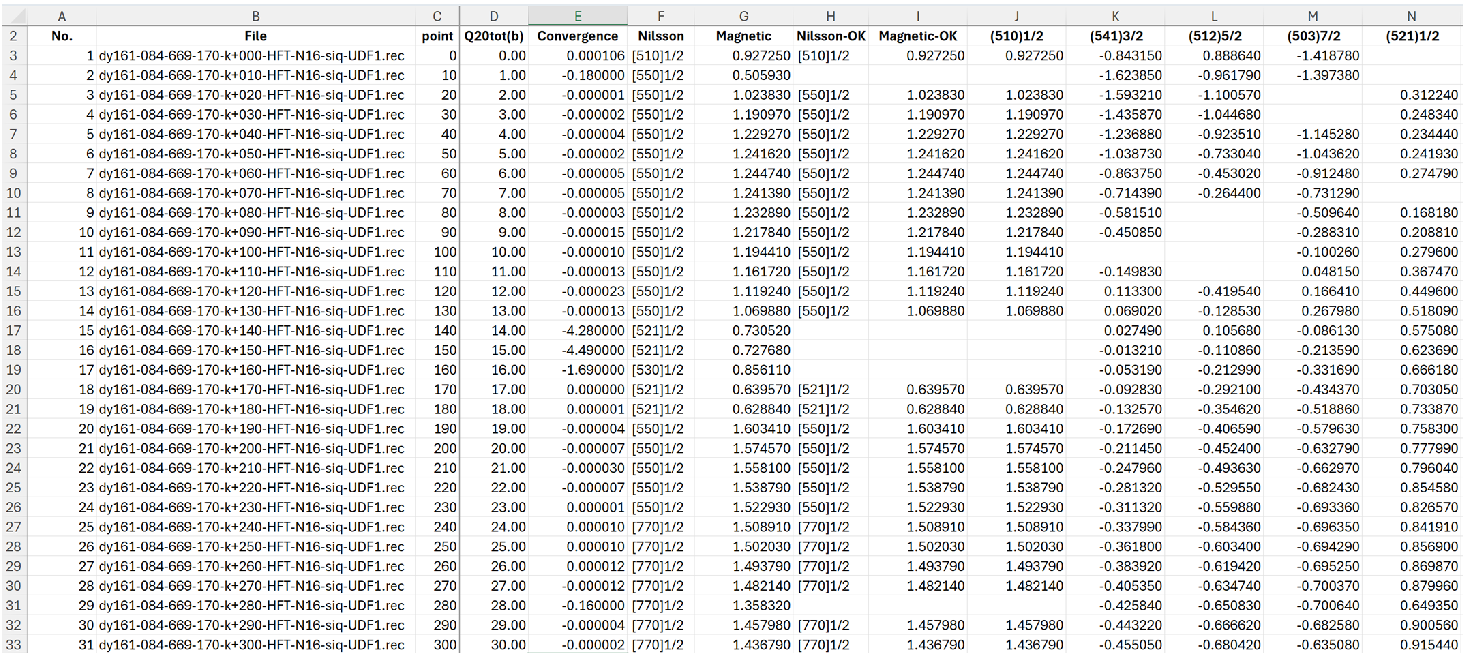}
\end{center}
\vspace*{-5mm}
\caption{Example of the records of Excel databases with deformation-dependence results, available in the raw data repository~\cite{rep-GdOs}, see text.
\vspace*{-5mm}
}
\label{dy161-ttt-671-170-k+zz0-HFT-AMP-N16-siq-UDF1.333b.magnetic.tag}
\end{figure*}

\newpage
\noindent The headers of the columns in the Excel databases containing the deformation-dependence results are as follows.\label{page2}

\vspace*{2mm}\hspace*{20mm}
\begin{minipage}{15.5cm}\begin{itemize}
\item[{\bf No.             }] Predefined consecutive number of the record, see description on page~\pageref{page-no.2}.\vspace*{-2.5mm}
\item[{\bf File            }] The name of the corresponding file stored in~\cite{rep-GdOs}.\vspace*{-2.5mm}
\item[{\bf point           }] Indices of deformations, from 0 to 30, to identify the computer files stored in~\cite{rep-GdOs}.\vspace*{-2.5mm}
\item[{\bf Q20tot(b)       }] Total axial intrinsic constrained quadrupole moments $Q_{20}^{\text{tot}}, Eq.~(\protect\ref{Qtot})$ (in barn).\vspace*{-2.5mm}
\item[{\bf Convergence     }] Convergence indicator $E_{\text{conv}}$ (in MeV) equal to the stability energy defined in Eq.~(37) of Ref.~\cite{(Dob97b)}.\vspace*{-2.5mm}
\item[{\bf Nilsson         }] Self-consistent Nilsson labels $[N_0n_z\Lambda]K$ of the blocked quasiparticle state.\vspace*{-2.5mm}
\item[{\bf Energy(HFB)     }] The total HFB energy (in MeV). Applies to files with 'OBSERVABLE' equal 'energy'.\vspace*{-2.5mm}
\item[{\bf Magnetic        }] Spectroscopic magnetic dipole moment $\mu$ (in $\mu_N$). Applies to files with 'OBSERVABLE' equal 'magnetic'.\vspace*{-2.5mm}
\item[{\bf Quadrupole      }] Spectroscopic electric quadrupole moment $Q$ (in barn). We remind the reader that for $\Omega=1/2$ band heads, the values listed in the Excel databases pertain to the $I=3/2$ members of the corresponding rotational bands. Applies to files with 'OBSERVABLE' equal 'quadrupole '.\vspace*{-2.5mm}
\item[{\bf IntrinsicQ      }] Effective intrinsic electric quadrupole moment $Q^{\text{intr}}_{\text{eff}}$ (in barn), Eq.~(\ref{Qrot2eff}). Applies to files with 'OBSERVABLE' equal 'quadrupole '.\vspace*{-2.5mm}
\item[{\bf IntrinsicQ/Q20  }] Scaled effective intrinsic electric quadrupole moment, column {\bf IntrinsicQ} divided by column {\bf Q20tot(b)}. Applies to files with 'OBSERVABLE' equal 'quadrupole '.\vspace*{-2.5mm}
\item[{\bf Nilsson-OK      }] Value of {\bf Nilsson} for energy-converged calculation defined by $|E_{\text{conv}}|<0.01$\,MeV.\vspace*{-2.5mm}
\item[{\bf Energy(HFB)-OK  }] Value of {\bf Energy(HFB)} for energy-converged calculation defined by $|E_{\text{conv}}|<0.01$\,MeV.\vspace*{-2.5mm}
\item[{\bf Magnetic-OK     }] Value of {\bf Magnetic} for energy-converged calculation defined by $|E_{\text{conv}}|<0.01$\,MeV.\vspace*{-2.5mm}
\item[{\bf IntrinsicQ/Q20-OK     }] Value of {\bf IntrinsicQ/Q20} for energy-converged calculation defined by $|E_{\text{conv}}|<0.01$\,MeV.\vspace*{-2.5mm}
\item[]                       Multiple columns follow.\vspace*{-2.5mm}
\item[{\boldmath{$(N_0n_z\Lambda)K$}\unboldmath}] Prolate tag state's Nilsson label. Applies to files with 'NILSSON' equal 'tag'.\vspace*{-2.5mm}
\item[{\boldmath{$[N_0n_z\Lambda]K$}\unboldmath}] Self-consistent Nilsson label. Applies to files with 'NILSSON' equal 'def'.\vspace*{-2.5mm}
\end{itemize}\end{minipage}

\vspace*{-2mm}
\section{Plots of results obtained for all elements considered in this study, shown in a consistent style (start on the next page).}
\end{widetext}

\begin{figure*}
\begin{center}
\includegraphics[width=0.93\textwidth]{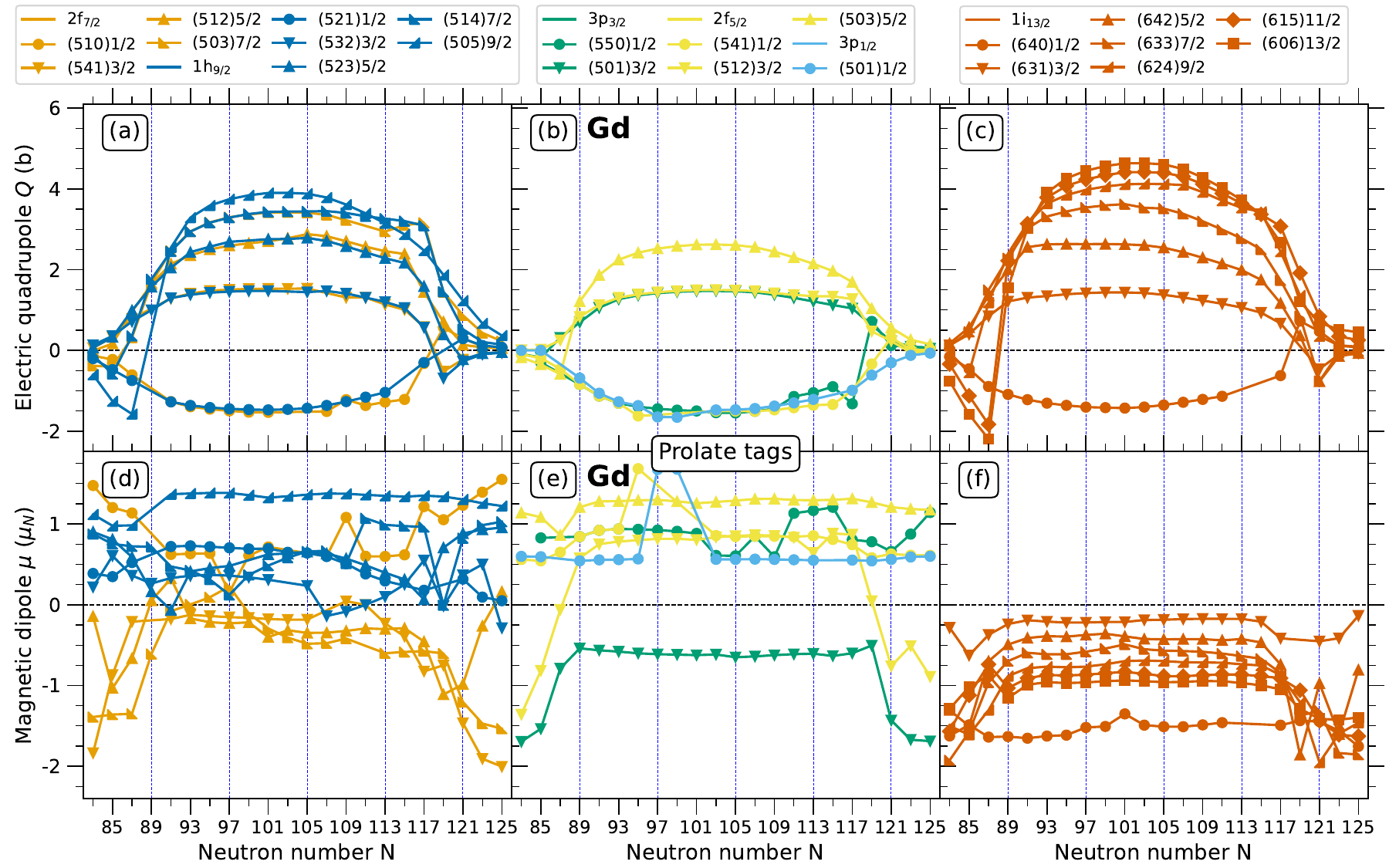}\vspace*{-1mm}
\end{center}
\caption{Same as in Fig.~\protect\ref{fig:Q2,mu_Dy_prol_2} but for the gadolinium isotopes.
}
\label{fig:Q2,mu_Gd_prol_2}
\begin{center}
\includegraphics[width=0.93\textwidth]{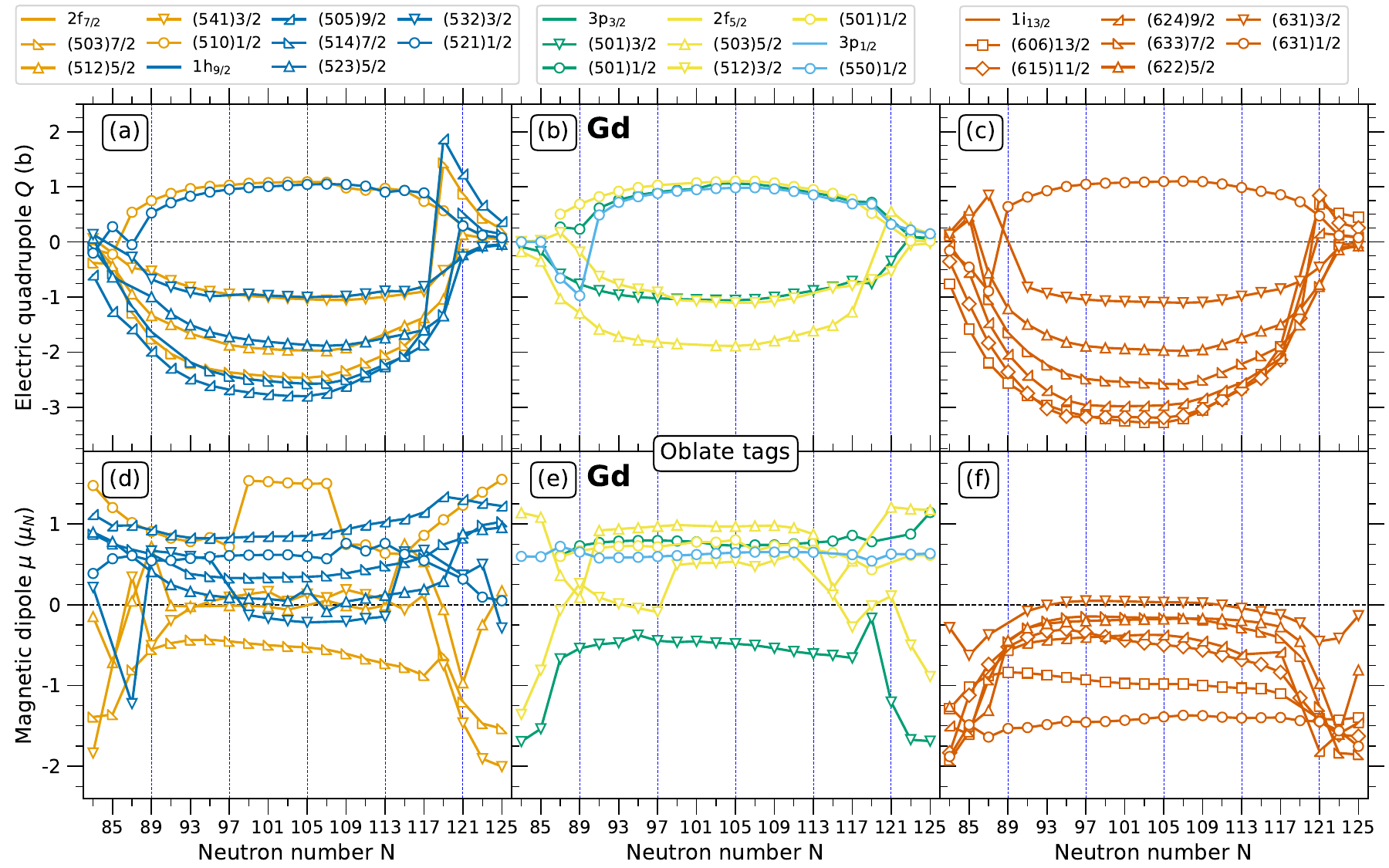}\vspace*{-1mm}
\end{center}
\caption{Same as in Fig.~\protect\ref{fig:Q2,mu_Dy_obl_2} but for the gadolinium isotopes.
}
\label{fig:Q2,mu_Gd_obl_2}
\end{figure*}

\begin{figure*}
\begin{center}
\includegraphics[width=0.93\textwidth]{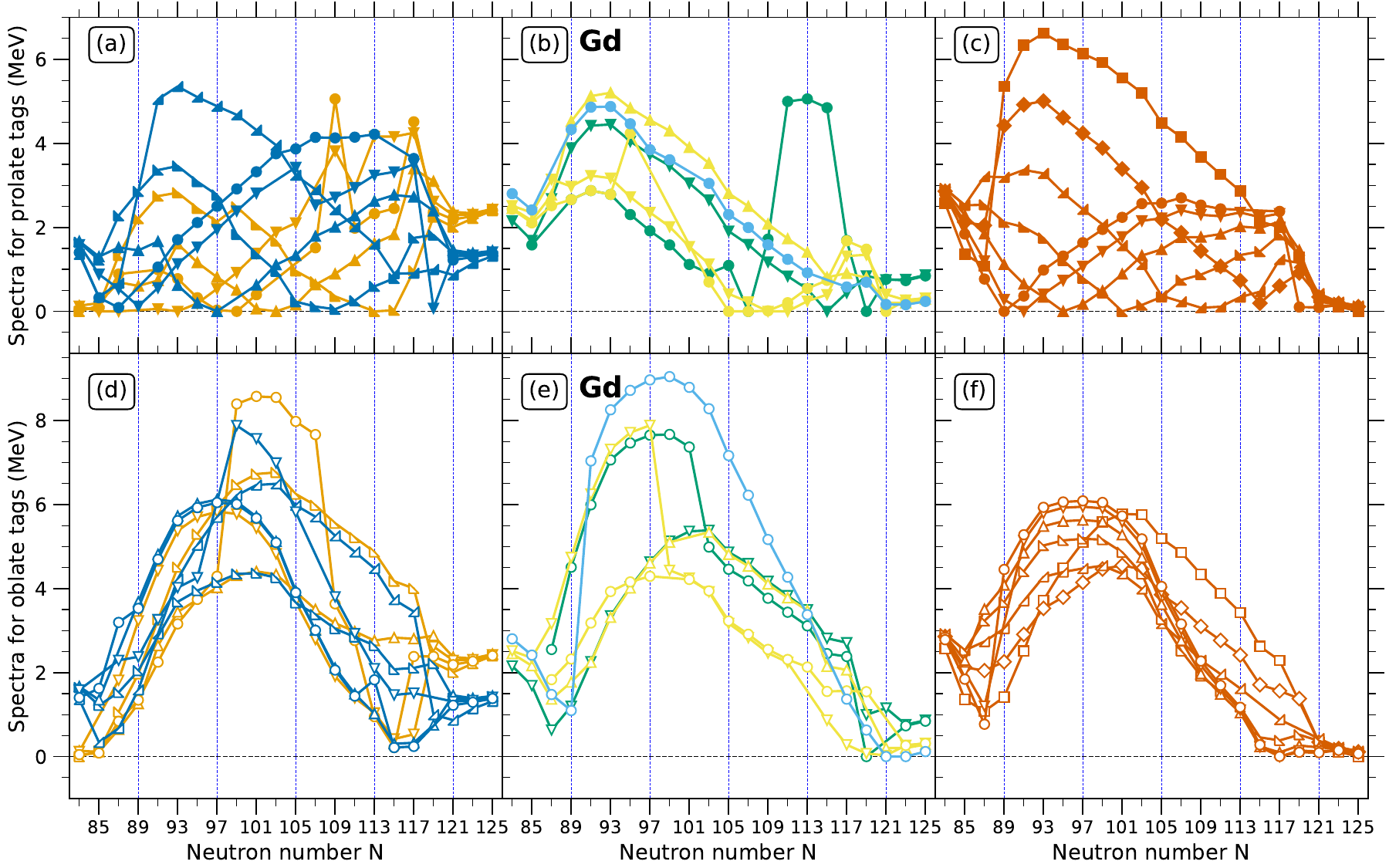}\vspace*{-5mm}
\end{center}
\caption{Same as in Fig.~\protect\ref{fig:Spect_Dy} but for the gadolinium isotopes.
}
\label{fig:Spect_Gd}
\begin{center}
\includegraphics[width=0.98\textwidth]{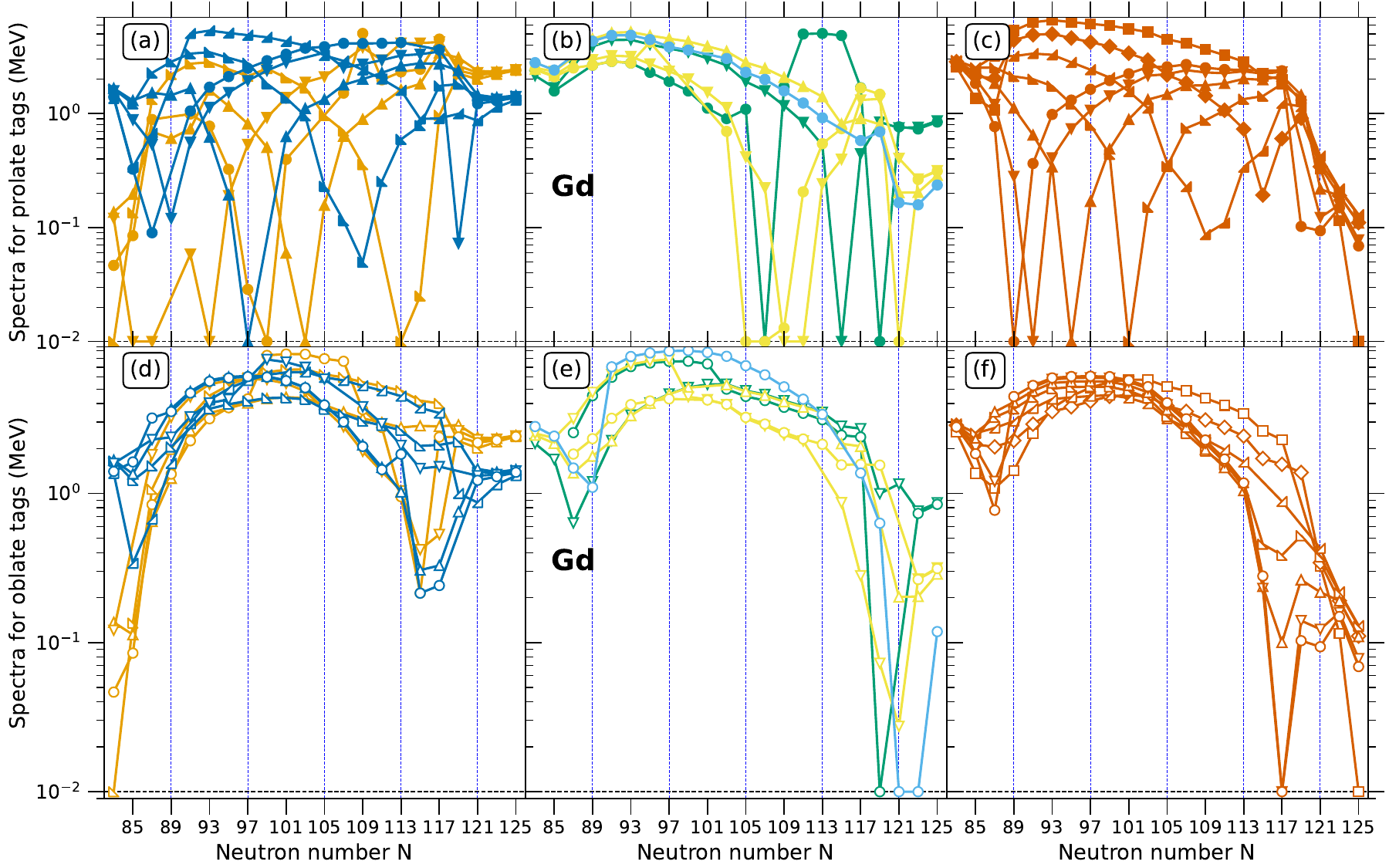}\vspace*{-5mm}
\end{center}
\caption{Same as in Fig.~\protect\ref{fig:Spect_Gd} but plotted in a logarithmic scale with  $E_{\text{exc}}=0$ (ground states) plotted artificially at $E_{\text{exc}}=0.01$\,MeV.
\label{fig:Spect_log_Gd}
}
\end{figure*}


\begin{figure*}
\begin{center}
\includegraphics[width=0.93\textwidth]{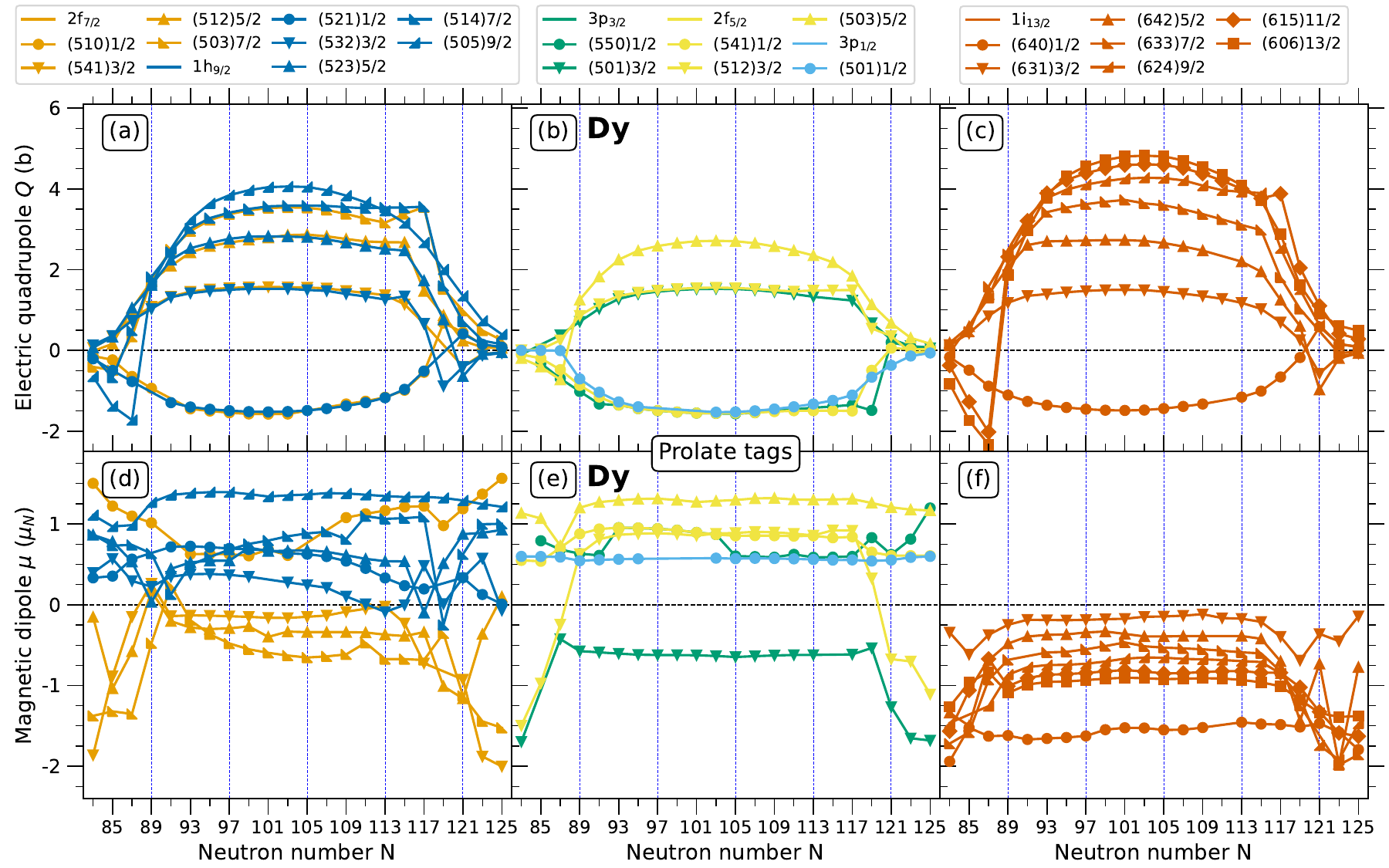}\vspace*{-1mm}
\end{center}
\caption{Figure~\protect\ref{fig:Q2,mu_Dy_prol_2} repeated in a consistent style of the Supplemental Material.
}
\label{fig:Q2,mu_Dy2_prol_2}
\begin{center}
\includegraphics[width=0.93\textwidth]{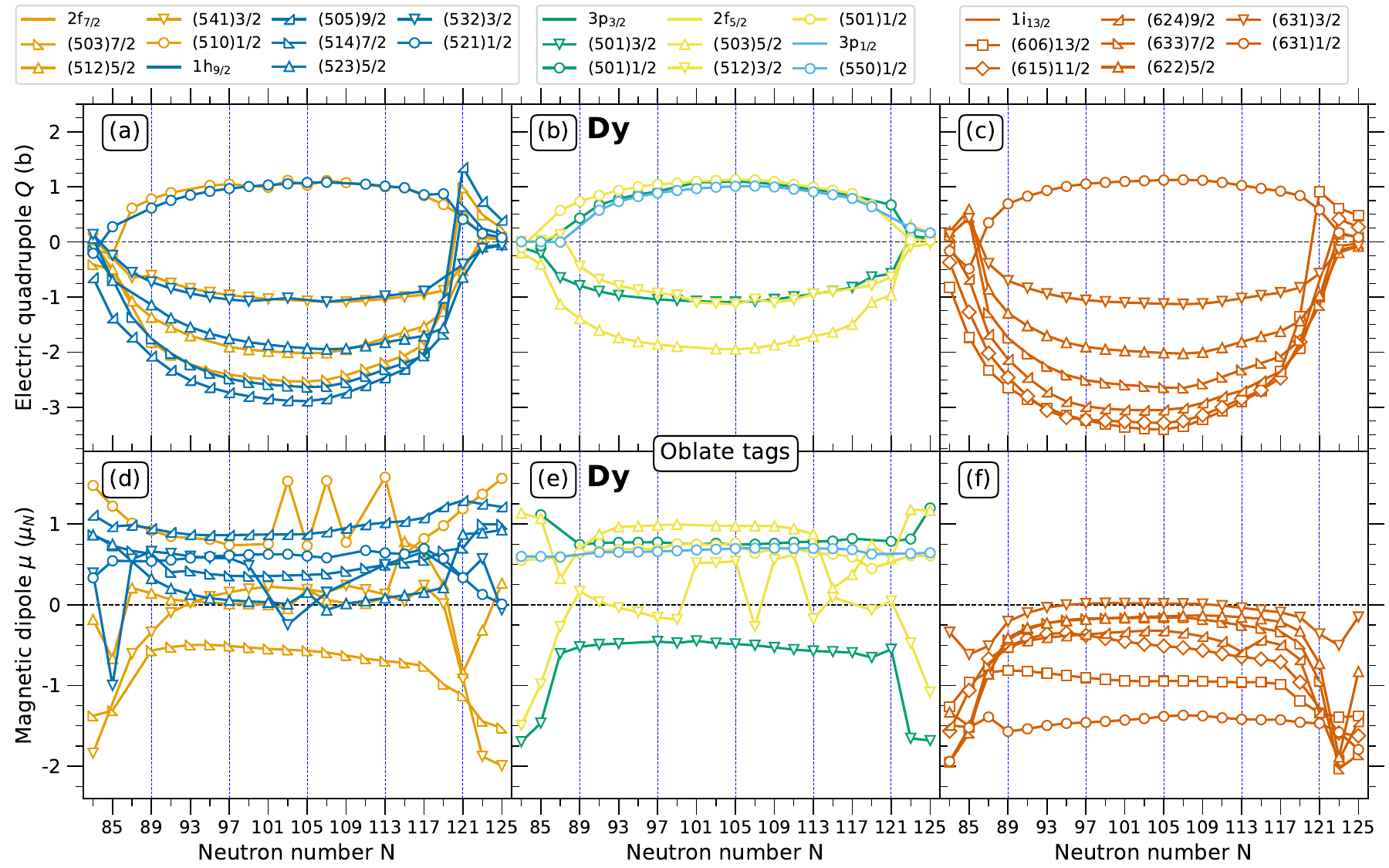}\vspace*{-1mm}
\end{center}
\caption{Figure~\protect\ref{fig:Q2,mu_Dy_obl_2} repeated in a consistent style of the Supplemental Material.
}
\label{fig:Q2,mu_Dy2_obl_2}
\end{figure*}

\begin{figure*}
\begin{center}
\includegraphics[width=0.93\textwidth]{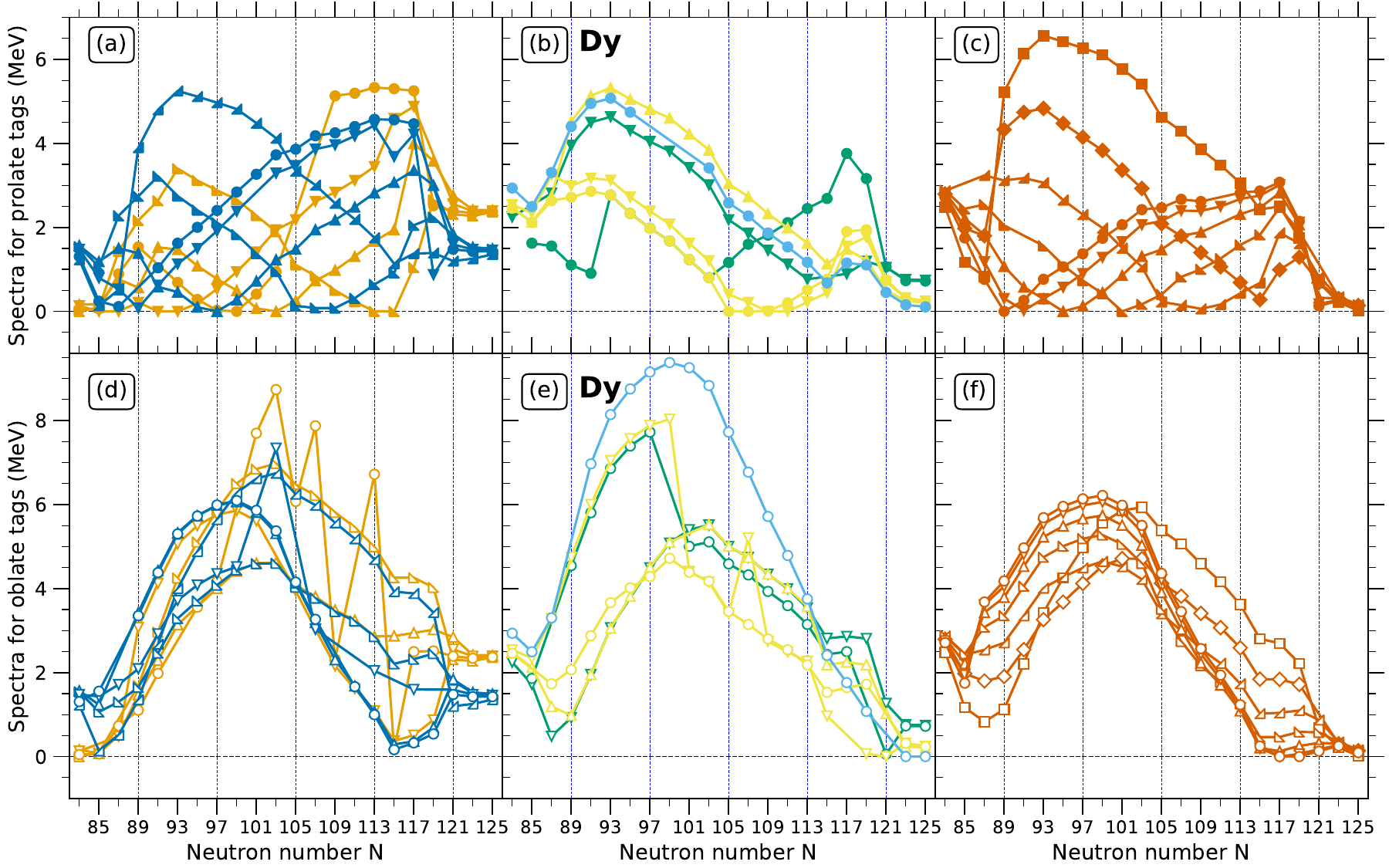}\vspace*{-5mm}
\end{center}
\caption{Figure~\protect\ref{fig:Spect_Dy} repeated in a consistent style of the Supplemental Material.
}
\label{fig:Spect_Dy2}
\begin{center}
\includegraphics[width=0.98\textwidth]{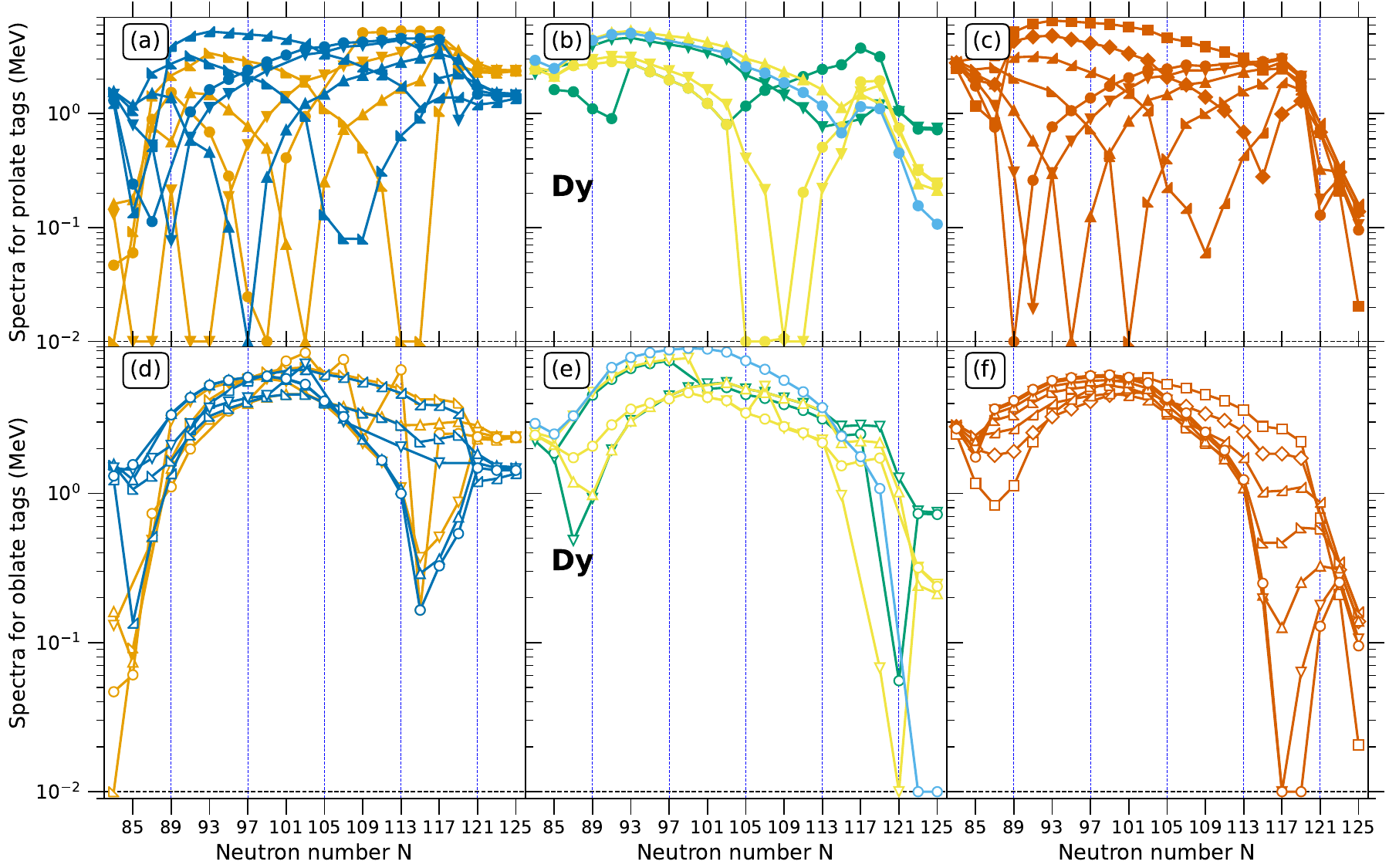}\vspace*{-5mm}
\end{center}
\caption{Figure~\protect\ref{fig:Spect_Dy2} but plotted in a logarithmic scale with  $E_{\text{exc}}=0$ (ground states) plotted artificially at $E_{\text{exc}}=0.01$\,MeV.
\label{fig:Spect_log_Dy2}
}
\end{figure*}


\begin{figure*}
\begin{center}
\includegraphics[width=0.93\textwidth]{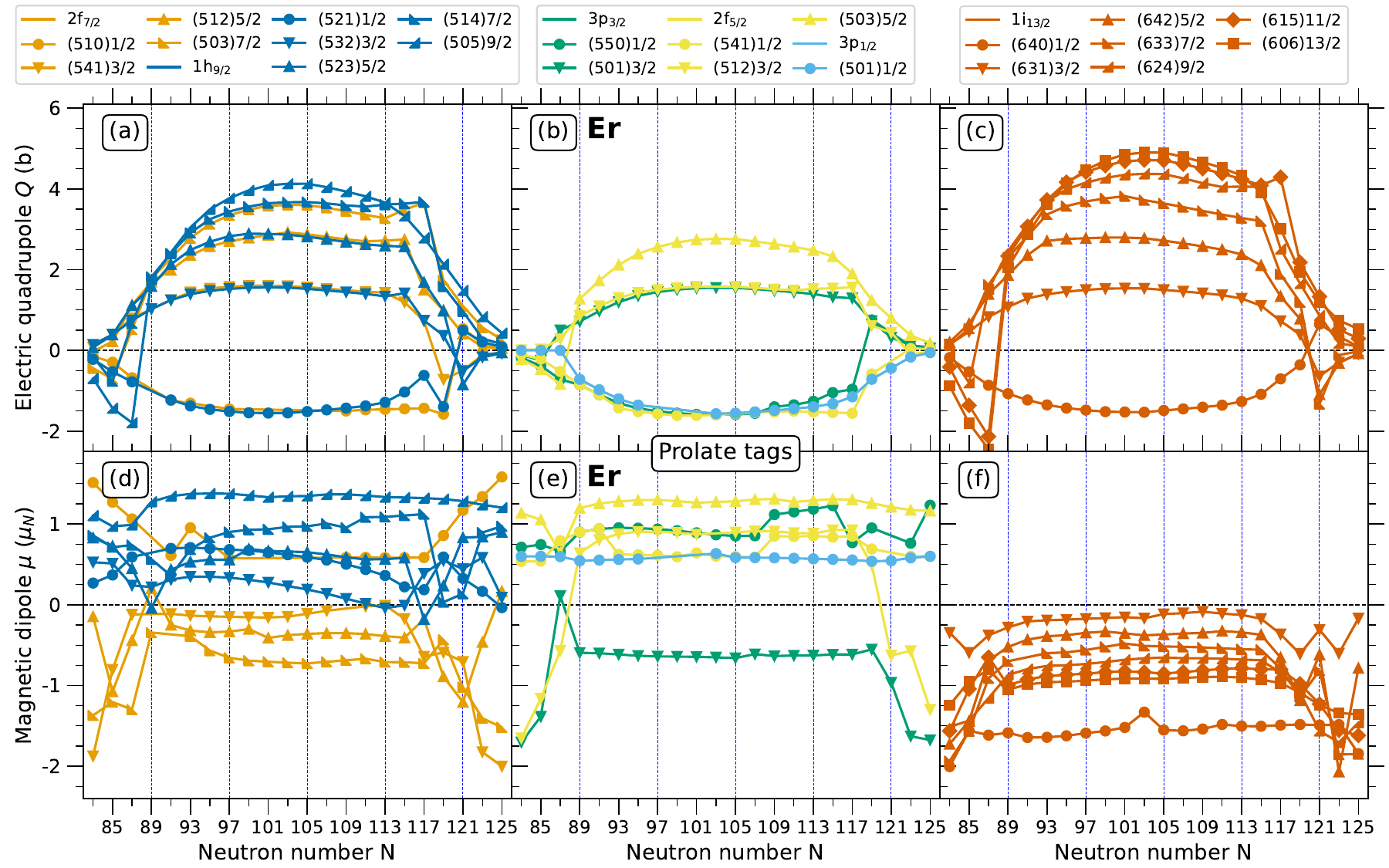}\vspace*{-1mm}
\end{center}
\caption{Same as in Fig.~\protect\ref{fig:Q2,mu_Dy_prol_2} but for the erbium isotopes.
}
\label{fig:Q2,mu_Er_prol_2}
\begin{center}
\includegraphics[width=0.93\textwidth]{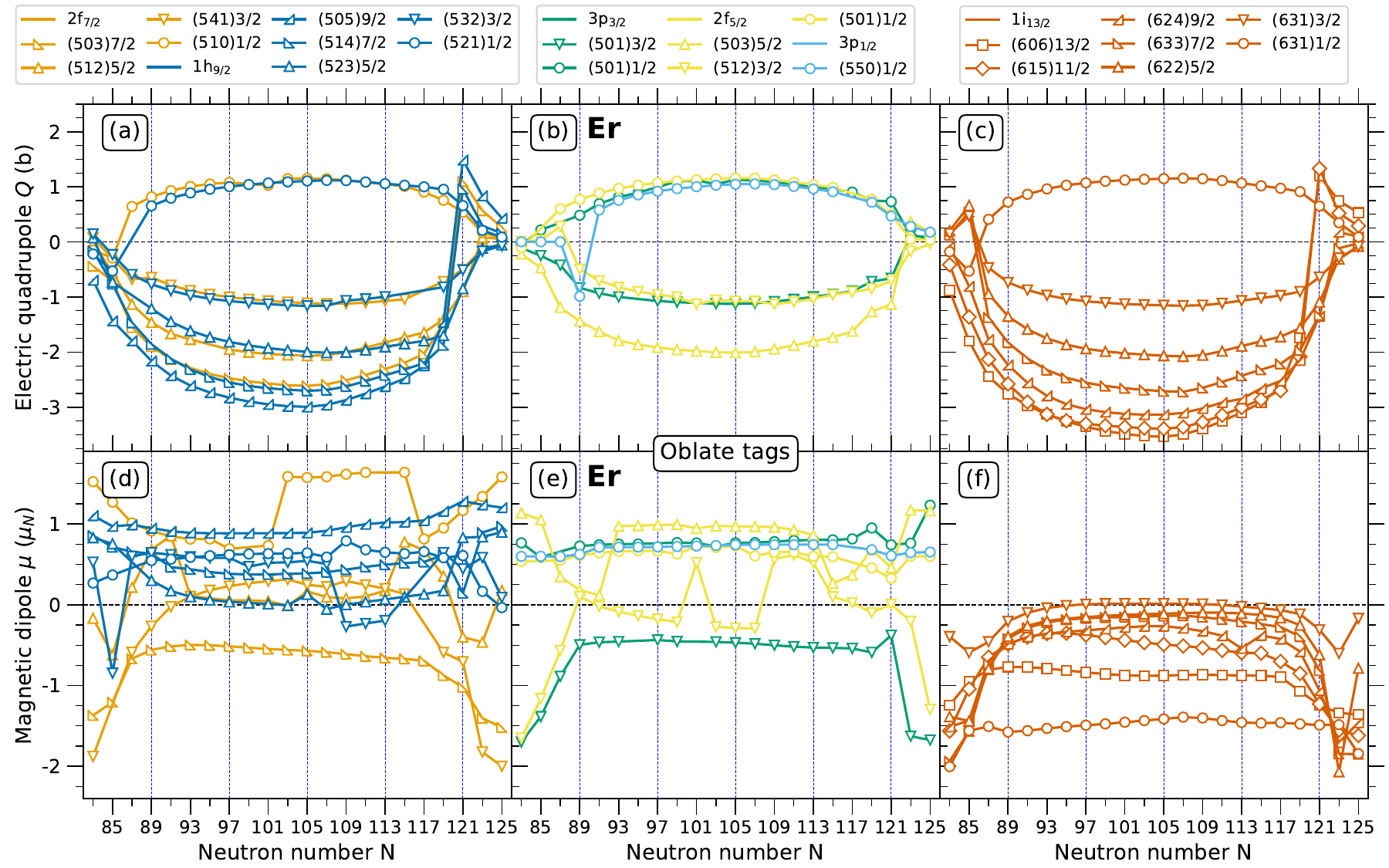}\vspace*{-1mm}
\end{center}
\caption{Same as in Fig.~\protect\ref{fig:Q2,mu_Dy_obl_2} but for the erbium isotopes.
}
\label{fig:Q2,mu_Er_obl_2}
\end{figure*}

\begin{figure*}
\begin{center}
\includegraphics[width=0.93\textwidth]{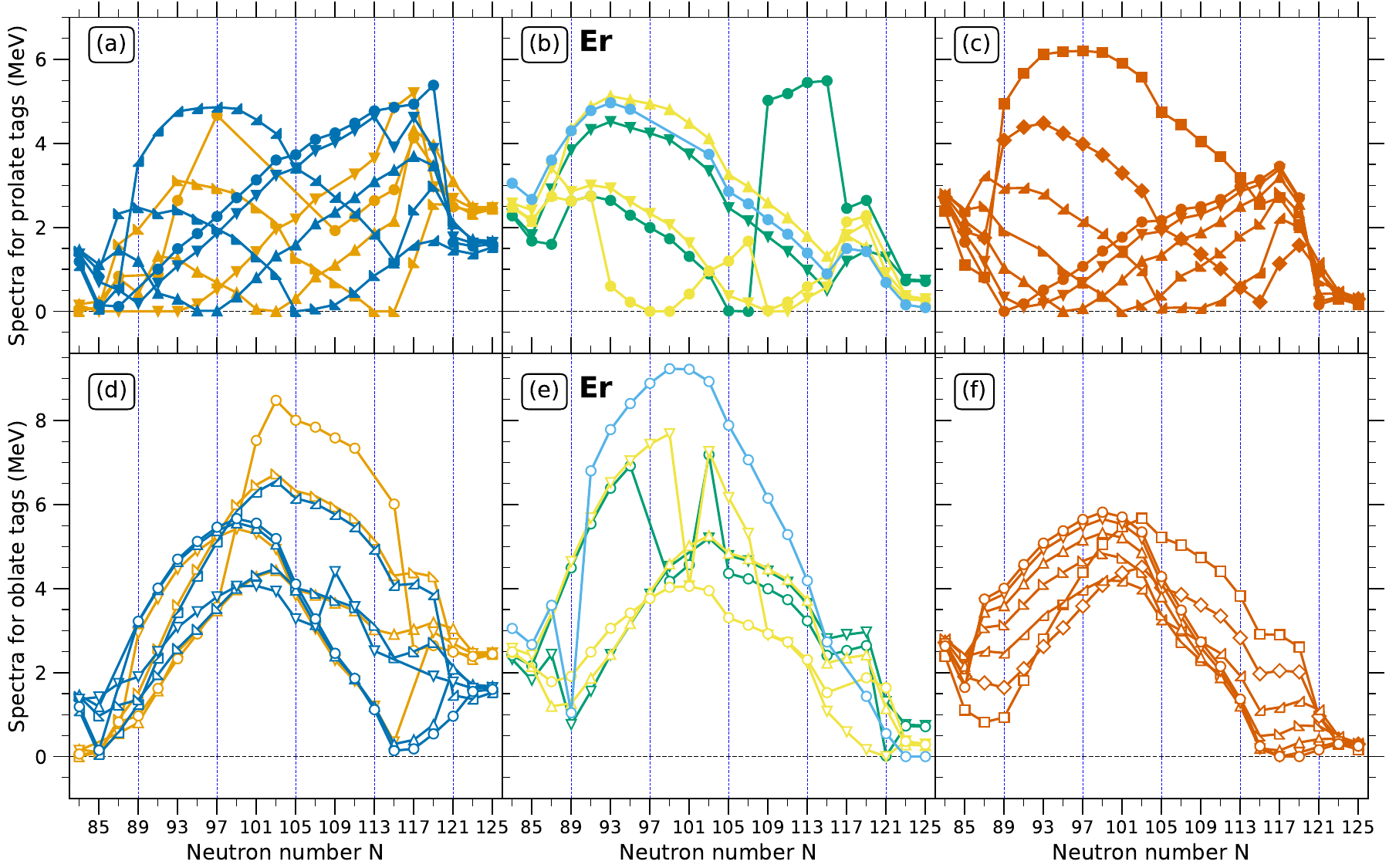}\vspace*{-5mm}
\end{center}
\caption{Same as in Fig.~\protect\ref{fig:Spect_Dy} but for the erbium isotopes.
}
\label{fig:Spect_Er}
\begin{center}
\includegraphics[width=0.98\textwidth]{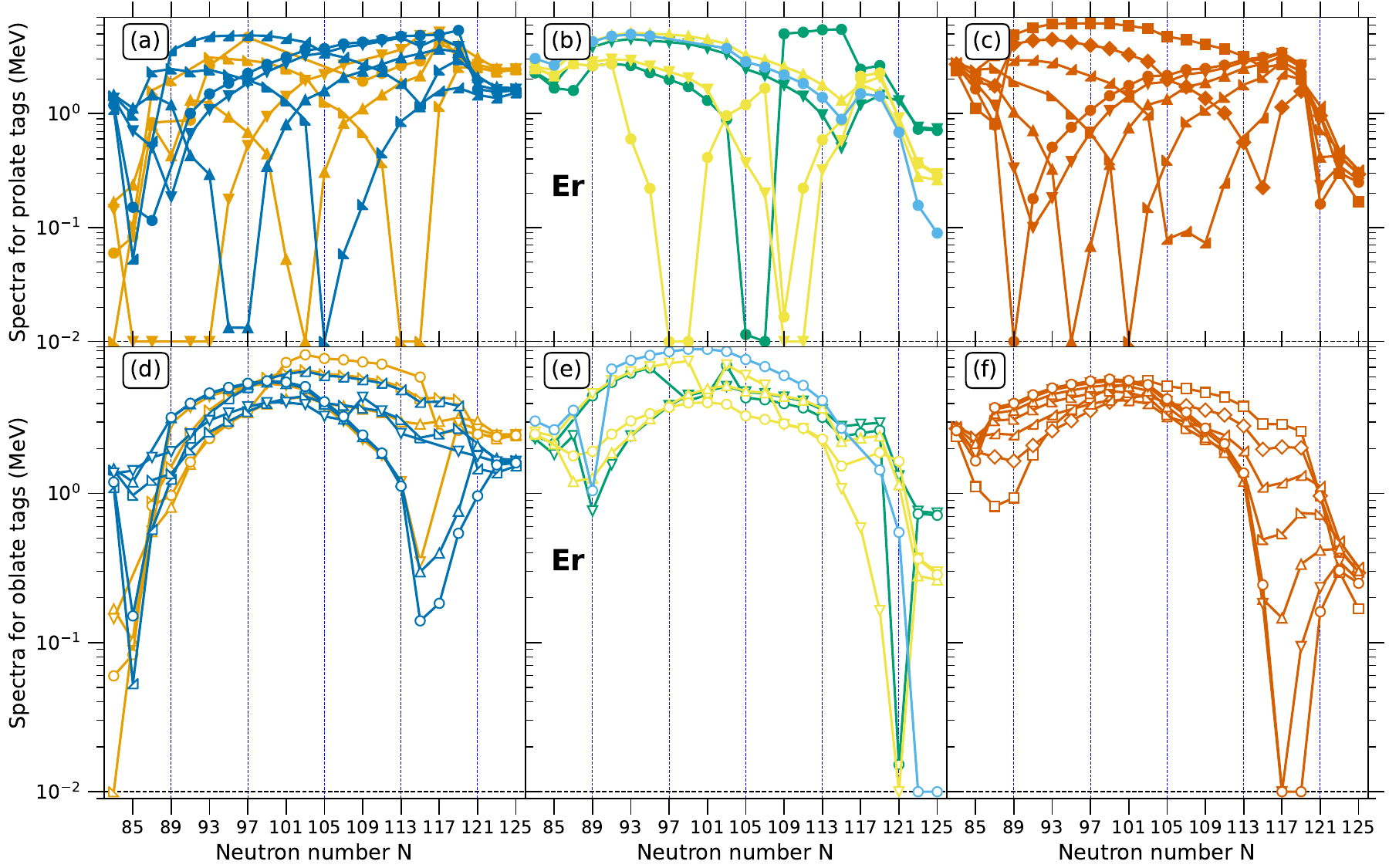}\vspace*{-5mm}
\end{center}
\caption{Same as in Fig.~\protect\ref{fig:Spect_Er} but plotted in a logarithmic scale with  $E_{\text{exc}}=0$ (ground states) plotted artificially at $E_{\text{exc}}=0.01$\,MeV.
\label{fig:Spect_log_Er}
}
\end{figure*}


\begin{figure*}
\begin{center}
\includegraphics[width=0.93\textwidth]{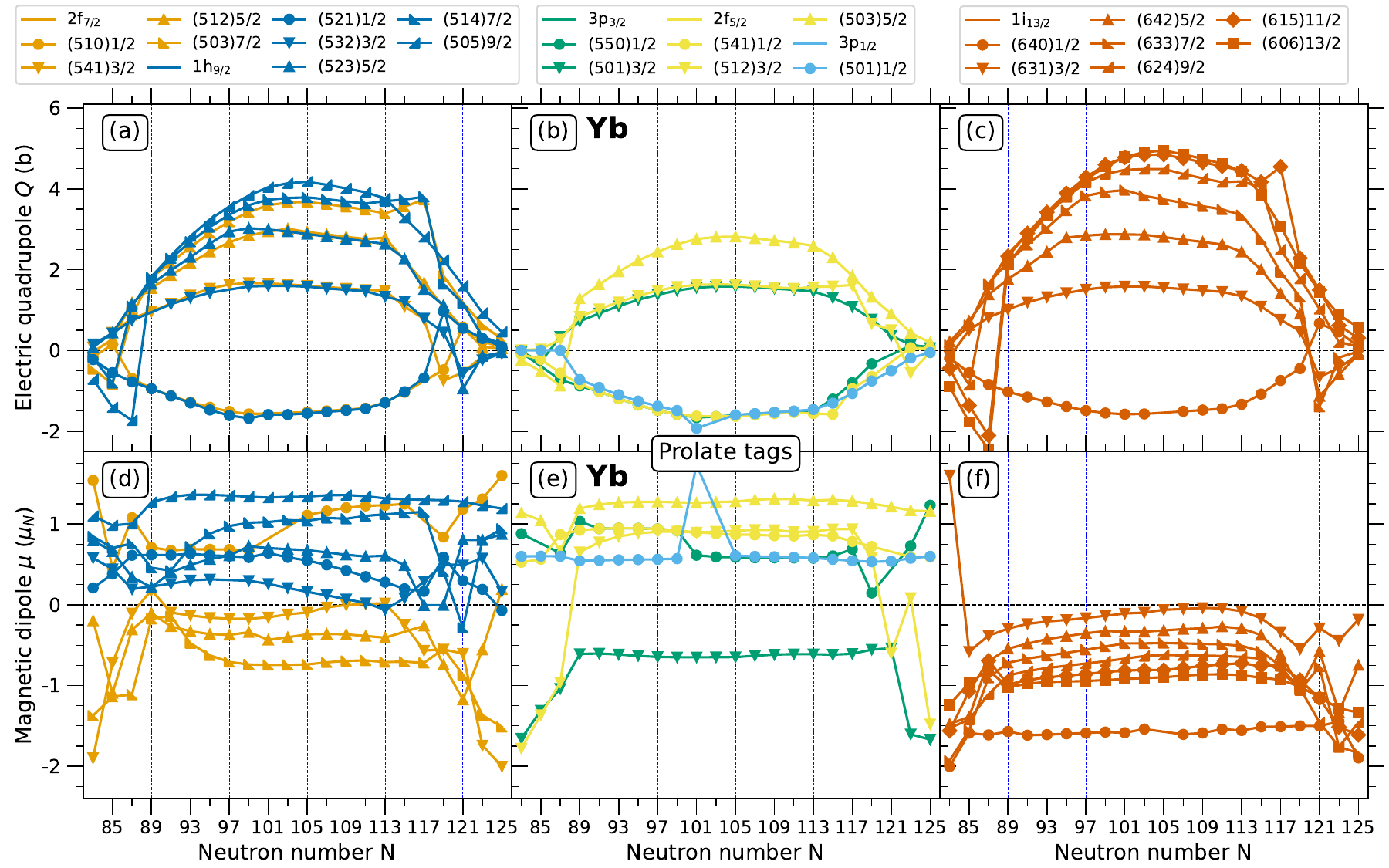}\vspace*{-1mm}
\end{center}
\caption{Same as in Fig.~\protect\ref{fig:Q2,mu_Dy_prol_2} but for the ytterbium isotopes.
}
\label{fig:Q2,mu_Yb_prol_2}
\begin{center}
\includegraphics[width=0.93\textwidth]{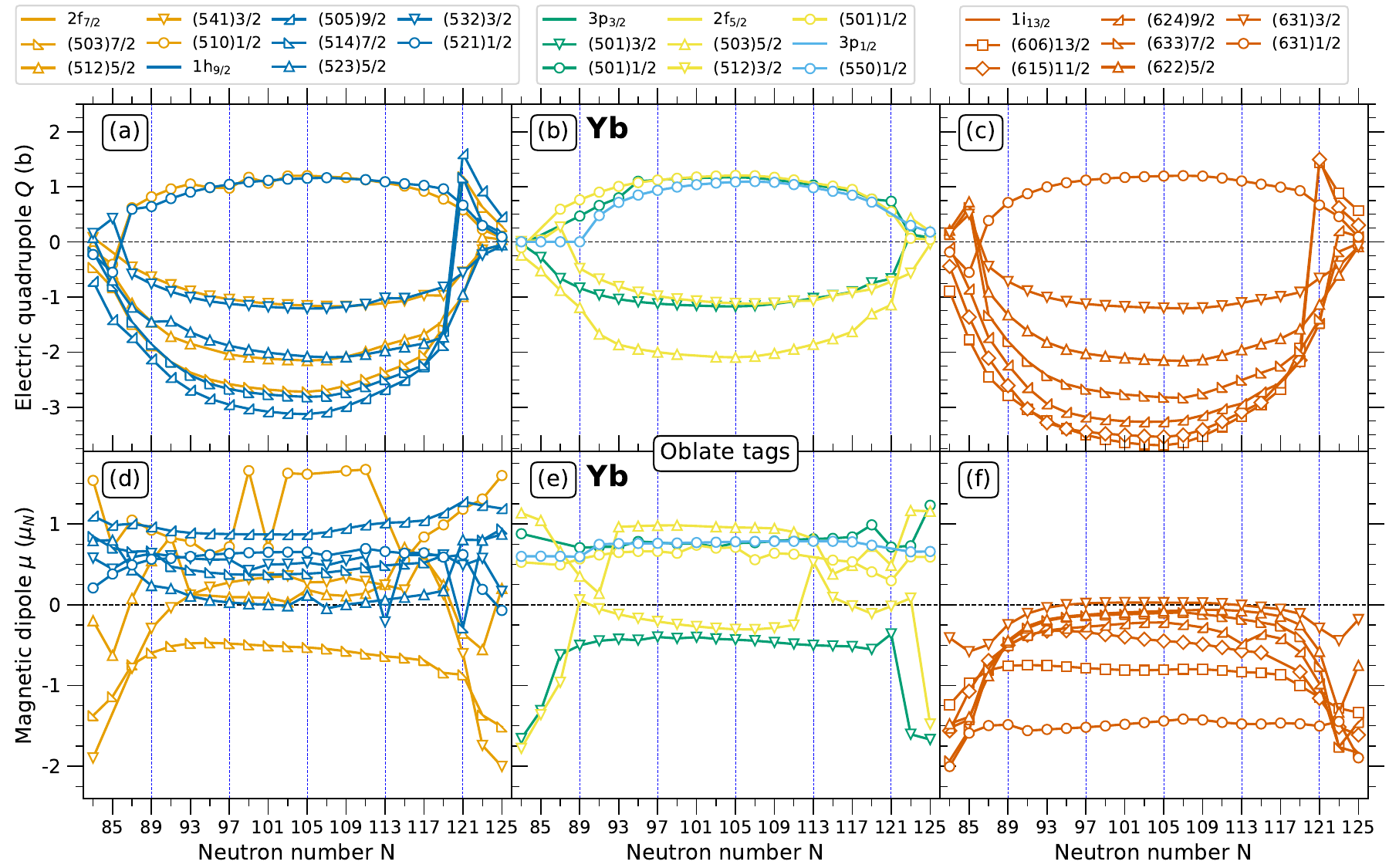}\vspace*{-1mm}
\end{center}
\caption{Same as in Fig.~\protect\ref{fig:Q2,mu_Dy_obl_2} but for the ytterbium isotopes.
}
\label{fig:Q2,mu_Yb_obl_2}
\end{figure*}

\begin{figure*}
\begin{center}
\includegraphics[width=0.93\textwidth]{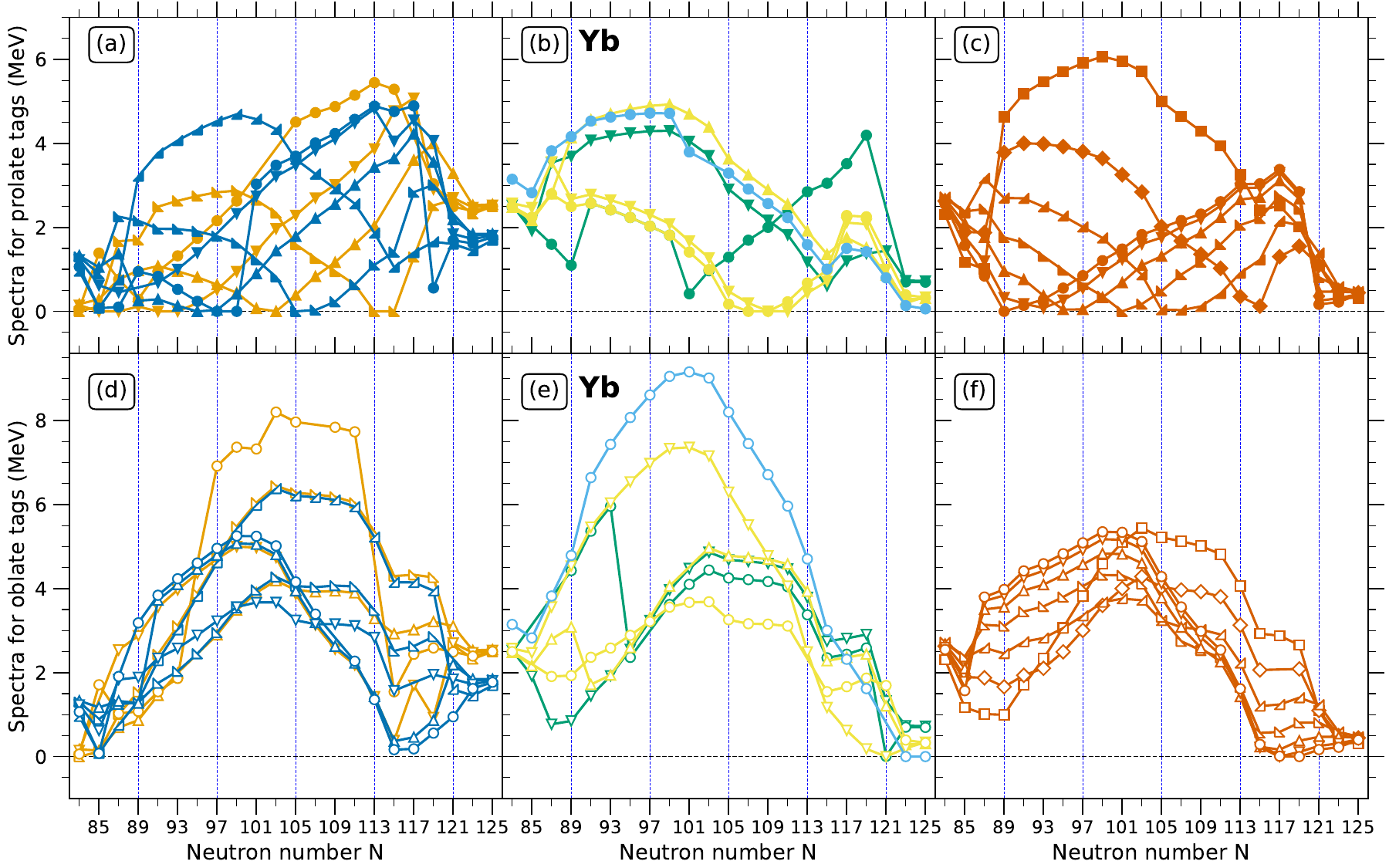}\vspace*{-5mm}
\end{center}
\caption{Same as in Fig.~\protect\ref{fig:Spect_Dy} but for the ytterbium isotopes.
}
\label{fig:Spect_Yb}
\begin{center}
\includegraphics[width=0.98\textwidth]{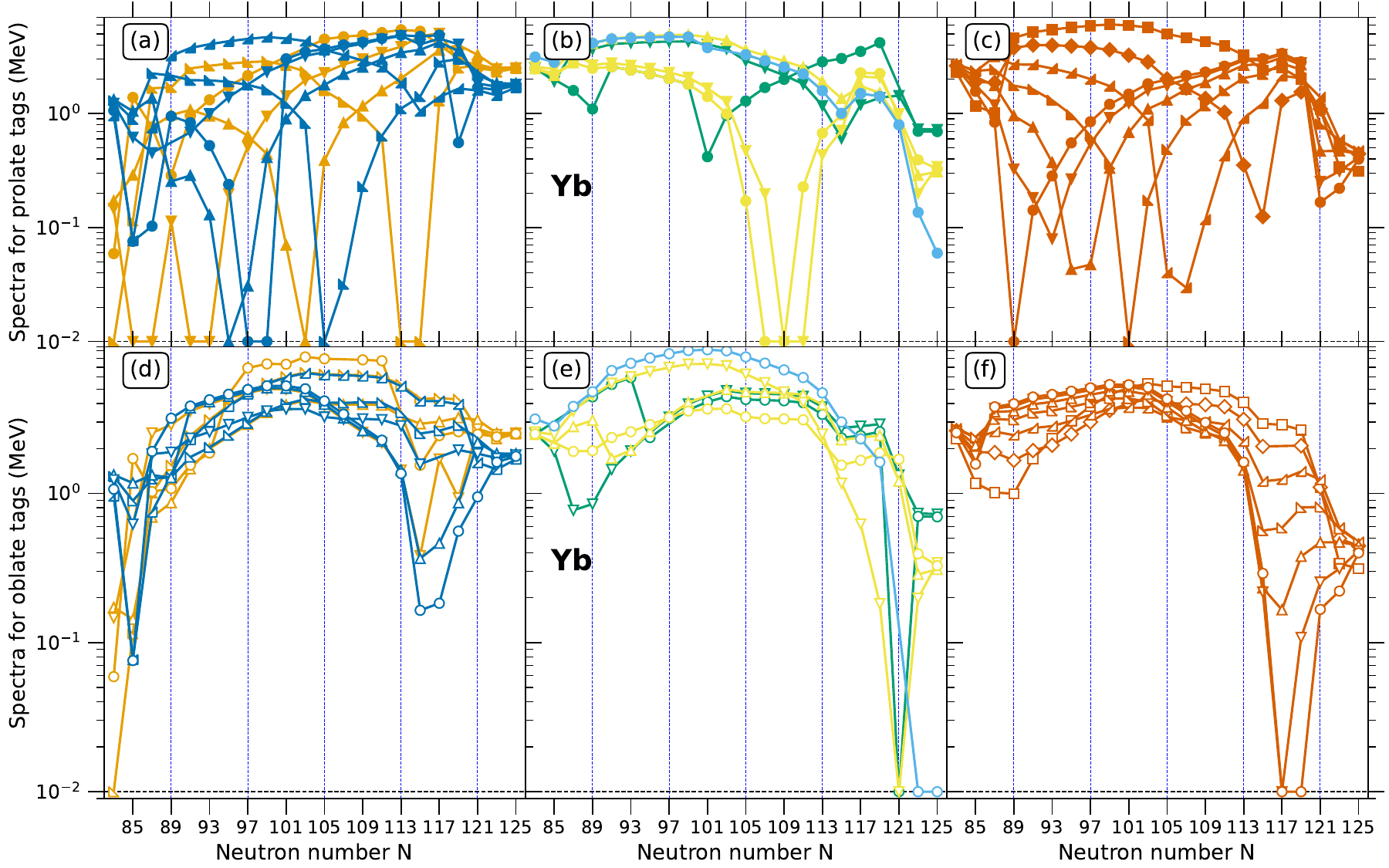}\vspace*{-5mm}
\end{center}
\caption{Same as in Fig.~\protect\ref{fig:Spect_Yb} but plotted in a logarithmic scale with  $E_{\text{exc}}=0$ (ground states) plotted artificially at $E_{\text{exc}}=0.01$\,MeV.
\label{fig:Spect_log_Yb}
}
\end{figure*}


\begin{figure*}
\begin{center}
\includegraphics[width=0.93\textwidth]{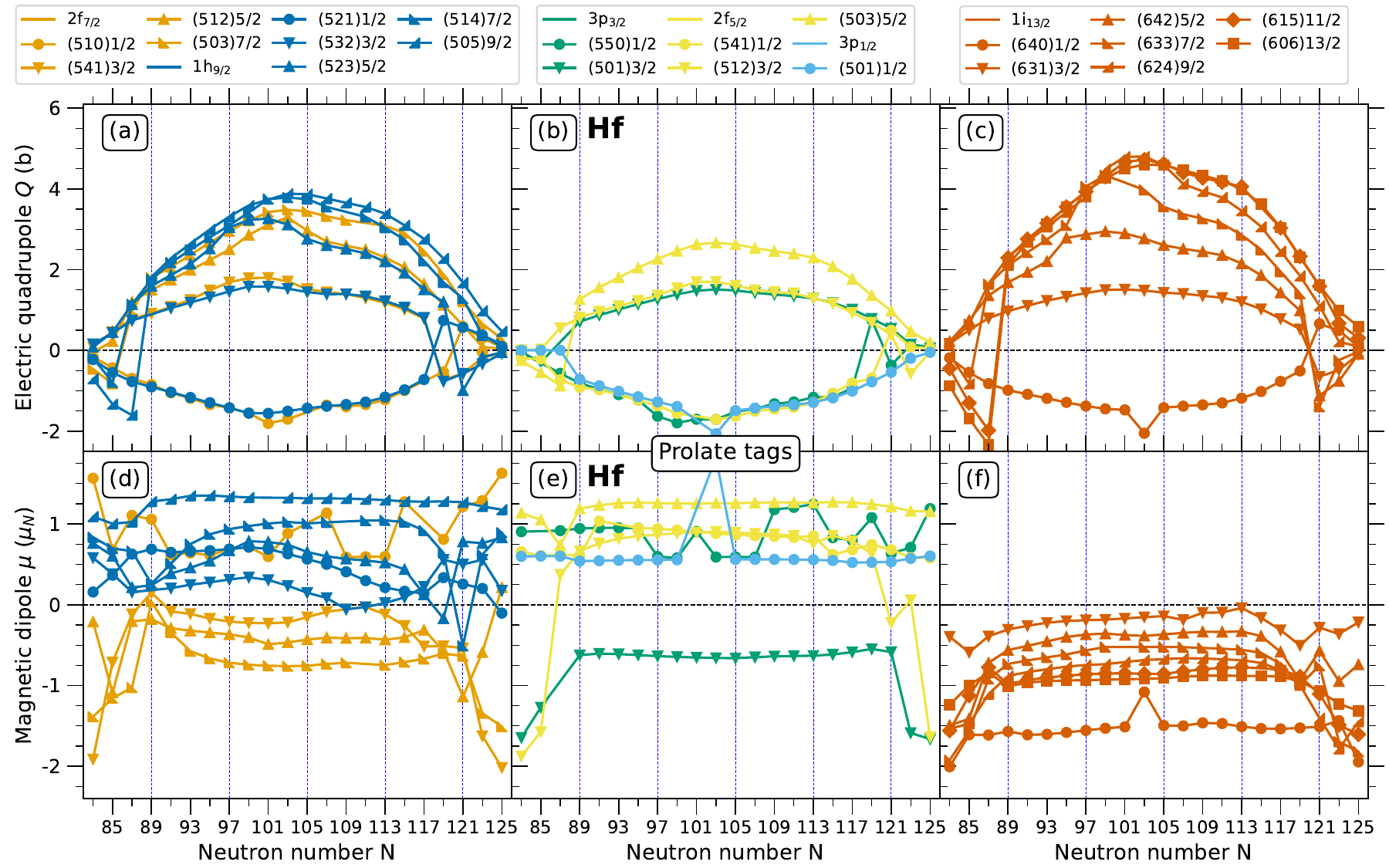}\vspace*{-1mm}
\end{center}
\caption{Same as in Fig.~\protect\ref{fig:Q2,mu_Dy_prol_2} but for the hafnium isotopes.
}
\label{fig:Q2,mu_Hf_prol_2}
\begin{center}
\includegraphics[width=0.93\textwidth]{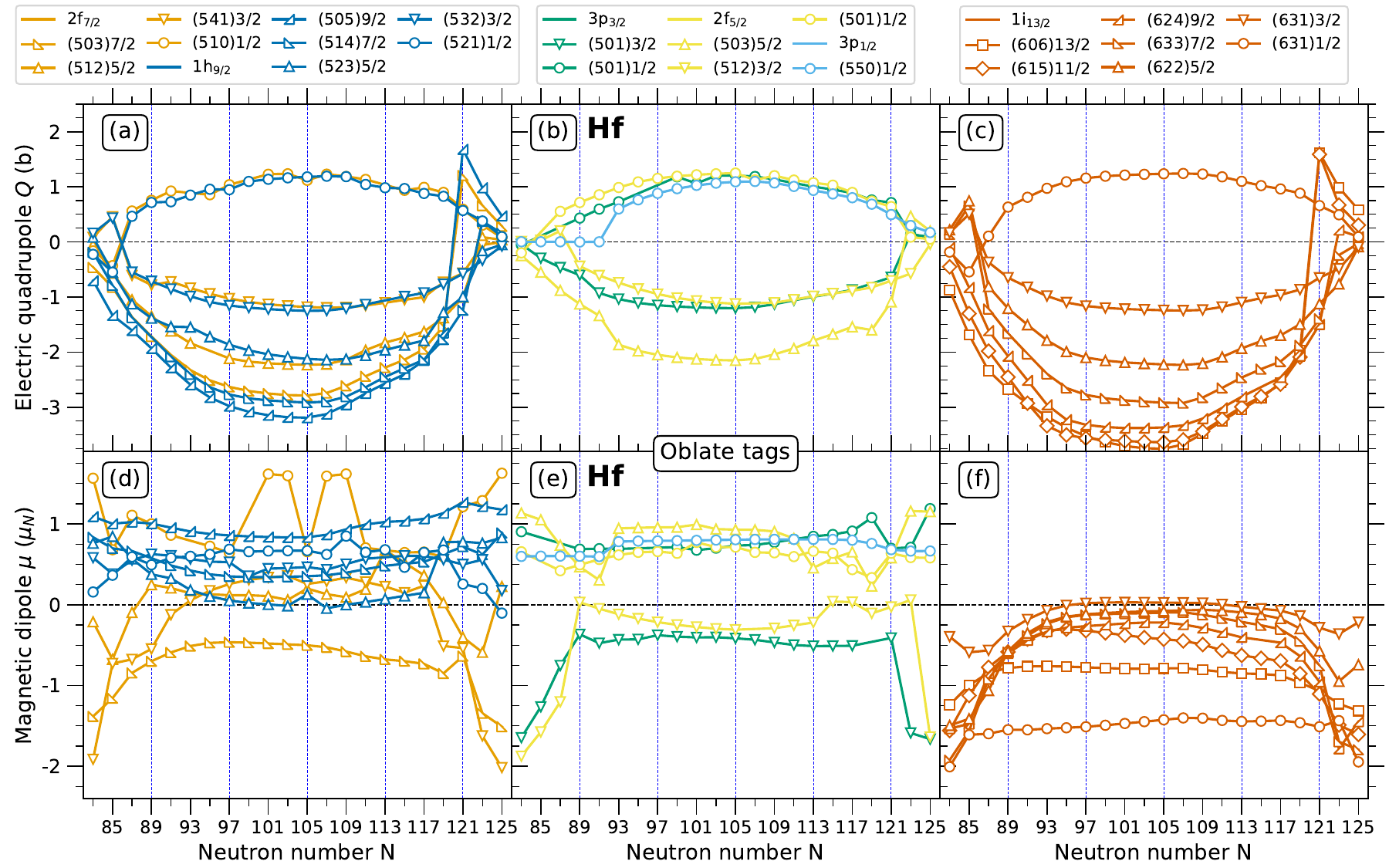}\vspace*{-1mm}
\end{center}
\caption{Same as in Fig.~\protect\ref{fig:Q2,mu_Dy_obl_2} but for the hafnium isotopes.
}
\label{fig:Q2,mu_Hf_obl_2}
\end{figure*}

\begin{figure*}
\begin{center}
\includegraphics[width=0.93\textwidth]{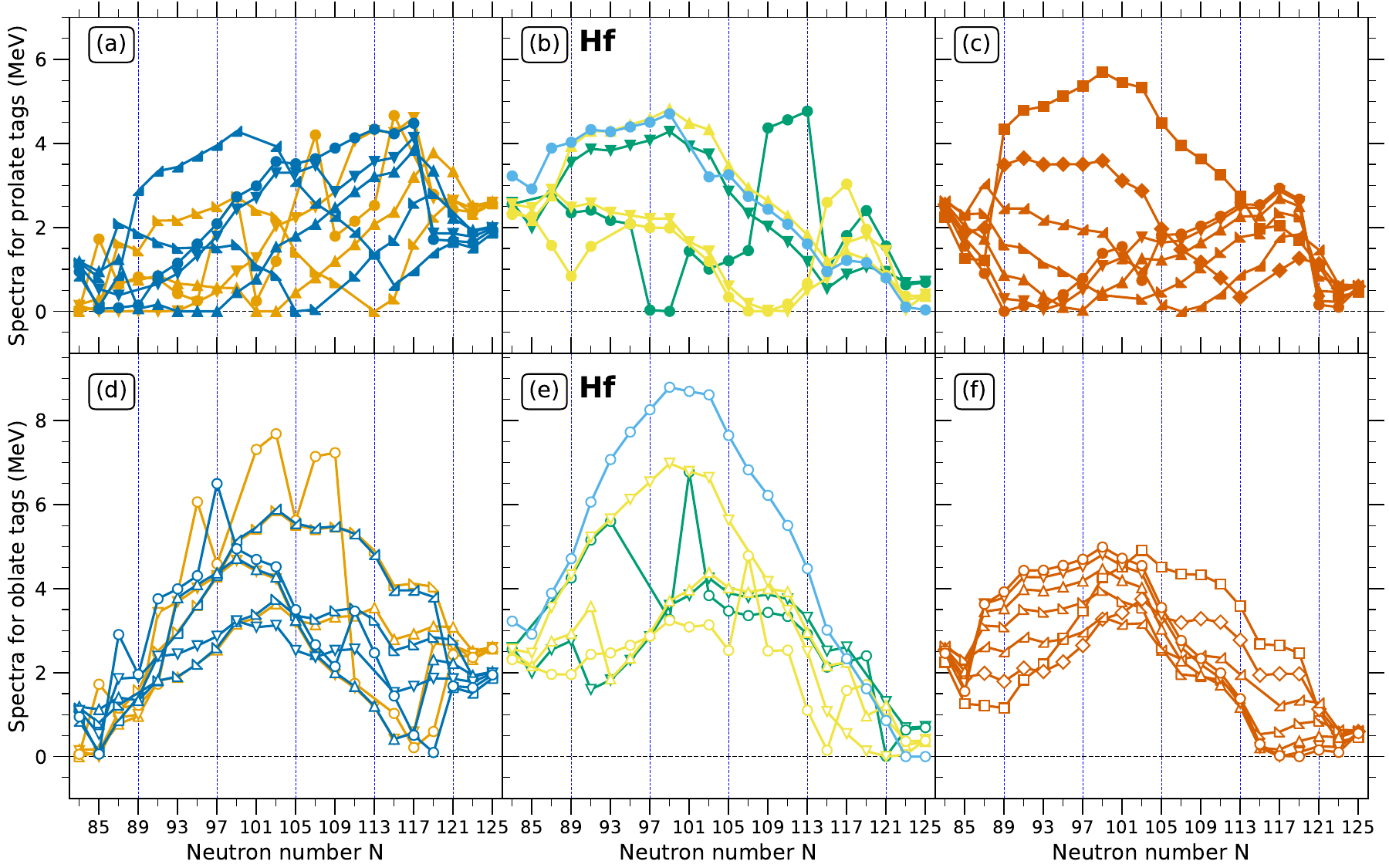}\vspace*{-5mm}
\end{center}
\caption{Same as in Fig.~\protect\ref{fig:Spect_Dy} but for the hafnium isotopes.
}
\label{fig:Spect_Hf}
\begin{center}
\includegraphics[width=0.98\textwidth]{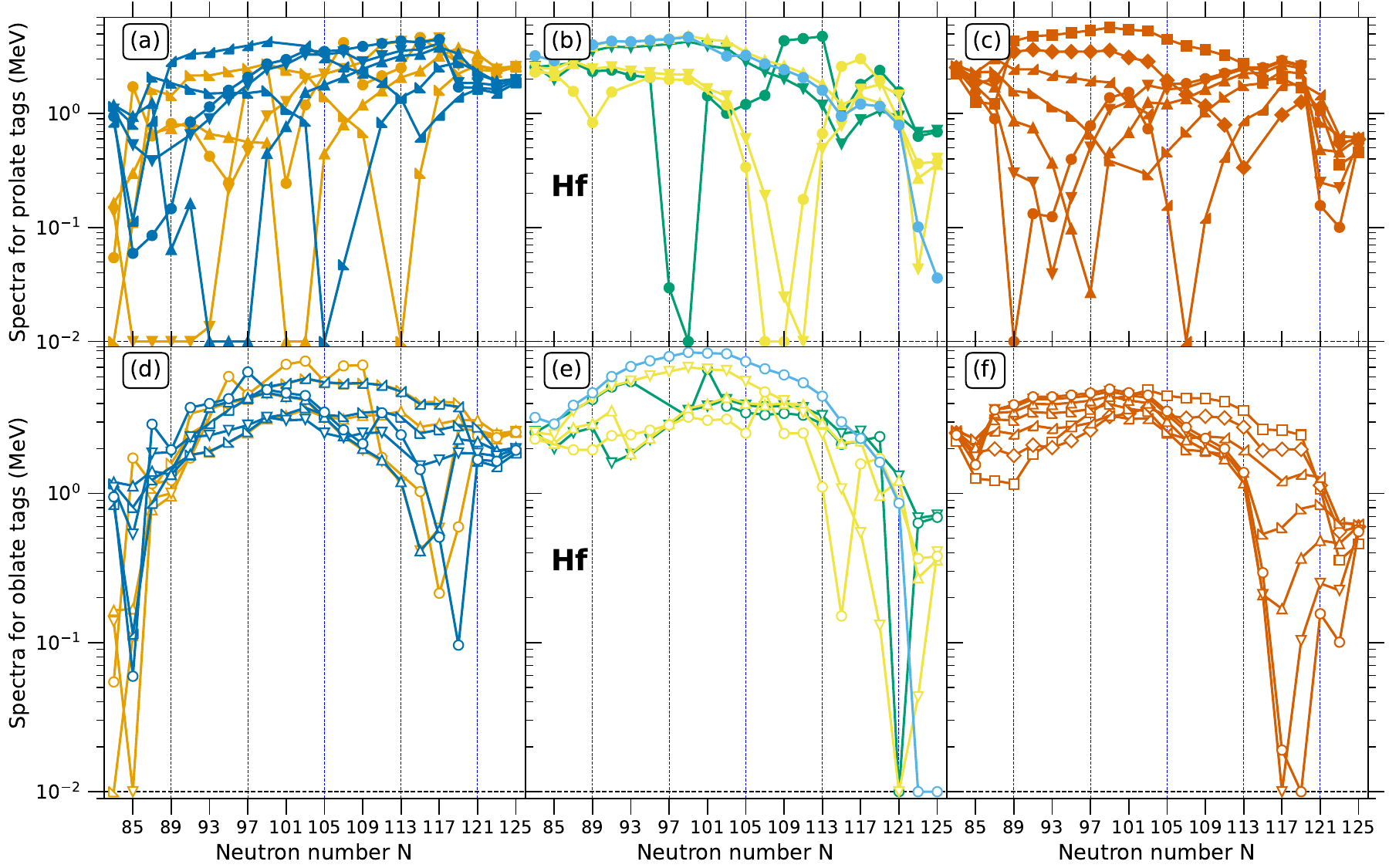}\vspace*{-5mm}
\end{center}
\caption{Same as in Fig.~\protect\ref{fig:Spect_Hf} but plotted in a logarithmic scale with  $E_{\text{exc}}=0$ (ground states) plotted artificially at $E_{\text{exc}}=0.01$\,MeV.
\label{fig:Spect_log_Hf}
}
\end{figure*}


\begin{figure*}
\begin{center}
\includegraphics[width=0.93\textwidth]{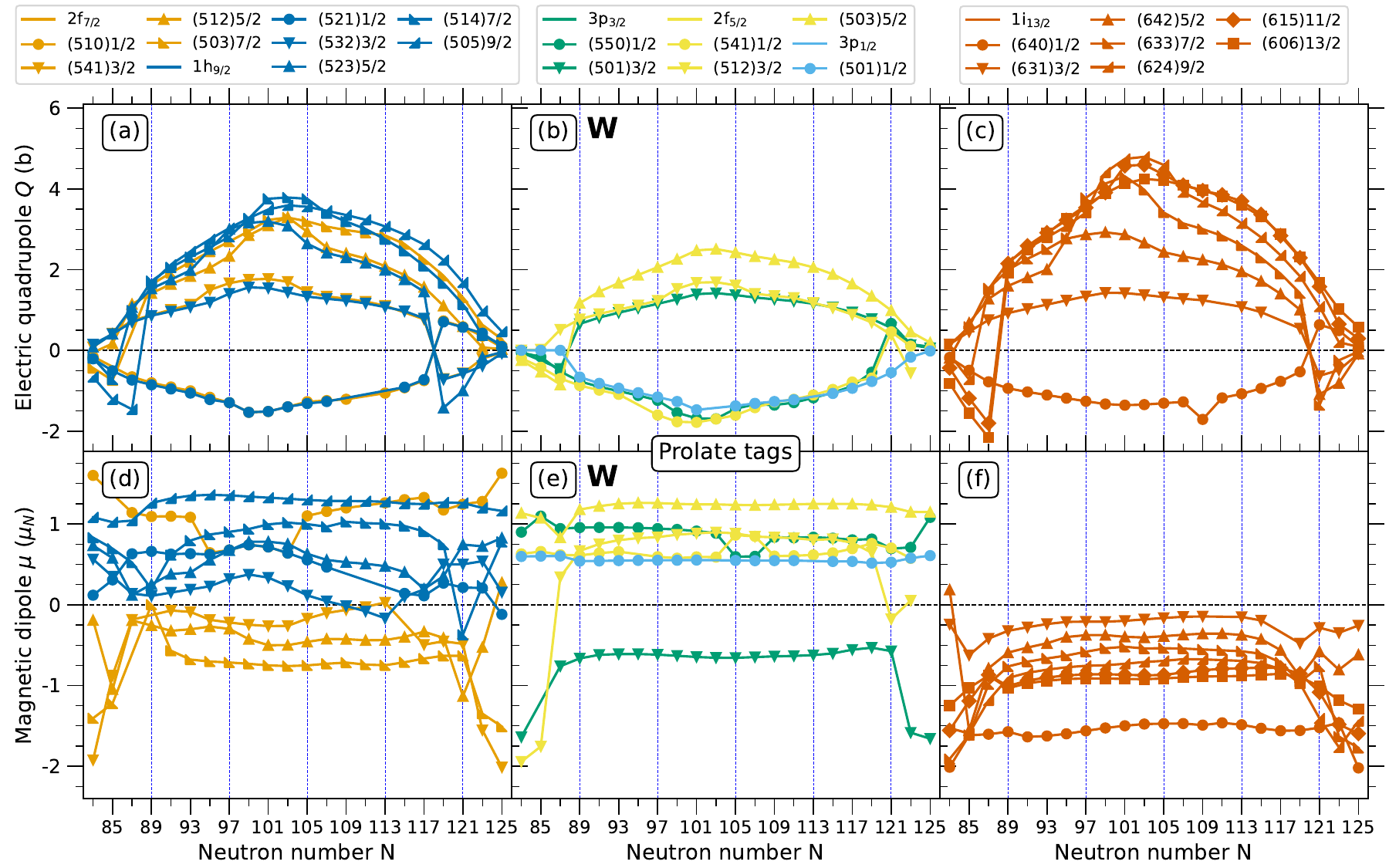}\vspace*{-1mm}
\end{center}
\caption{Same as in Fig.~\protect\ref{fig:Q2,mu_Dy_prol_2} but for the tungsten isotopes.
}
\label{fig:Q2,mu_W_prol_2}
\begin{center}
\includegraphics[width=0.93\textwidth]{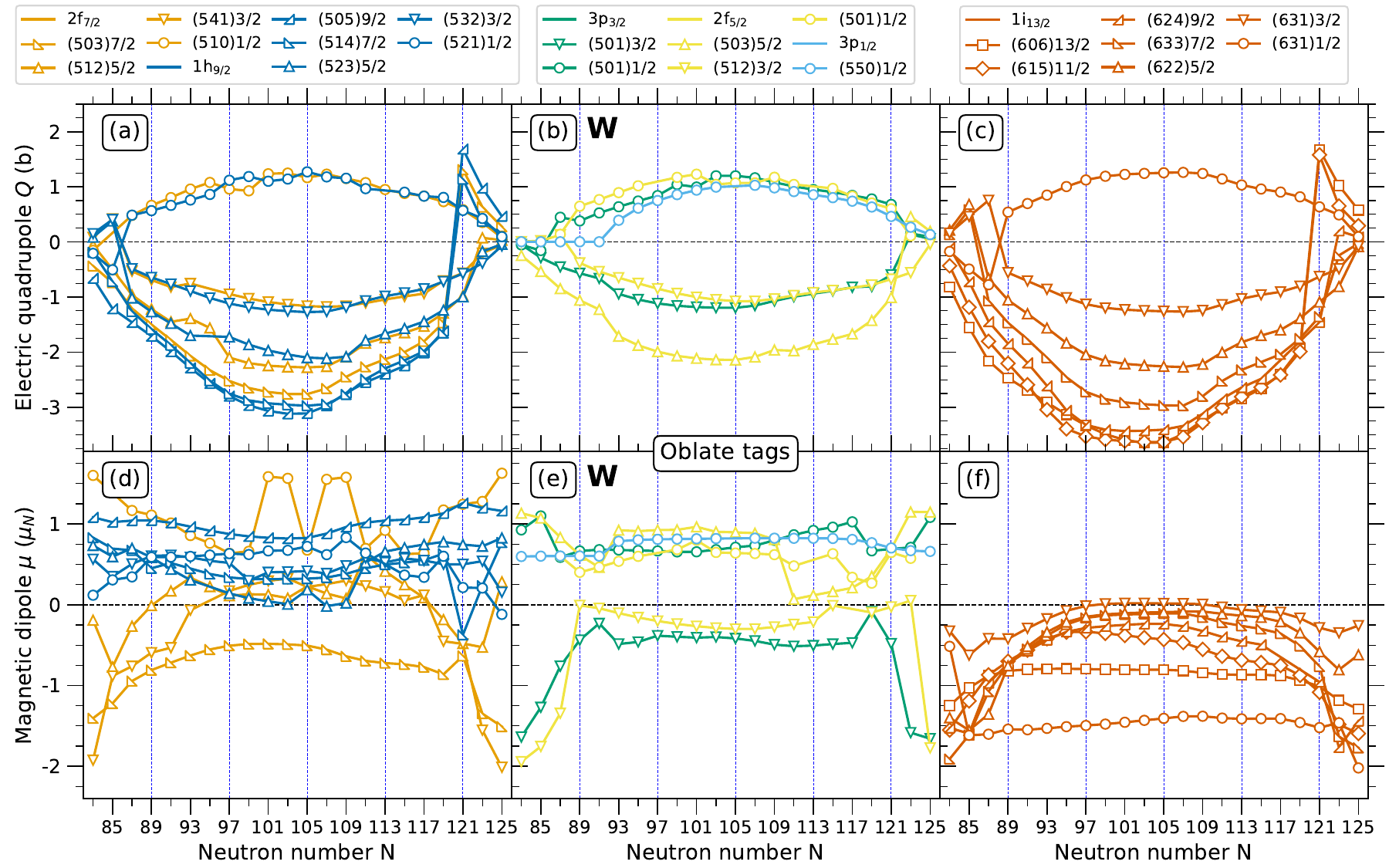}\vspace*{-1mm}
\end{center}
\caption{Same as in Fig.~\protect\ref{fig:Q2,mu_Dy_obl_2} but for the tungsten isotopes.
}
\label{fig:Q2,mu_W_obl_2}
\end{figure*}

\begin{figure*}
\begin{center}
\includegraphics[width=0.93\textwidth]{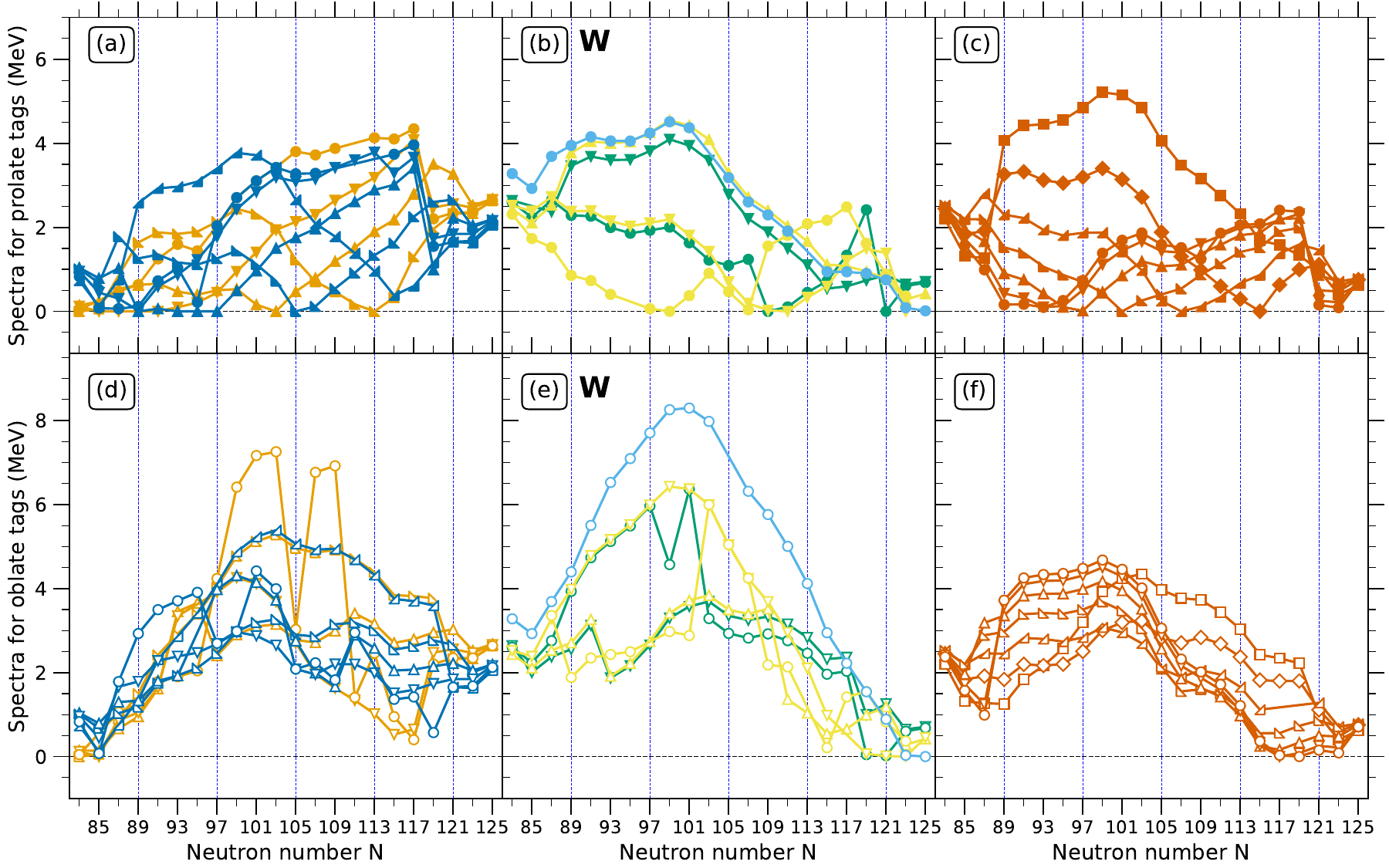}\vspace*{-5mm}
\end{center}
\caption{Same as in Fig.~\protect\ref{fig:Spect_Dy} but for the tungsten isotopes.
}
\label{fig:Spect_W}
\begin{center}
\includegraphics[width=0.98\textwidth]{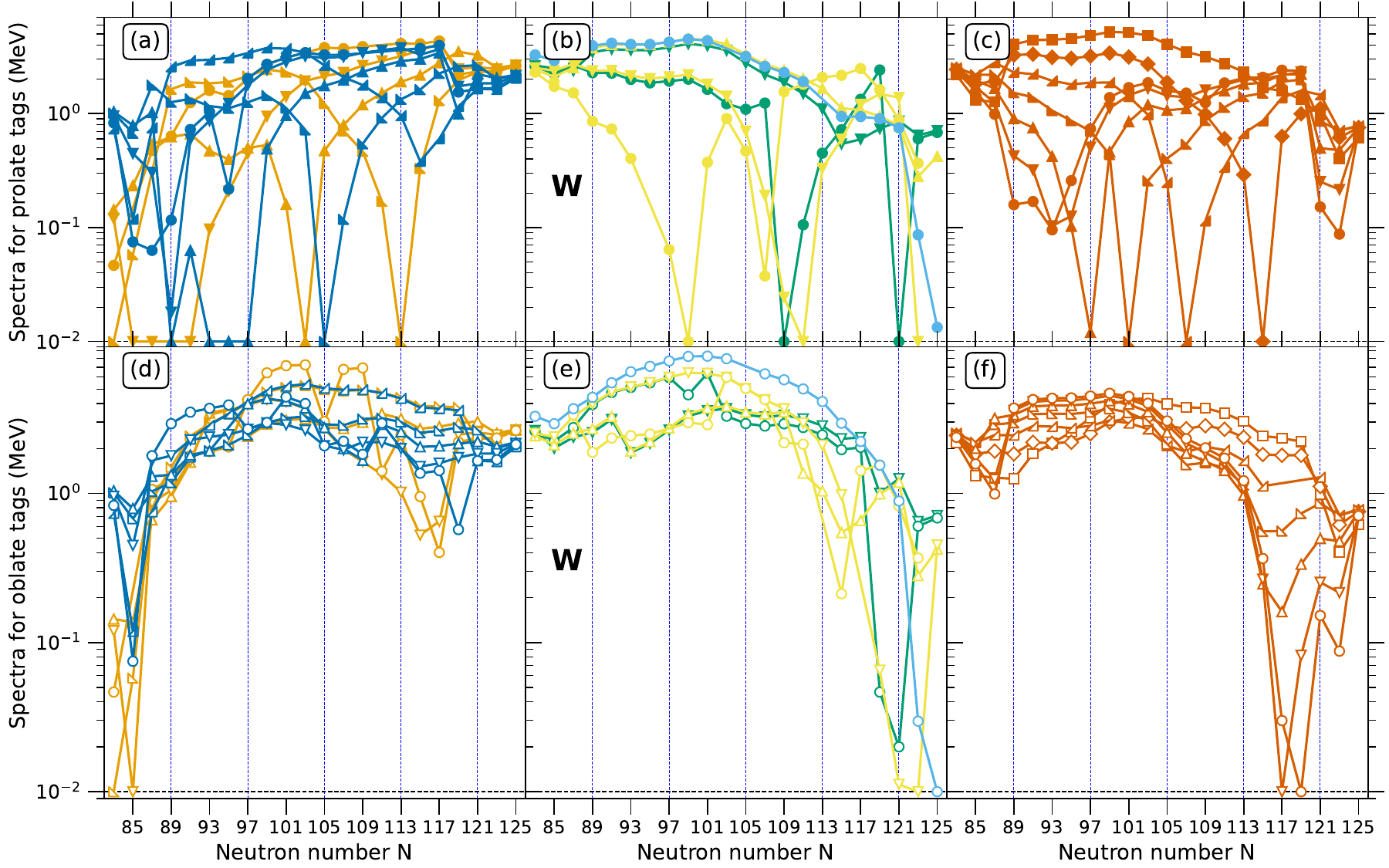}\vspace*{-5mm}
\end{center}
\caption{Same as in Fig.~\protect\ref{fig:Spect_W} but plotted in a logarithmic scale with  $E_{\text{exc}}=0$ (ground states) plotted artificially at $E_{\text{exc}}=0.01$\,MeV.
\label{fig:Spect_log_W}
}
\end{figure*}


\begin{figure*}
\begin{center}
\includegraphics[width=0.93\textwidth]{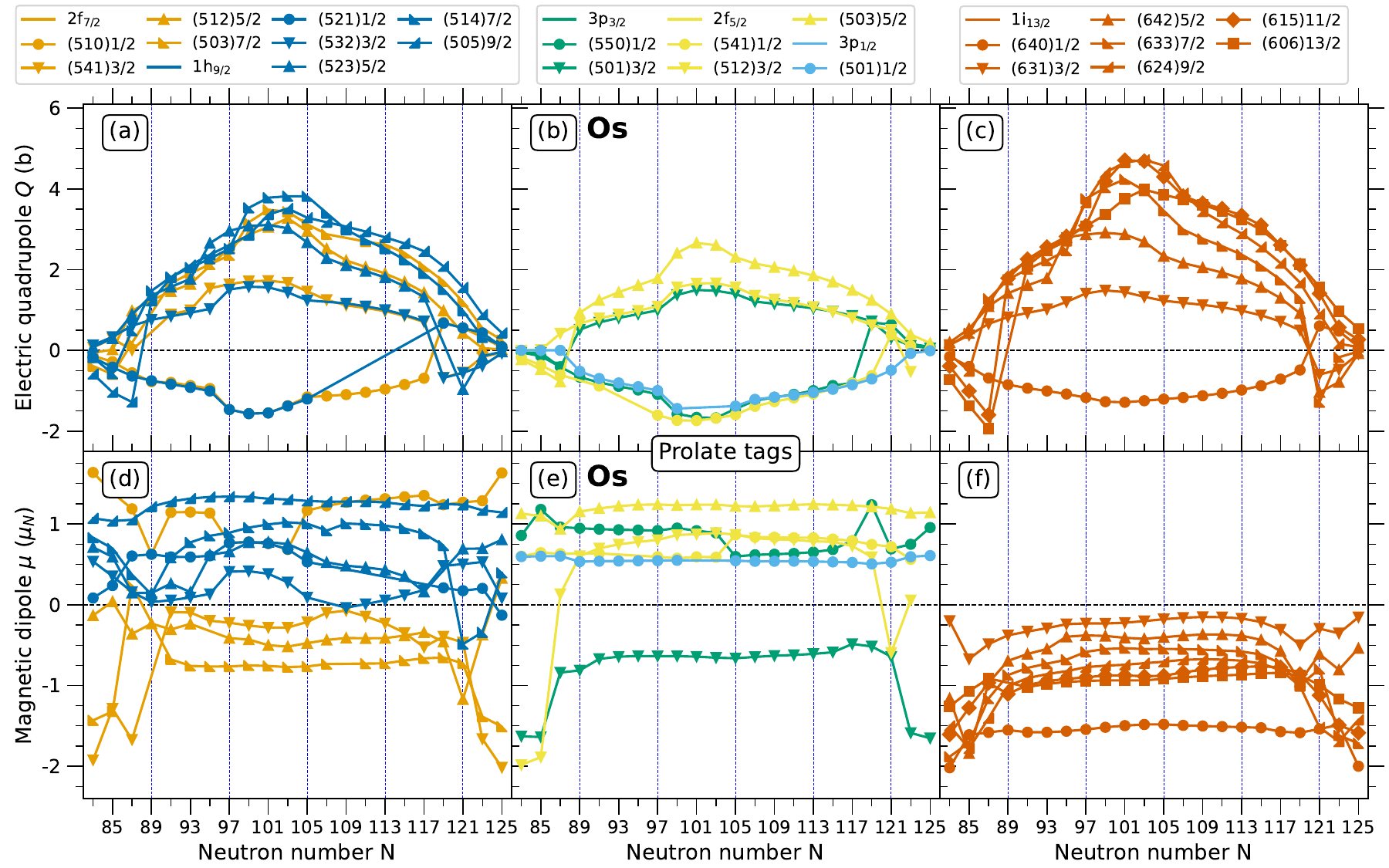}\vspace*{-1mm}
\end{center}
\caption{Same as in Fig.~\protect\ref{fig:Q2,mu_Dy_prol_2} but for the osmium isotopes.
}
\label{fig:Q2,mu_Os_prol_2}
\begin{center}
\includegraphics[width=0.93\textwidth]{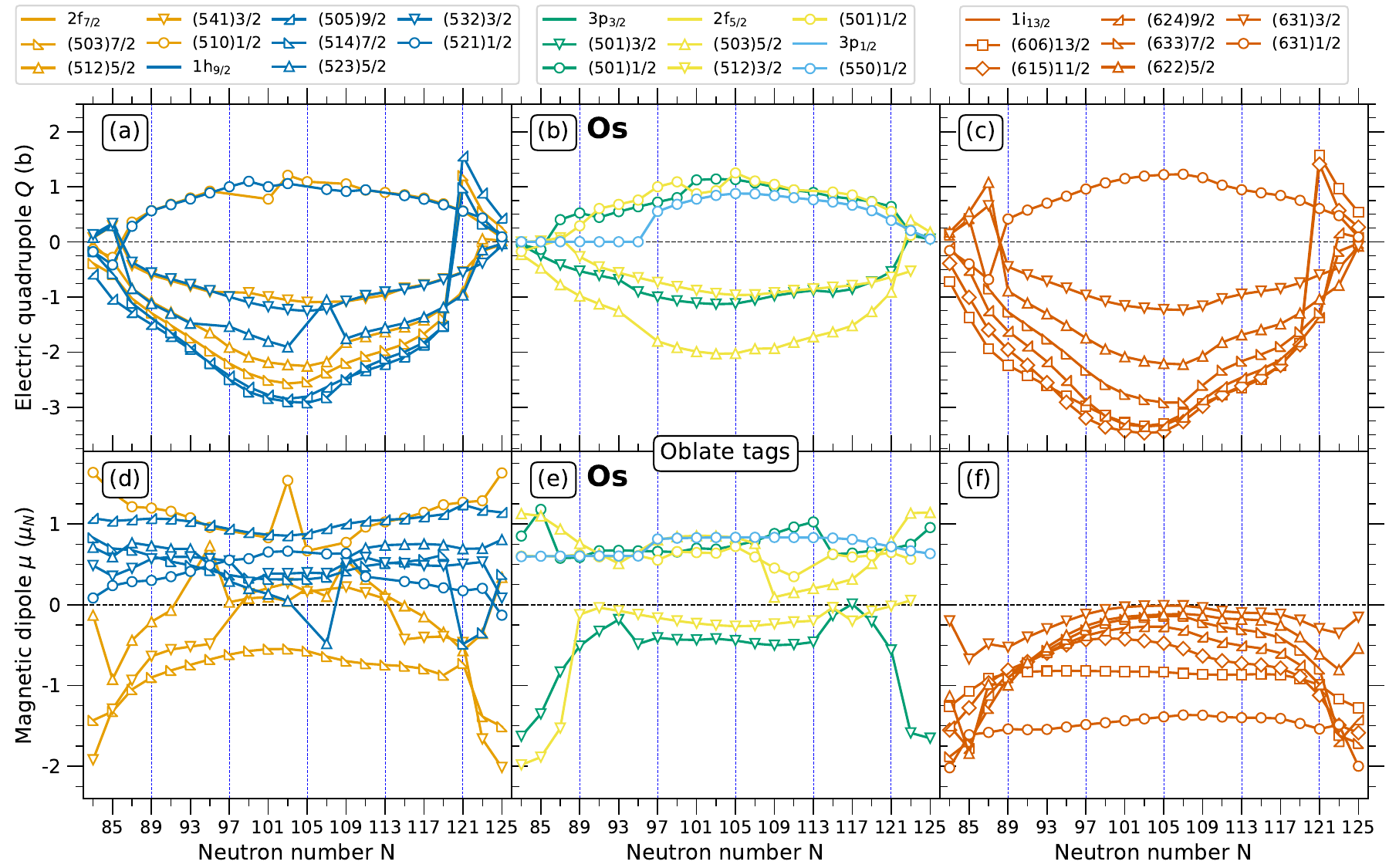}\vspace*{-1mm}
\end{center}
\caption{Same as in Fig.~\protect\ref{fig:Q2,mu_Dy_obl_2} but for the osmium isotopes.
}
\label{fig:Q2,mu_Os_obl_2}
\end{figure*}

\end{document}